\PassOptionsToPackage{hypertexnames=false}{hyperref}
\documentclass[11pt]{article}

\usepackage[margin=1in]{geometry}
\usepackage{amsmath,amssymb,amsthm,mathtools}
\usepackage{booktabs}
\usepackage{enumitem}
\usepackage{microtype}
\usepackage{algorithm}
\usepackage{algpseudocode}
\usepackage[hidelinks]{hyperref}
\usepackage[capitalise,noabbrev]{cleveref}  
\usepackage{thmtools}
\usepackage{thm-restate}
\usepackage{xcolor}
\usepackage[most]{tcolorbox} 

\usepackage{color-edits}[showdeletions]

\def\opt{\mathrm{OPT}}

\def\optmcutap{\mathrm{OPT}^{\mathrm{MC\text{-}UTAP}}}

\def\optutap{\mathrm{OPT}^{\mathrm{UTAP}}}

\def\mccap{{\mathrm{MC\text{-}CAP}}}
\def\mctap{{\mathrm{MC\text{-}TAP}}}
\def\mcutap{{\mathrm{MC\text{-}UTAP}}}
\def\mcrap{{\mathrm{MC\text{-}RAP}}}
\def\tap{\mathrm{TAP}}
\def\utap{\mathrm{UTAP}}
\newcommand{\mcraplp}{\textsf{MC-RAP-LP}}
\newcommand{\mcutaplp}{\textsf{MC-UTAP-LP}}
\newcommand{\mccaplp}{\textsf{MC-CAP-LP}}
\newcommand{\utaplp}{\textsf{UTAP-LP}}
\newcommand{\dutaplp}{\textsf{Dual-UTAP-LP}}
\newcommand{\mcecsslp}{\textsf{MC-$k$-ECSS-LP}}

\theoremstyle{plain}
\newtheorem{theorem}{Theorem}[section]
\newtheorem{lemma}[theorem]{Lemma}
\newtheorem{claim}[theorem]{Claim}

\theoremstyle{definition}
\newtheorem{definition}[theorem]{Definition}

\newtheorem{observation}[theorem]{Observation}
\newtheorem{remark}[theorem]{Remark}

\crefname{theorem}{Theorem}{Theorems}
\crefname{lemma}{Lemma}{Lemmas}
\crefname{claim}{Claim}{Claims}
\crefname{proposition}{Proposition}{Propositions}
\crefname{corollary}{Corollary}{Corollaries}
\crefname{fact}{Fact}{Facts}
\crefname{definition}{Definition}{Definitions}
\crefname{problem}{Problem}{Problems}
\crefname{remark}{Remark}{Remarks}
\crefname{observation}{Observation}{Observations}
\crefname{algorithm}{Algorithm}{Algorithms}

\newcommand{\E}{\mathbb{E}}
\newcommand{\Prb}{\mathbb{P}}
\newcommand{\QP}{\mathrm{DTIME}\bigl(n^{\mathrm{polylog}\,n}\bigr)}
\newcommand{\NP}{\mathsf{NP}}

\usepackage{utap-tikz}

\newcommand{\mcutapLPBox}[1]{
  \begin{equation*}
    \boxed{
      \begin{aligned}
        \mcutaplp\qquad
        &\begin{aligned}[t]
          \min \quad & Z \\[3pt]
          \text{s.t.}\quad
          & \sum_{a: e \in P(a)} x_a \geq 1
            && \forall e \in E(T) \\
          & \sum_{a \in L^{\uparrow}} x_a c_i(a) \leq Z
            && \forall i \in [\ell] \\
          & Z, x \geq 0
        \end{aligned}
      \end{aligned}
    }
  \end{equation*}
}

\newcommand{\utapLPBox}[1]{
  \begin{equation*}
\boxed{
\begin{aligned}
\utaplp\qquad
&\begin{aligned}[t]
     \min \quad & \sum_{a \in L^{\uparrow}} x_a c(a) \\
     \text{s.t.}\quad
     &\sum_{a: e \in P(a)} x_a \geq 1 && \forall e \in E(T) \\
     &x \geq 0
\end{aligned}
\end{aligned}
}
\end{equation*}
}

\newcommand{\utapDualLPBox}[1]{
  \begin{equation*}
\boxed{
\begin{aligned}
\dutaplp\qquad
\max \quad & \sum_{e\in E(T)} y_e\\
\text{s.t.}\quad
&\sum_{e\in P(a)}y_e\leq c(a)
&&\forall a\in L^{\uparrow},\\
&y_e\geq 0
&&\forall e\in E(T).
\end{aligned}
}
\end{equation*}
}

\newtcolorbox{introresultbox}{
  enhanced,
  breakable,
  colback=gray!6,
  colframe=gray!38,
  boxrule=0.55pt,
  arc=2.5mm,
  boxsep=0pt,
  left=11pt,
  right=11pt,
  top=9pt,
  bottom=9pt,
  before skip=10pt,
  after skip=10pt,
  drop fuzzy shadow=black!9
}

\newcommand{\mccapLPBox}[1]{
  \begin{equation*}
\boxed{
\begin{aligned}
\mccaplp \qquad \min\quad & Z\\
\text{s.t.}\quad
& \sum_{a\in\delta_L(S)}x_a\ge1
&& \forall S\in\mathcal C,\\
& \sum_{a\in L}c_i(a)x_a\le Z
&& \forall i\in[\ell],\\
& Z , x\ge0.
\end{aligned}
}
\end{equation*}
}

\newcommand{\mcecssLPBox}[1]{
\begin{equation*}
\boxed{
\begin{aligned}
\mcecsslp \qquad  \min\quad & Z\\
\text{s.t.}\quad
& \sum_{e\in\delta_G(S)}x_e\ge k
&& \forall\varnothing\ne S\subsetneq V,\\
& \sum_{e\in E}c_i(e)x_e\le Z
&& \forall i\in[\ell],\\
& 0\le x_e\le1
&& \forall e\in E,\\
& Z\ge0.
\end{aligned}
}
\end{equation*}
}

\title{Approximation Algorithms for Multi-Criteria Network Design Problems}
\author{%
  Mehrshad Taziki\thanks{\texttt{mtaziki@ethz.ch}}\\
  {\normalsize ETH Z\"urich}
  \and
  Ali Vakilian\thanks{\texttt{a.vakilian@rutgers.edu}}\\
  {\normalsize Rutgers University}
}
\date{}

\begin{document}

\maketitle

\begin{abstract}
Min-max multi-criteria optimization asks for a single feasible
solution minimizing the maximum of its costs under $\ell$ nonnegative cost functions, a formulation that captures both fairness across groups and robustness across scenarios. It has proved hard to approximate even for basic problems: for robust shortest paths, polylogarithmic approximations are known only in quasi-polynomial time. For network design, the best known guarantee is an $\ell$-approximation, which is tight for the natural LP relaxation.

We show that exponentially better and nearly optimal guarantees are achievable for two fundamental network design problems. For multi-criteria connectivity augmentation (MC-CAP), which asks to increase the edge connectivity of a given undirected graph by one, we give a randomized polynomial-time $O(\log \ell)$-approximation algorithm. Applying our augmentation procedure iteratively and analyzing its randomized choices jointly across iterations yields an $O(\log k \log \ell)$-approximation for the multi-criteria $k$-edge-connected spanning subgraph (MC-$k$-ECSS) problem, for every $k \geq 2$. We complement these algorithms with a nearly matching lower bound: for every fixed $\varepsilon > 0$, MC-CAP admits no polynomial-time $O(\log^{1-\varepsilon} \ell)$-approximation, even when the underlying graph is a star, unless $\NP\subseteq\mathsf{DTIME}\bigl(n^{\log^{O(1)}n}\bigr)$. The same hardness extends to the MC-$k$-ECSS problem.

\end{abstract}

\thispagestyle{empty}
\newpage
\setcounter{page}{1}
\section{Introduction}
\label{sec:introduction}

A central theme in combinatorial optimization is to select a single feasible object, such as a path, a spanning tree, a clustering, or a network, that minimizes its cost. 
In many settings, several stakeholders may disagree about what constitutes a good solution; moreover, in optimization under uncertainty, the cost of a fixed decision depends on which future scenario materializes. A network may be evaluated by several stakeholders with different priorities; a clustering may need to serve several demographic groups; and a path may incur different costs under different scenarios; for example, travel times can differ across hours of the day and days of the week as traffic patterns change. Optimizing an aggregate objective can conceal a poor outcome for one of these criteria. Likewise, optimizing for a single predicted scenario can leave the chosen solution vulnerable when conditions differ from the prediction. This motivates the \emph{multi-criteria} or \emph{min-max} formulation: given a feasible family $\mathcal F$ and nonnegative cost functions $c_1,\ldots,c_\ell$, find
\begin{equation}
    \label{eq:intro-minmax}
    \min_{F\in\mathcal F}\;\max_{i\in[\ell]}c_i(F).
\end{equation}
For additive costs, $c_i(F)=\sum_{e\in F}c_i(e)$, this asks for one feasible combinatorial solution whose worst cost across the $\ell$ criteria is as small as possible. Depending on the application, the criteria can represent agents, groups, objectives, or possible scenarios. The common feature is that the solution must be shared: optimizing separately for each criterion is not an option.

The min-max perspective has led to a growing body of work on otherwise classical optimization problems. Examples include socially fair clustering, where one limits the maximum clustering cost among groups~\cite{AbbasiBhaskaraVenkatasubramanian2021,GhadiriSamadiVempala2021,MakarychevVakilian2021,ChlamtacMakarychevVakilian2022,anand_makarychev_shan} including constant-factor algorithms with bicriteria or parameter-dependent guarantees~\cite{GhadiriSinghVempala2022}, clustering in varying metrics, where the same centers are evaluated under multiple distance functions~\cite{ChakrabartyConroySarkar2025}, fair facility location with group-sensitive objectives~\cite{GuptaMoondraSingh2023}, and shortest paths with vector-valued edge costs, where the goal is to control the cost of a single path across scenarios~\cite{LiXuZhang2024,MakarychevOvsiankinTani2024,CarlsonMakarychevMosenzon2026}. Recent work has also investigated multi-criteria matroid bases from a parameterized-complexity perspective~\cite{MittalMakarychevHajiaghayi2027}. These settings illustrate a recurring algorithmic difficulty: a combinatorial structure that admits efficient optimization, or a strong approximation guarantee, under one linear cost function can become substantially harder when evaluated by the maximum of many such functions.

Network design is a natural setting for this phenomenon. Infrastructure shared by several users must satisfy global connectivity requirements, yet users may assign very different costs to its edges. For a single criterion, survivable network design, connectivity augmentation, and $k$-edge-connected spanning subgraph have been central problems in approximation algorithms, with powerful techniques based on cut relaxations, iterative rounding, and structural graph representations~\cite{Jain2001,CTZ21,TraubZenklusenWeightedCAP, hershkowitz2024ghost, cohen2026strong}.  In the multi-criteria setting, Laddha, Singh, and Vempala~\cite{LADDHA2022536FairNetworkDesign} gave an $\ell$-approximation for generalized Steiner network design and showed that the factor $\ell$ is tight for the natural LP relaxation. This raises a basic question: 
\begin{quote}
    \emph{Can one exploit the additional structure of connectivity problems to obtain an approximation guarantee that depends only logarithmically on the number of criteria, despite the linear integrality gap of the standard relaxation?}
\end{quote}

We answer this question affirmatively for the connectivity augmentation problem, and use the resulting algorithm to obtain multi-criteria guarantees for higher edge connectivity.

\subsection{Our Results}
\label{sec:intro-results}

In the \emph{multi-criteria connectivity augmentation problem} ($\mccap$), we are given an undirected graph $G=(V,E)$, a set of available links $L$, and $\ell$ nonnegative link-cost functions $c_1,\ldots,c_\ell$. The goal is to select a set $A\subseteq L$ that increases the edge connectivity of $G$ by one, while minimizing
\[
    \mathrm{cost}(A):=\max_{i\in[\ell]}\sum_{a\in A}c_i(a).
\]
Equivalently, if $\mathcal C$ denotes the family of minimum cuts of $G$, the selected links must cross every cut in $\mathcal C$. This problem captures a basic step in designing robust networks as it enables augmenting the connectivity of an existing network without allowing any one cost criterion to become disproportionately expensive.

Our first result (\cref{thm:mc-cap}) is a logarithmic approximation for $\mccap$.
\begin{introresultbox}
\textbf{Result 1 (Multi-criteria connectivity augmentation; \cref{thm:mc-cap}).}
There is a randomized polynomial-time $O(\log\ell)$-approximation algorithm for $\mccap$, with the guarantee holding in expectation.
\end{introresultbox}

The dependence on $\ell$ is nearly best possible under standard complexity assumptions. We establish an approximation-preserving reduction from the robust representatives selection problem~\cite{Deineko2013ComplexityAI,kasperski2015approximability}, giving the following hardness (\cref{thm:hardness-mccap}), even for very simple graphs.

\begin{introresultbox}
\textbf{Result 2 (Nearly tight hardness; \cref{thm:hardness-mccap}).}
For every fixed $\varepsilon>0$, $\mccap$ admits no polynomial-time $O(\log^{1-\varepsilon}\ell)$-approximation, even when $G$ is a star, unless every problem in $\NP$ has a deterministic quasi-polynomial-time algorithm. 
\end{introresultbox}

We next consider the \emph{multi-criteria $k$-edge-connected spanning subgraph problem} (MC-$k$-ECSS). Here, the input is an edge-weighted multigraph with $\ell$ cost functions, and the task is to find a spanning subgraph that remains connected after deleting any $k-1$ edges, while minimizing the maximum cost across the criteria. Applying our augmentation algorithm iteratively, together with a joint analysis of its randomized decisions, gives the following result (\cref{thm:k-ecss}).

\begin{introresultbox}
\textbf{Result 3 (Multi-criteria $k$-edge connectivity; \cref{thm:k-ecss}).}
For every $k\geq2$, there is a randomized polynomial-time $O(\log k\log\ell)$-approximation algorithm for MC-$k$-ECSS, with the guarantee holding in expectation.
\end{introresultbox}

In particular, for every fixed $k\geq2$, the guarantee is $O(\log\ell)$. The hardness from Result~2 transfers to MC-$2$-ECSS by assigning zero cost to the edges of the underlying tree, showing that the logarithmic dependence on the number of criteria is essentially unavoidable already for two-edge connectivity. The same holds for any constant $k$ too.

\subsection{Related Work}
\label{sec:intro-related}

\paragraph{Multi-criteria combinatorial optimization.}
Min--max optimization captures both robustness across scenarios and fairness across groups. Socially fair clustering was independently introduced by Abbasi, Bhaskara, and Venkatasubramanian~\cite{AbbasiBhaskaraVenkatasubramanian2021} and Ghadiri, Samadi, and Vempala~\cite{GhadiriSamadiVempala2021}. Subsequent work has investigated nearly tight polynomial-time approximation bounds~\cite{MakarychevVakilian2021}, tight FPT approximations~\cite{GoyalJaiswal2023,anand_makarychev_shan}, bicriteria algorithms~\cite{GhadiriSinghVempala2022}, extensions to general cascaded norm objectives~\cite{ChlamtacMakarychevVakilian2022}, and portfolios of solutions for different fairness measures~\cite{GuptaMoondraSingh2023,GuptaMoondraSingh2025}. Unlike portfolio-based approaches, our setting seeks a single solution that simultaneously controls all criteria. A related multi-criteria formulation is clustering in varying metrics~\cite{ChakrabartyConroySarkar2025}, which evaluates the same centers under multiple distance metrics rather than across client groups. It admits constant-factor approximations for two general metrics, whereas for three or more general metrics no finite-factor polynomial-time approximation is possible.

For paths and other network structures with vector costs, Li, Xu, and Zhang~\cite{LiXuZhang2024} obtained polylogarithmic approximations for robust $s$--$t$ paths in quasi-polynomial time. More precisely, their factor is $O(\log n\log\ell)$ for an $n$-vertex graph with $\ell$ scenarios; for bounded-treewidth graphs, their algorithm runs in polynomial time. They also show that an $O(\log^{1-\varepsilon}\ell)$-approximation for any $\varepsilon > 0$ in polynomial time would imply $\NP\subseteq\QP$. Makarychev, Ovsiankin, and Tani~\cite{MakarychevOvsiankinTani2024} gave polylogarithmic approximations for $\ell_p$-shortest paths and related vector-cost network problems, in quasi-polynomial time on general graphs and polynomial time on series--parallel graphs. Carlson, Makarychev, and Mosenzon~\cite{CarlsonMakarychevMosenzon2026} established nearly matching hardness bounds for shortest paths, including $\widetilde{\Omega}(\log^2 n)$ hardness for the $\ell_\infty$ objective. From a different angle, Chekuri, Vondr\'ak, and Zenklusen~\cite{VondrakChekuriZenklusenMinMax} developed dependent-rounding tools for matroid polytopes, including min-max packing applications.

More recently, Mittal, Makarychev, and Hajiaghayi~\cite{MittalMakarychevHajiaghayi2027} studied multi-criteria matroid bases through fixed-parameter algorithms and lower bounds. The present paper concerns the same broad min-max objective, but exploits the cut structure of connectivity augmentation rather than matroid exchange or path decompositions.

Multi-criteria optimization is often substantially more challenging than its classical single-criterion counterpart, both in terms of computational complexity and approximability.

\paragraph{Connectivity augmentation and survivable network design.} 
Network design is a central area of combinatorial optimization and graph algorithms. In the classical setting with a single cost function, Jain's seminal iterative-rounding algorithm~\cite{Jain2001} provides a factor-$2$ approximation for the general survivable network design problem.

Connectivity augmentation and the $k$-edge-connected spanning subgraph problem are two fundamental problems in this class. In particular, the unweighted connectivity augmentation problem has been the subject of an extensive line of research~\cite{kortsarz2015simplified, AdjiashviliTAP2017, cheriyan2018approximating,grandoni2018improved,byrka2020breaching,CTZ21,traub2022better,traub2022local,TraubZenklusenWeightedCAP,cohen2026strong}, culminating in the state-of-the-art $1.393$-approximation algorithm of Cecchetto, Traub, and Zenklusen~\cite{CTZ21}.
For the weighted setting, the long-standing factor-$2$ approximation barrier was first broken by Traub and Zenklusen~\cite{TraubZenklusenWeightedCAP,traub2022local}, who developed a local-search-based algorithm achieving an approximation ratio of $(1.5+\varepsilon)$. This bound was recently improved for the weighted tree augmentation problem to a randomized $1.49$-approximation by Cohen-Addad, Drygala, Klein, and Svensson~\cite{cohen2026strong}, using a stronger LP relaxation inspired by the Sherali--Adams hierarchy.


Connectivity augmentation has also benefited from reductions through cactus representations of minimum cuts~\cite{DKL76}; see, for example,~\cite{CTZ21, TraubZenklusenWeightedCAP}. In the multi-criteria setting, Laddha, Singh, and Vempala~\cite{LADDHA2022536FairNetworkDesign} obtained an $\ell$-approximation for the survivable network design problem via iterative rounding and demonstrated a matching integrality gap for its natural LP. 


\section{Our Techniques}
\label{sec:our-techniques}

Our algorithm for $\mccap$ combines a structural reduction with a randomized rounding procedure tailored to the multi-criteria objective. We first reduce the rounding problem for connectivity augmentation in a general graph to rounding two fractional uplink tree augmentation instances, with only a constant-factor increase in fractional cost under each criterion. For these instances, we apply independent randomized rounding followed by a completion step. A characterization of optimal dual solutions for single-criterion uplink tree augmentation allows us to show that the expected completion cost decays exponentially with the sampling rate. Choosing a rate of $\Theta(\log \ell)$ offsets the factor of $\ell$ incurred by aggregating the cost functions when computing the completion.

Although the canonical multi-criteria LP has integrality gap $\Theta(\ell)$, we obtain a logarithmic approximation by proving a rounding bound that depends on both the fractional objective value and the maximum individual link cost. We then guess the largest link cost in an optimal solution and restrict the available links accordingly. For the correct guess, both quantities are bounded by the integral optimum, yielding an $O(\log \ell)$-approximation for $\mccap$.

We also extend this framework to MC-$k$-ECSS by applying our reduction and rounding procedure over successive connectivity augmentation iterations. Here, analyzing the steps separately would charge the maximum-link-cost term repeatedly. A joint analysis of the sampling decisions across all steps avoids this loss and yields an $O(\log k\log\ell)$-approximation. We describe these ideas below.

\paragraph{Reducing connectivity augmentation to uplink tree augmentation (\cref{sec:reduction}).}
Let $\mathcal C$ be the family of minimum cuts of $G$, and let $\delta_L(S)$ denote the links in $L$ with exactly one endpoint in $S$. Increasing the edge connectivity of $G$ by one is equivalent to covering every cut in $\mathcal C$. The canonical LP relaxation of $\mccap$ is therefore:

\mccapLPBox

Given a feasible fractional solution $x$ of value $Z$, we replace
these cut constraints by the edge-covering constraints of two trees, losing only a constant factor. We represent every minimum cut of $G$ by a tree edge in one of the two trees, and map each link to one uplink in each tree. The mapping ensures that an uplink covers the tree edge representing a cut only if its original link covers that cut. 
Thus, mapping solutions of the two tree instances back to $G$ and
taking their union gives a feasible augmentation. The converse can
fail, since a link may cover a cut without its uplink covering the
corresponding tree edge, but a constant multiple of $x$ still
fractionally covers every tree edge; this scaling accounts for the
constant-factor loss in the approximation guarantee.

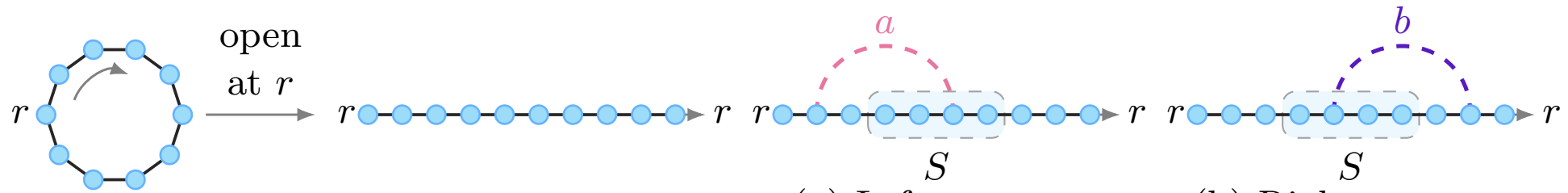
\begin{figure}[!htbp]
\centering
\begingroup

\definecolor{mcoVertexFill}{rgb}{0.61338,0.86345,0.97537}
\definecolor{mcoVertexDraw}{rgb}{0.4,0.7,1.0}
\definecolor{mcoInk}{rgb}{0.15,0.15,0.15}
\definecolor{mcoMagenta}{rgb}{0.91472,0.46078,0.62707}
\definecolor{mcoPurple}{rgb}{0.34749,0.1,0.75250}
\definecolor{mcoPale}{rgb}{0.92587,0.97171,0.99509}

\tikzset{
  mco/picture/.style={
    x=1cm,
    y=1cm,
    font=\footnotesize,
    every node/.style={inner sep=1.5pt}
  },
  mco/edge/.style={
    draw=mcoInk,
    line width=.7pt
  },
  mco/vertex/.style={
    circle,
    draw=mcoVertexDraw,
    fill=mcoVertexFill,
    line width=.5pt,
    minimum size=4.3pt,
    inner sep=0pt,
    outer sep=0pt
  },
  mco/pink/.style={
    draw=mcoMagenta,
    line width=.95pt,
    dashed
  },
  mco/purple/.style={
    draw=mcoPurple,
    line width=.95pt,
    dashed
  },
  mco/region/.style={
    draw=black!36,
    fill=mcoPale,
    dashed,
    line width=.4pt,
    rounded corners=4pt
  },
  mco/arrow/.style={
    draw=black!50,
    line width=.55pt,
    -{Latex[length=1.5mm]}
  },
  mco/title/.style={
    align=center,
    inner sep=0pt
  }
}

\long\def\mcoFit#1{%
  \sbox0{#1}%
  \ifdim\wd0>\linewidth
    \resizebox{\linewidth}{!}{\usebox0}%
  \else
    \usebox0%
  \fi
}

\mcoFit{%
\begin{tikzpicture}[mco/picture]
  \def\step{.28}

  \foreach \j in {0,...,9} {
    \coordinate (r\j) at
      ({.82+.56*cos(180-36*\j)},
       {.56*sin(180-36*\j)});
  }

  \draw[mco/edge]
    (r0)--(r1)--(r2)--(r3)--(r4)--(r5)
    --(r6)--(r7)--(r8)--(r9)--cycle;

  \foreach \j in {0,...,9} {
    \node[mco/vertex] at (r\j) {};
  }

  \node[left=2pt] at (r0) {$r$};

  \draw[mco/arrow]
    ({.82+.35*cos(160)},{.35*sin(160)})
    arc[start angle=160,end angle=70,radius=.35];

  \draw[mco/arrow]
    (1.57,0)--(2.47,0)
    node[midway,above=3pt,align=center] {open\\at $r$};

  \node[mco/title] at (.82,-.76) {};

  \foreach \tag/\dx/\iscut in {B/2.90/0,C/6.30/1,D/9.70/1} {
    \begin{scope}[xshift=\dx cm]

      \foreach \i in {0,...,9} {
        \coordinate (\tag\i) at ({\i*\step},0);
      }

      \ifnum\iscut=1
        \draw[mco/region,rounded corners=3pt]
          ({2.5*\step},-.19)
          rectangle
          ({6.5*\step},.19);

        \node at ({4.5*\step},-.42) {$S$};
      \fi

      \draw[mco/edge] (\tag0)--(\tag9);

      \draw[mco/arrow,densely dashed]
        (\tag9)--++(.25,0)
        node[right=0pt] {$r$};

      \foreach \i in {0,...,9} {
        \node[mco/vertex] at (\tag\i) {};
      }

      \node[left=1pt] at (\tag0) {$r$};

    \end{scope}
  }

  \node[mco/title] at (4.18,-.76)
    {};

  \draw[mco/pink]
    (C1) .. controls +(0,.75) and +(0,.75) .. (C5)
    node[midway,above=1pt,text=mcoMagenta] {$a$};

  \draw[mco/purple]
    (D4) .. controls +(0,.75) and +(0,.75) .. (D8)
    node[midway,above=1pt,text=mcoPurple] {$b$};

  \foreach \tag/\i in {C/1,C/5,D/4,D/8} {
    \node[mco/vertex] at (\tag\i) {};
  }

  \node[mco/title] at (7.58,-.76)
    {(a) Left coverage};

  \node[mco/title] at (10.98,-.76)
    {(b) Right coverage};

\end{tikzpicture}%
}

\endgroup
\caption{Opening the ring at $r$ gives a vertex order in which each
relevant cut is an interval $S$. The pink link $a$ covers $S$ from
the left, and the purple link $b$ covers it from the right.}
\label{fig:techniques-left-right}
\end{figure}

We begin by using the cactus representation of minimum cuts to transform the instance into a {\em ring augmentation} problem with a specified family of cuts \cite{DKL76,CTZ21}. Fixing a root and opening the ring at this root makes every relevant cut an interval. Each link covering an interval does so from either the left or the right. Since the total fractional coverage is at least one, at least one of these directions contributes $x$-mass at least $1/2$ (see \cref{fig:techniques-left-right}).

\begin{figure}[!htbp]
\centering
\begingroup

\definecolor{mcoVertexFill}{rgb}{0.61338,0.86345,0.97537}
\definecolor{mcoVertexDraw}{rgb}{0.4,0.7,1.0}
\definecolor{mcoInk}{rgb}{0.15,0.15,0.15}

\tikzset{
  mco/edge/.style={
    draw=mcoInk,
    line width=.7pt
  },
  mco/vertex/.style={
    circle,
    draw=mcoVertexDraw,
    fill=mcoVertexFill,
    line width=.5pt,
    minimum size=4.3pt,
    inner sep=0pt,
    outer sep=0pt
  },
  mco/blocklabel/.style={
    fill=white,
    inner sep=1pt
  },
  mco/sidenote/.style={
    text=black!60
  }
}

\long\def\mcoFit#1{%
  \sbox0{#1}%
  \ifdim\wd0>\linewidth
    \resizebox{\linewidth}{!}{\usebox0}%
  \else
    \usebox0%
  \fi
}

\mcoFit{%
\begin{tikzpicture}[
  x=1cm,
  y=.34cm, 
  font=\footnotesize,
  every node/.style={inner sep=1.5pt}
]

\node[mco/sidenote] at (-5.70,1.1) {layer $i$};
\node[mco/sidenote] at (-4.45,1.1) {inserted};

\foreach \i in {0,...,12} {
  \node[mco/sidenote] at (-5.70,-\i) {$\i$};

  \draw[draw=black!11,line width=.3pt,densely dotted]
    (-3.75,-\i)--(3.75,-\i);
}

\foreach \i in {1,...,12} {
  \node[mco/sidenote] at (-4.45,-\i) {$a_{\i}$};
}

\node[mco/vertex] (v01)  at ( 0,  0) {};
\node[mco/vertex] (v11)  at ( 0, -1) {};
\node[mco/vertex] (v21)  at ( 0, -2) {};
\node[mco/vertex] (v31)  at ( 0, -3) {};
\node[mco/vertex] (v41)  at ( 0, -4) {};

\node[mco/vertex] (v51)  at (-2, -5) {};
\node[mco/vertex] (v52)  at ( 2, -5) {};
\node[mco/vertex] (v61)  at (-2, -6) {};
\node[mco/vertex] (v62)  at ( 2, -6) {};
\node[mco/vertex] (v71)  at (-2, -7) {};
\node[mco/vertex] (v72)  at ( 2, -7) {};

\node[mco/vertex] (v81)  at (-2, -8) {};
\node[mco/vertex] (v82)  at ( 1, -8) {};
\node[mco/vertex] (v83)  at ( 3, -8) {};
\node[mco/vertex] (v91)  at (-2, -9) {};
\node[mco/vertex] (v92)  at ( 1, -9) {};
\node[mco/vertex] (v93)  at ( 3, -9) {};

\node[mco/vertex] (v101) at (-3,-10) {};
\node[mco/vertex] (v102) at (-1,-10) {};
\node[mco/vertex] (v103) at ( 1,-10) {};
\node[mco/vertex] (v104) at ( 3,-10) {};

\node[mco/vertex] (v111) at (-3,-11) {};
\node[mco/vertex] (v112) at (-1,-11) {};
\node[mco/vertex] (v113) at ( 1,-11) {};
\node[mco/vertex] (v114) at ( 3,-11) {};

\node[mco/vertex] (v121) at (-3,-12) {};
\node[mco/vertex] (v122) at (-1,-12) {};
\node[mco/vertex] (v123) at ( 1,-12) {};
\node[mco/vertex] (v124) at ( 3,-12) {};

\draw[mco/edge] (v01)--(v11);
\draw[mco/edge] (v11)--(v21);
\draw[mco/edge] (v21)--(v31);
\draw[mco/edge] (v31)--(v41);

\draw[mco/edge] (v41)--(v51);
\draw[mco/edge] (v41)--(v52);

\draw[mco/edge] (v51)--(v61);
\draw[mco/edge] (v52)--(v62);
\draw[mco/edge] (v61)--(v71);
\draw[mco/edge] (v62)--(v72);

\draw[mco/edge] (v71)--(v81);
\draw[mco/edge] (v72)--(v82);
\draw[mco/edge] (v72)--(v83);

\draw[mco/edge] (v81)--(v91);
\draw[mco/edge] (v82)--(v92);
\draw[mco/edge] (v83)--(v93);

\draw[mco/edge] (v91)--(v101);
\draw[mco/edge] (v91)--(v102);
\draw[mco/edge] (v92)--(v103);
\draw[mco/edge] (v93)--(v104);

\draw[mco/edge] (v101)--(v111);
\draw[mco/edge] (v102)--(v112);
\draw[mco/edge] (v103)--(v113);
\draw[mco/edge] (v104)--(v114);

\draw[mco/edge] (v111)--(v121);
\draw[mco/edge] (v112)--(v122);
\draw[mco/edge] (v113)--(v123);
\draw[mco/edge] (v114)--(v124);

\node[above=3pt] at (v01) {$\varnothing$};

\node[
  mco/blocklabel,
  right=8pt,
  yshift=4pt,
  inner sep=0pt
] at (v41) {$\{a_2,a_1,a_4,a_3\}$};

\node[mco/blocklabel,left=5pt]
  at (v51) {$\{a_2,a_5\}$};

\node[mco/blocklabel,right=5pt]
  at (v52) {$\{a_1,a_4,a_3\}$};

\node[below=5pt] at (v121)
  {$\{a_7,a_{11},a_2\}$};

\node[below=5pt] at (v122)
  {$\{a_9,a_5,a_{10}\}$};

\node[below=5pt] at (v123)
  {$\{a_1,a_{12},a_8\}$};

\node[below=5pt] at (v124)
  {$\{a_4,a_3,a_6\}$};

\node[
  anchor=north west,
  align=left,
  fill=white,
  inner sep=3pt,
  text=black!75
] at (1.15,1.1) {
};

\end{tikzpicture}%
}

\endgroup
\caption{The tree records the partition history as
$a_1,\ldots,a_{12}$ are inserted. Each layer has one vertex per block.
A block has one child if it continues without splitting, and two
children if it splits. On inserting $a_5$, the layer-$4$ block
$\{a_2,a_1,a_4,a_3\}$ gains $a_5$ and splits into
$\{a_2,a_5\}$ and $\{a_1,a_4,a_3\}$.}
\label{fig:techniques-partition-tree}
\end{figure}
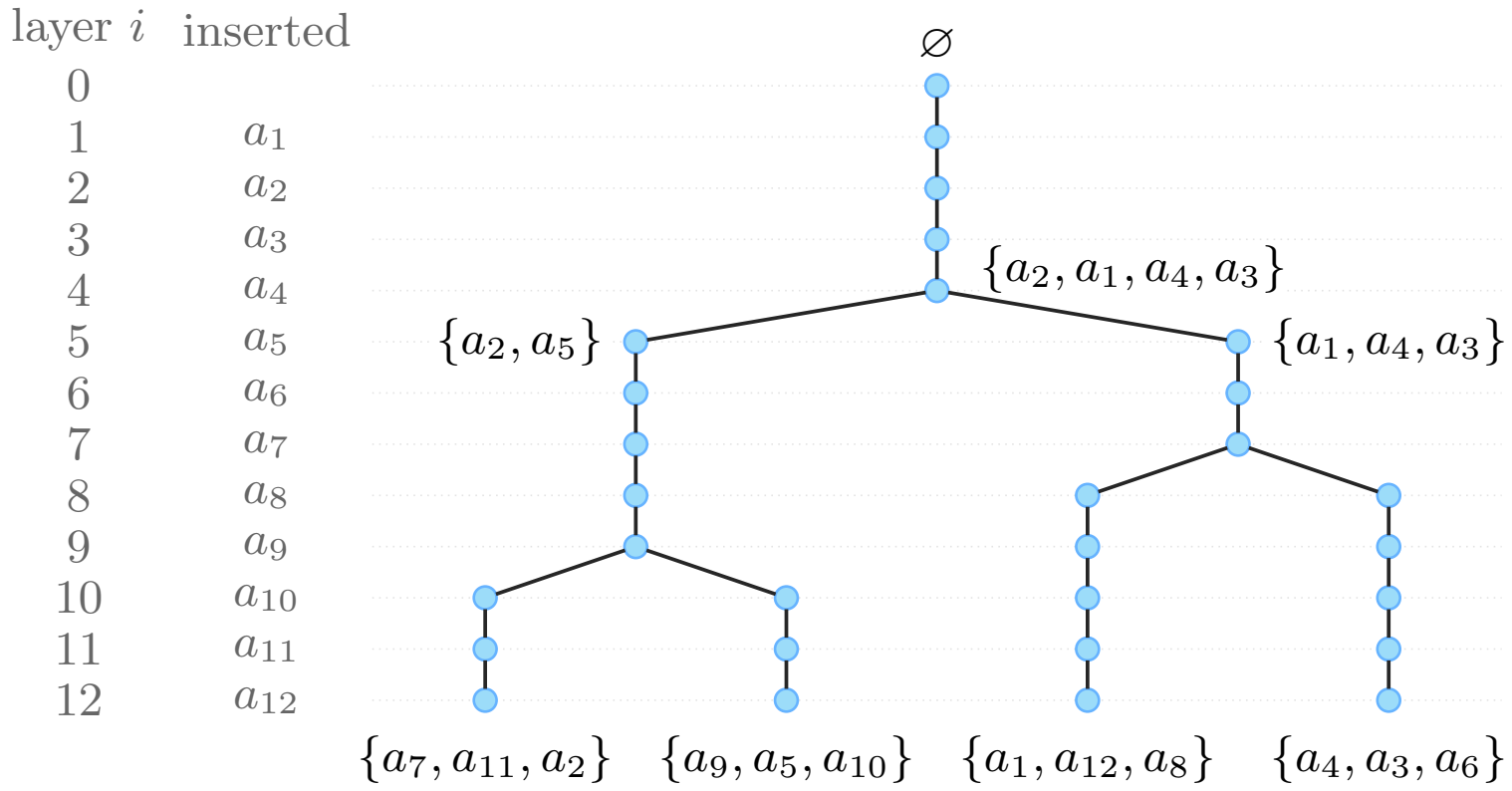

We assign each cut to one such direction and construct a tree for each of the two resulting families. Consider the cuts assigned to the left. We process links in the order of their left endpoints, maintaining a partition of the links seen so far into consecutive blocks in the order of their right endpoints. Whenever a block becomes too large in fractional mass, we split it into two consecutive blocks. Apart from inserting a new link at each iteration, the partitions evolve only by splitting blocks. This gives the blocks an \emph{almost laminar} structure, which we use to encode their history as a rooted tree (see \cref{fig:techniques-partition-tree}).

The blocks are constructed so that, for every cut assigned to the left, there is an entire block consisting of links that cover that cut and carries a constant amount of fractional mass. The tree edge associated with this block represents the cut. The construction for cuts assigned to the right is symmetric.
More precisely, we first split each link into parallel copies with
equal fractional values, so that no copy has a large fractional value, and continue to write $L$ and $x$ for the resulting multiset and its fractional solution. For a cut $S$ represented by the tree edge $e_S$, let $I(S)$ be the set of links whose uplinks cover $e_S$. The construction guarantees that
\begin{equation}
\label{eq:tech-cutmap}
    I(S)\subseteq\delta_L(S),
\qquad
x(I(S))\ge\frac{1}{10}.
\end{equation}
Several cuts may be represented by the same tree edge, while each tree has a one-to-one correspondence between its uplinks and the original multiset of links. For each link $a\in L$, we assign its corresponding uplink the fractional value $\min\{1,10x_a\}$ and leave all its costs unchanged. By~\eqref{eq:tech-cutmap}, both resulting fractional solutions are feasible. Since costs are preserved, each solution has cost at most $10Z$ under every objective. Mapping the integral solutions back to $G$ and taking their union then yields a feasible connectivity augmentation for $G$.

\paragraph{Why completion is inexpensive: the dual structure of $\utap$ (\cref{sec:dual-prop}).}
In a rooted tree $T$, an uplink $a$ joins a vertex to one of
its ancestors and covers the tree path $P(a)$ between them. Given a set $L^\uparrow$ of uplinks with nonnegative costs $c$, $\utap$ asks for a cheapest set of uplinks covering every edge of $T$. Its canonical LP and its dual are as follows.

\noindent
\begin{minipage}[t]{0.49\linewidth}
\centering
\small
\begin{equation*}
\boxed{
\begin{gathered}
\utaplp \\[3pt]
\begin{aligned}
\min \quad & \sum_{a \in L^\uparrow} x_a c(a) \\
\text{s.t.}\quad
& \sum_{a:e\in P(a)} x_a \geq 1
&& \forall e \in E(T) \\
& x \geq 0
\end{aligned}
\end{gathered}
}
\end{equation*}
\end{minipage}%
\hfill
\begin{minipage}[t]{0.49\linewidth}
\centering
\small
\begin{equation*}
\boxed{
\begin{gathered}
\dutaplp \\[3pt]
\begin{aligned}
\max \quad & \sum_{e\in E(T)} y_e \\
\text{s.t.}\quad
& \sum_{e\in P(a)} y_e \leq c(a)
&& \forall a\in L^\uparrow \\
& y_e \geq 0
&& \forall e\in E(T)
\end{aligned}
\end{gathered}
}
\end{equation*}
\end{minipage}
\par\addvspace{\baselineskip}

Given a feasible solution $x$ to $\utaplp$, we sample each uplink
independently with probability $p_a=\min\{1,\gamma x_a\}$ for some
$\gamma\ge1$. Every tree edge is then left uncovered with probability
at most $e^{-\gamma}$. One would like the expected cost of completing
the sample to shrink by the same factor, but this does not follow
directly. An uplink can cover many tree edges, so completing the sample
by adding each link of a fixed solution whose path contains an
uncovered edge may add that link with probability much larger than
$e^{-\gamma}$.

Instead, we make the sampled uplinks free. The cheapest cover of $T$
under these modified costs is then exactly the cheapest completion of
the sample. Since $\utaplp$ is integral~\cite{AdjiashviliTAP2017}, this
cost equals the optimal value of $\utaplp$ under the modified costs
and, by LP duality, the optimal value of $\dutaplp$. We therefore study
how sampling changes the optimum of $\dutaplp$.

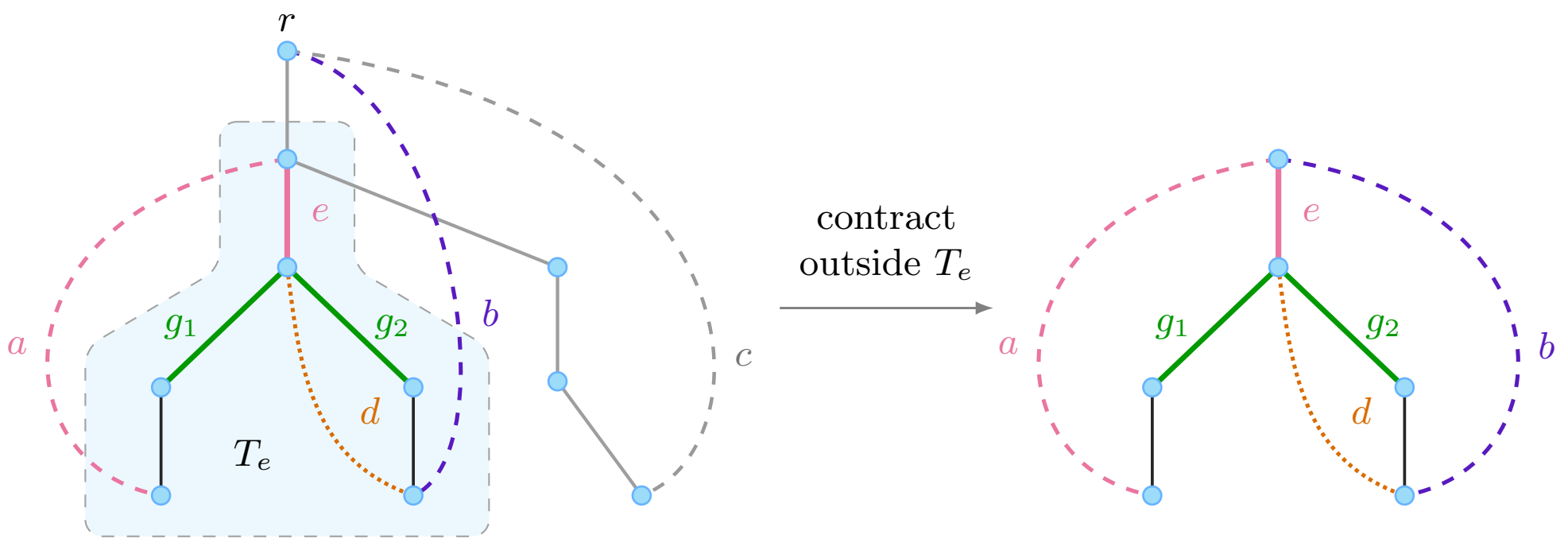
\begin{figure}[!htbp]
\centering
\begingroup

\definecolor{mcoVertexFill}{rgb}{0.61338,0.86345,0.97537}
\definecolor{mcoVertexDraw}{rgb}{0.4,0.7,1.0}
\definecolor{mcoInk}{rgb}{0.15,0.15,0.15}
\definecolor{mcoMagenta}{rgb}{0.91472,0.46078,0.62707}
\definecolor{mcoPurple}{rgb}{0.34749,0.1,0.75250}
\definecolor{mcoOrange}{rgb}{0.85,0.425,0}
\definecolor{mcoGreen}{rgb}{0,0.6,0}
\definecolor{mcoPale}{rgb}{0.92587,0.97171,0.99509}

\tikzset{
  mco/picture/.style={
    x=1cm,
    y=1cm,
    font=\footnotesize,
    every node/.style={inner sep=1.5pt}
  },
  mco/edge/.style={
    draw=mcoInk,
    line width=.7pt
  },
  mco/vertex/.style={
    circle,
    draw=mcoVertexDraw,
    fill=mcoVertexFill,
    line width=.5pt,
    minimum size=4.3pt,
    inner sep=0pt,
    outer sep=0pt
  },
  mco/pink/.style={
    draw=mcoMagenta,
    line width=.95pt,
    dashed
  },
  mco/purple/.style={
    draw=mcoPurple,
    line width=.95pt,
    dashed
  },
  mco/orange/.style={
    draw=mcoOrange,
    line width=.95pt,
    densely dotted
  },
  mco/faintlink/.style={
    draw=black!40,
    line width=.85pt,
    dashed
  },
  mco/child/.style={
    draw=mcoGreen,
    line width=1.25pt
  },
  mco/focus/.style={
    draw=mcoMagenta,
    line width=1.35pt
  },
  mco/contract/.style={
    draw=black!38,
    line width=.8pt
  },
  mco/region/.style={
    draw=black!36,
    fill=mcoPale,
    dashed,
    line width=.4pt,
    rounded corners=4pt
  },
  mco/arrow/.style={
    draw=black!50,
    line width=.55pt,
    -{Latex[length=1.5mm]}
  },
  mco/title/.style={
    align=center,
    inner sep=0pt
  }
}

\long\def\mcoFit#1{%
  \sbox0{#1}%
  \ifdim\wd0>\linewidth
    \resizebox{\linewidth}{!}{\usebox0}%
  \else
    \usebox0%
  \fi
}

\mcoFit{%
\begin{tikzpicture}[mco/picture]
  \def\dx{8.25}

  \def\allnodes{
    r/0/0,
    u/0/-.90,
    v/0/-1.80,
    s/2.25/-1.80,
    x/-1.05/-2.80,
    y/1.05/-2.80,
    z/2.25/-2.75,
    p/-1.05/-3.70,
    q/1.05/-3.70,
    t/2.95/-3.70
  }

  \def\subnodes{
    u/0/-.90,
    v/0/-1.80,
    x/-1.05/-2.80,
    y/1.05/-2.80,
    p/-1.05/-3.70,
    q/1.05/-3.70
  }

  \def\subedges{u/v,v/x,v/y,x/p,y/q}

  \foreach \n/\x/\y in \allnodes {
    \coordinate (A\n) at (\x,\y);
  }

  \foreach \n/\x/\y in \subnodes {
    \coordinate (B\n) at ({\x+\dx},\y);
  }


  \draw[mco/region]
    (-.56,-.59)
    --(.56,-.59)
    --(.56,-1.81)
    --(1.68,-2.47)
    --(1.68,-4.04)
    --(-1.68,-4.04)
    --(-1.68,-2.47)
    --(-.56,-1.81)
    --cycle;

  \foreach \p/\q in {r/u,u/s,s/z,z/t} {
    \draw[mco/contract] (A\p)--(A\q);
  }

  \foreach \p/\q in \subedges {
    \draw[mco/edge] (A\p)--(A\q);
  }

  \draw[mco/focus] (Au)--(Av);
  \draw[mco/child] (Av)--(Ax) (Av)--(Ay);

  %

  \draw[mco/pink]
    (Au)
    .. controls (-2.45,-1.25) and (-2.45,-3.45) ..
    (Ap)
    node[pos=.53,left=3pt,text=mcoMagenta] {$a$};

  \draw[mco/purple]
    (Ar)
    .. controls (1.40,-.28) and (1.85,-3.35) ..
    (Aq)
    node[pos=.57,right=4pt,text=mcoPurple] {$b$};

  \draw[mco/faintlink]
    (Ar)
    .. controls (4.02,-.50) and (4.02,-3.30) ..
    (At)
    node[pos=.64,right=3pt,text=black!55] {$c$};

  \draw[mco/orange]
    (Av)
    .. controls (.10,-2.60) and (.18,-3.40) ..
    (Aq);

  \node[text=mcoOrange] at (.69,-2.99) {$d$};

  \path (Au)--node[right=4pt,text=mcoMagenta] {$e$} (Av);
  \path (Av)--node[left=4pt,text=mcoGreen] {$g_1$} (Ax);
  \path (Av)--node[right=4pt,text=mcoGreen] {$g_2$} (Ay);

  \node at (-.28,-3.36) {$T_e$};

  \foreach \n/\x/\y in \allnodes {
    \node[mco/vertex] at (A\n) {};
  }

  \node[above=3pt] at (Ar) {$r$};

  \node[mco/title] at (.55,-4.84)
    {(a) Original instance};

  \draw[mco/arrow]
    (4.10,-2.14)--(5.88,-2.14)
    node[midway,above=5pt,align=center]
      {contract\\outside $T_e$};

  %

  \foreach \p/\q in \subedges {
    \draw[mco/edge] (B\p)--(B\q);
  }

  \draw[mco/focus] (Bu)--(Bv);
  \draw[mco/child] (Bv)--(Bx) (Bv)--(By);

  \draw[mco/pink]
    (Bu)
    .. controls ({\dx-2.45},-1.25)
            and ({\dx-2.45},-3.45) ..
    (Bp)
    node[pos=.53,left=3pt,text=mcoMagenta] {$a$};

  \draw[mco/purple]
    (Bu)
    .. controls ({\dx+2.45},-1.25)
            and ({\dx+2.45},-3.45) ..
    (Bq)
    node[pos=.53,right=3pt,text=mcoPurple] {$b$};

  \draw[mco/orange]
    (Bv)
    .. controls ({\dx+.10},-2.60)
            and ({\dx+.18},-3.40) ..
    (Bq);

  \node[text=mcoOrange] at ({\dx+.69},-2.99) {$d$};

  \path (Bu)--node[right=4pt,text=mcoMagenta] {$e$} (Bv);
  \path (Bv)--node[left=4pt,text=mcoGreen] {$g_1$} (Bx);
  \path (Bv)--node[right=4pt,text=mcoGreen] {$g_2$} (By);


  \foreach \n/\x/\y in \subnodes {
    \node[mco/vertex] at (B\n) {};
  }

  \node[mco/title] at (\dx,-4.84)
    {(b) Instance on $T_e$};

\end{tikzpicture}%
}

\endgroup
\caption{The shaded subtree $T_e$ contains $e$ and all edges below it,
with $C(e)=\{g_1,g_2\}$. Contracting the gray tree edges gives the
instance on the right.}
\label{fig:techniques-utap}
\end{figure}

We think of $y_e$ as a price for covering the edge $e$. Each dual constraint (in $\dutaplp$) requires that the total price of the edges on a link's path be at most the cost of that link. To describe an optimal choice of prices, let $T_e$
consist of $e$ and all edges below it, let $C(e)$ be the set of edges
immediately below $e$, and let $O_e$ be the optimal cost of covering
$T_e$ after contracting the rest of the tree
(see \cref{fig:techniques-utap}).

We show that
\begin{equation}
    \label{eq:tech-dual-structure}
    y_e=O_e-\sum_{g\in C(e)}O_g
\end{equation}
defines an optimal dual solution. Intuitively, $y_e$ measures the marginal cost of covering $e$ beyond the cost of covering its child subtrees. Its nonnegativity follows because an uplink cannot cover edges in two distinct child subtrees. Dual feasibility follows by telescoping along each uplink path, while summing the dual variables over the whole tree telescopes to the optimal covering cost

To exploit this formula after sampling, define $c^S$ by assigning cost zero to sampled links and retaining the original costs on all other links. The optimal $\utap$ value under $c^S$ is exactly the minimum additional cost of \emph{completing} $S$. Let $y^S$ be the optimal dual solution given by \eqref{eq:tech-dual-structure}, and let $L(e)$ be the uplinks covering $e$. If an edge $e$ is already covered by $S$, then $y_e^S=0$. This establishes
\[
    y_e^S
    =\mathbf 1_{\{S\cap L(e)=\varnothing\}}\,
      y_e^{S\setminus L(e)}.
\]
The indicator and the second factor depend on disjoint families of independent sampling decisions. Therefore,
\[
    \mathbb E[y_e^S]
    \leq e^{-\gamma}\,\mathbb E[y_e^{S\setminus L(e)}].
\]
It remains to control the sum of these residual dual contributions. Expanding \eqref{eq:tech-dual-structure} and reindexing the child-subtree terms produces a telescoping sum where we obtain the following completion bound:
\begin{equation}
    \label{eq:tech-patching}
    \mathbb E\bigl[\optutap(T,c^S)\bigr]
    \leq 2e^{-\gamma}\sum_{a\in L^\uparrow}c(a)p_a
    \leq 2\gamma e^{-\gamma}
         \sum_{a\in L^\uparrow}c(a)x_a.
\end{equation}
Thus, for uplink instances, the small probability of leaving an edge uncovered also leads to a small expected completion cost and the optimal completion cost becomes exponentially cheaper as the sampling rate grows.

\paragraph{Multi-criteria rounding and the integrality-gap obstacle (\cref{sec:rand-compl,sec:analysis-mccap}).}
Now let $x$ be a feasible fractional solution of value $Z$ to the standard LP relaxation of $\mcutap$: it has the same covering constraints as $\utaplp$, and its cost under every criterion is at most $Z$, that is,
$c_i(x)=\sum_{a\in L^\uparrow}c_i(a)\,x_a\le Z$ for all $i\in[\ell]$. Choose $\gamma=\lceil\log\ell+1\rceil$, we independently sample uplinks as above, then complete the sampled set $S$ using the aggregate cost function $c(a)=\sum_{i=1}^\ell c_i(a)$. Since $\utaplp$ is integral, this min-cost completion can be found in polynomial time. If $H$ denotes the additional links, \eqref{eq:tech-patching} gives
\[
    \mathbb E\bigl[\max_{i\in[\ell]}c_i(H)\bigr]
    \leq \mathbb E[c(H)]
    \leq 2\gamma e^{-\gamma}
           \sum_{a\in L^\uparrow}\sum_{i=1}^\ell c_i(a)x_a
    \leq 2\gamma e^{-\gamma}\ell Z
    \leq 2\gamma Z.
\]
The aggregate cost could ordinarily be a factor $\ell$ larger than the maximum criterion cost. Here, however, the $e^{-\gamma}$ decay of the completion cost offsets that factor when $\gamma=\Theta(\log\ell)$.


Bounding the maximum cost of the sample $S$ needs a different
argument. Each criterion has small expected cost,
$\mathbb E[c_i(S)]\le\gamma Z$, but the criterion attaining the maximum depends on the outcome: in general,
$\mathbb E[\max_i c_i(S)]\ge\max_i\mathbb E[c_i(S)]$, and the gap can be as large as a factor of $\ell$. If every \emph{available} uplink has $c_i(a)\le B$ for all $i$, an exponential-moment argument gives $\mathbb E[\max_{i\in[\ell]}c_i(S)]\le B\log\ell+2\gamma Z$. Combining the two stages with $\gamma=\Theta(\log\ell)$, we obtain
\begin{equation}
    \label{eq:tech-rounding-guarantee}
    \mathbb E\bigl[\max_{i\in[\ell]}c_i(S\cup H)\bigr]
    \leq B\log\ell+4\gamma Z
    =O(\log\ell)(Z+B).
\end{equation}
Together with the constant-loss reduction, the same type of guarantee holds for $\mccap$.


The dependence on $B$ is unavoidable for the canonical relaxation:
$\mccaplp$ has integrality gap $\Theta(\ell)$ even on a single edge. Take $\ell$ parallel links $a_1,\ldots,a_\ell$ across the edge, with $c_i(a_j)=1$ if $i=j$ and $0$ otherwise. Every integral solution has cost at least $1$, whereas $x_{a_j}=1/\ell$ for all $j$ is feasible with value $1/\ell$. Hence no $O(\log\ell)$ guarantee holds relative to the unrestricted fractional optimum alone. 
We overcome this obstacle by comparing with an optimal
\emph{integral} solution $O^*$, of value
$\mathrm{OPT}=\max_{i\in[\ell]}c_i(O^*)$. For each candidate threshold $B$, namely each of the at most $|L|$ values $\max_i c_i(a)$, we discard every link $a$ with $\max_i c_i(a)>B$, solve the restricted LP, and round its solution. This yields one solution per threshold, and we return the one with the smallest maximum cost $\max_{i\in[\ell]}c_i(\cdot)$. 
For the threshold $B^*=\max_{a\in O^*}\max_{i\in[\ell]}c_i(a)$, every link of $O^*$ survives, so the restricted LP has value $Z\le\mathrm{OPT}$; moreover, $B^*\le\mathrm{OPT}$. Applying~\eqref{eq:tech-rounding-guarantee}
to this threshold, the rounded solution has expected cost at most
$O(\log\ell)(Z+B^*) = O(\log\ell)\cdot\mathrm{OPT}$. The returned
solution is the cheapest over all thresholds, so it costs no more than this one, and the algorithm is an $O(\log\ell)$-approximation in expectation. This does not contradict the $\Theta(\ell)$ integrality gap, because we compare with $\mathrm{OPT}$ rather than with the value of the unrestricted LP.


\paragraph{Hardness of approximation (\cref{sec:mccap-hardness}).}
To complement our approximation guarantee, we give an approximation-preserving reduction from the {\em min-max robust representatives selection} problem~\cite{Deineko2013ComplexityAI,kasperski2015approximability}to $\mcutap$ on a star. We introduce one leaf for each group and represent each member of that group by an uplink from the center to the corresponding leaf, preserving its cost under each criterion. Covering the edge incident to a leaf then amounts to choosing at least one representative from its group. Consequently, for every fixed $\varepsilon>0$, there is no
polynomial-time $O(\log^{1-\varepsilon}\ell)$-approximation algorithm for $\mcutap$, and hence none for the more general $\mccap$, unless $\NP\subseteq\mathsf{DTIME}\bigl(n^{\log^{O(1)}n}\bigr)$.

\paragraph{Extending the framework to MC-$k$-ECSS (\cref{sec:ecss}).}
We extend our framework to MC-$k$-ECSS through a sequence of
augmentation phases, each of which increases the edge connectivity by one. As before, we guess a threshold $B$ and discard every edge $e$ with $\max_i c_i(e)>B$. Let $x$ be an optimal solution, of value $Z$, to the LP relaxation of MC-$k$-ECSS on the remaining edges: it minimizes $\max_i c_i(x)$ subject to $0\le x_e\le1$ and $x(\delta_G(S))\ge k$ for every nontrivial cut $S$. For the correct guess, $Z,B\le\mathrm{OPT}$.

We use a standard argument that increases the connectivity one phase at a time. The algorithm builds a chain of subgraphs
$H^{(1)}\subseteq H^{(2)}\subseteq\cdots\subseteq H^{(k)}$, where
$H^{(j)}$ is $j$-edge-connected and each step from $H^{(j)}$ to
$H^{(j+1)}$ is one phase of the augmentation framework. We start with a multi-criteria spanning tree $H^{(1)}$ of cost
$O(\log\ell)\cdot\mathrm{OPT}$, computed by the algorithm
of~\cite{VondrakChekuriZenklusenMinMax}. In phase $j$, if $H^{(j)}$
is already $(j+1)$-edge-connected, we set $H^{(j+1)}=H^{(j)}$.
Otherwise, every minimum cut $S$ of $H^{(j)}$ is crossed by exactly $j$ edges of $H^{(j)}$. Since $x(\delta_G(S))\ge k$ and $x_e\le1$,
\[
x\bigl(\delta_G(S)\setminus E(H^{(j)})\bigr)\ge k-j.
\]
Hence $x$, restricted to the edges outside $H^{(j)}$ and scaled by
$1/(k-j)$, is a feasible fractional $\mccap$ solution of value at most $Z/(k-j)$, and rounding it yields the $(j+1)$-edge-connected graph $H^{(j+1)}$. Summed over all phases, these fractional values total $\sum_{j=1}^{k-1}Z/(k-j)=O(\log k)\,Z$.

The main difficulty is that applying~\eqref{eq:tech-rounding-guarantee} independently at each level would incur the additive $O(\log\ell)B$ term up to $k-1$ times. To avoid this loss, we analyze the completion and sampling costs separately. The expected completion costs can be summed directly, giving a total of $O(\log\ell\log k)Z$. For the sampling costs, however, the choices across iterations are dependent, since previously selected edges are no longer available. We overcome this by coupling the sampling decisions across all iterations with independent Bernoulli trials whose cumulative costs dominate the actual sampling costs under every criterion, for every realization. Applying a similar exponential-moment argument to these trials bounds the expected maximum total sampling cost by $B\log\ell+O(\log\ell\log k)Z$, paying the additive term involving $B$ only once. Together with the initial spanning tree and completion costs, this yields an $O(\log\ell\log k)$-approximation in expectation, which becomes $O(\log\ell)$ for fixed $k$. Moreover, the hardness for $\mccap$ on a tree transfers to MC-$2$-ECSS by making the original tree edges free, showing that the logarithmic dependence on $\ell$ is nearly tight already for $k=2$.

\section{Preliminaries}
\label{sec:preliminaries}

\paragraph{Notation.}
All graphs are finite undirected multigraphs and all cost functions are
nonnegative. For a positive integer $m$, let $[m]=\{1,\ldots,m\}$.
For a set of edges or links $F$ and a vertex set $U$, let $\delta_F(U)$
denote the elements of $F$ with exactly one endpoint in $U$. We write
$\delta_G(U) $ for $ \delta_{E(G)}(U)$ and $\delta_G(v) $for $\delta_G(\{v\})$, when $G$ is clear from context, we simply write $\delta(S)$ and $ \delta(v)$.
For a finite ground set $D$, a vector $x\in\mathbb R_{\ge0}^D$, and
$S\subseteq D$, write $x(S)$ for $\sum_{a\in S}x_a$. For a cost function $c$ on $D$, let $c(S)=\sum_{a\in S}c(a)$. Given $\ell$ cost functions
$\{c_i\}_{i\in[\ell]}$, define
\[
 \operatorname{cost}\bigl(S,\{c_i\}_{i\in[\ell]}\bigr)
 :=\max_{i\in[\ell]}\sum_{a\in S}c_i(a),
 \qquad
 \operatorname{cost}\bigl(x,\{c_i\}_{i\in[\ell]}\bigr)
 :=\max_{i\in[\ell]}\sum_{a\in D}c_i(a)x_a.
\]
When the cost functions are clear from context, we omit
$\{c_i\}_{i\in[\ell]}$ and write $\operatorname{cost}(S)$ and
$\operatorname{cost}(x)$. For a problem $\mathcal P$ and an instance $\mathcal I$, we write $\mathrm{OPT}^{\mathcal P}(\mathcal I)$ for the minimum objective value of a feasible integral solution. When clear from context, we omit instance data and write simply $\mathrm{OPT}^\mathcal{P}$.

We use log to denote the natural logarithm and write
$H_m=\sum_{j=1}^m 1/j\le 1+\log m$ for the $m$-th harmonic number. 

\subsection{Network Design Problems}
\label{subsec:network-design-problems}

\paragraph{Edge connectivity and connectivity augmentation.}
For a graph $G=(V,E)$ with at least two vertices, its edge connectivity is the minimum size of a cut, i.e. 
\[
 \kappa(G):=\min_{\emptyset\ne S\subsetneq V}|\delta_G(S)|.
\]
For an integer $k\ge1$, the graph is $k$-edge-connected if
$\kappa(G)\ge k$, equivalently, if it remains connected after the deletion
of any set of at most $k-1$ edges. We use $\mathcal C(G):=
 \{S \mid \emptyset\ne S\subsetneq V,\ |\delta_G(S)|=\kappa(G)\}$ to denote the family of min-cuts of $G$.

In the \emph{connectivity augmentation problem} (CAP), we are given a
graph $G=(V,E)$, a set $L$ of available links on $V$, and a cost function
$c:L\to\mathbb R_{\ge0}$. The links are distinct from the base edges,
but may have the same endpoints. For $A\subseteq L$, let $G \cup A$ denote
the graph obtained by adding the links in $A$ to $G$. The goal is to
minimize $c(A)$ subject to $\kappa(G \cup A)\ge\kappa(G)+1$. Equivalently, $A$ is feasible if and only for all min-cuts $S \in \mathcal{C}(G)$, $|A \cap \delta_L(S)| \geq 1$. 

In the \emph{multi-criteria connectivity augmentation problem}
($\mccap$), we are instead given $\ell$ cost functions $\{c_i:L\to\mathbb R_{\ge0}\}_{i\in[\ell]}$ and minimize
$\operatorname{cost}(A,\{c_i\}_{i\in[\ell]})$ over the same feasible sets. Throughout, we assume that the base graph of $\mccap$ is connected. When, $\kappa(G)=0$, contracting the connected components reduces $\mccap$ to the multi-criteria minimum spanning tree problem, which admits an $O(\log \ell)$-approximation for $\ell$ criteria \cite{VondrakChekuriZenklusenMinMax}.

\paragraph{Tree augmentation.}
The \emph{tree augmentation problem} ($\tap$) is the special case of CAP
in which the base graph is a tree $T=(V,E)$. A feasible solution is
therefore a set $A\subseteq L$ such that $T \cup A$ is $2$-edge-connected.
For a link $a=uv$, let $P_T(a)$ be the set of edges on the unique
$u,v$ path in $T$. When the tree is clear from context, we simply write $P(a)$. A link $a$ covers exactly the min-cuts corresponding to the deletions of the edges in $P(a)$. Therefore, we define $L(e):=\{a\in L:e\in P(a)\}$.

A solution $A$ is feasible
if and only if $A\cap L(e)\ne\emptyset$ for all $e \in E$. The \emph{multi-criteria tree augmentation problem} ($\mctap$) has the
same feasibility condition and minimizes $\operatorname{cost}(A)$
for the given cost functions $\{c_i\}_{i\in[\ell]}$.

\paragraph{Uplink tree augmentation and rooted-tree notation.}
Fix a root $r$ of $T$. A link is an \emph{uplink} if one endpoint is
an ancestor of the other. The \emph{uplink tree augmentation problem} ($\utap$) is
the restriction of TAP where all available links are uplinks. In this case, we denote the set of links by $L^\uparrow$ to distinguish the instance only consists of uplinks. Its
multi-criteria variant, $\mcutap$, minimizes $\operatorname{cost}(A)$ over
sets $A\subseteq L^\uparrow$ covering every tree edge. 

For a tree edge $e=uv$, where $u$ is the parent of $v$, let $ C(e):=\delta_T(v)\setminus\{e\}$ be the set of child edges of $e$. For
$e\in E(T)\setminus\delta_T(r)$, let $A(e)$ denote the unique parent
edge of $e$, so that $e\in C(A(e))$. Edges incident to $r$ have no
parent edge.  

\paragraph{Contracting tree edges.}
Let $D\subseteq E(T)$, and let $T/D$ be obtained by contracting the
edges in $D$. Each link $a=uv$ is mapped to a link $\bar a$ between
the images of $u$ and $v$, with its label and cost under every criterion
unchanged. For any link $a$ we have, $ P_{T/D}(\bar a)=P_T(a)\setminus D$. Thus, contracting tree edges and updating the link correspondingly produces
another TAP instance. Links that become loops cover no tree edges and may
be discarded. Rooting $T/D$ at the image of $r$, every image of an uplink is
again an uplink. Hence, contraction also preserves $\utap$ instances. 

\paragraph{Edge-connected spanning subgraphs.}
For an integer $k\ge1$, the \emph{$k$-edge-connected spanning subgraph
problem} ($k$-ECSS) takes a graph $G=(V,E)$ and a cost function
$c:E\to\mathbb R_{\ge0}$, and asks for a minimum-cost set
$F\subseteq E$ such that $(V,F)$ is $k$-edge-connected. Equivalently, for very cut $\emptyset\ne U\subsetneq V$, $\delta_F(S) \geq k$.

Each input edge can be selected at most once. In the
\emph{multi-criteria $k$-edge-connected spanning subgraph problem}
(MC-$k$-ECSS), the input has cost functions
$\{c_i:E\to\mathbb R_{\ge0}\}_{i\in[\ell]}$, and the objective is to
minimize $\operatorname{cost}(F)$. An instance is feasible
when the graph $G$ itself is $k$-edge-connected.

\section{Reducing $\mccap$ to $\mcutap$}
\label{sec:mccap-to-mcutap}
\label{sec:reduction}
In this section, we show that rounding a fractional $\mccap$ solution reduces to rounding two fractional $\mcutap$ solutions, each with cost at most an $O(1)$ factor of the original fractional cost. 
Before proceeding, we introduce a more structured version of the $\mccap$ called the Multi-Criteria Ring Augmentation Problem ($\mcrap$). In \cref{app:reduce-to-rap} we show any $\mccap$ instance can be formulated as a $\mcrap$ instance. Thus, it is sufficient to only show our reduction for rounding fractional $\mcrap$ solutions.

\begin{definition}
    Consider a cycle $G = (V,E)$, a family $\mathcal{W}$ of min-cuts in $G$, a multiset of links $L$ on $G$, and objective functions, $\{ c_i\}_{i \in [\ell]}$. A feasible solution to the $\mcrap$ is a set $H \subseteq L$ such that $H$ covers all cuts in $\mathcal{W}$, i.e., for all $S \in \mathcal{W}$, $|\delta_L(S) \cap H| \ge 1$. The cost of such a solution is defined as $\mathrm{cost}(H) = \max_{i \in [\ell]} \sum_{a \in H}c_i(a)$. The goal of $\mcrap$ is to find a feasible solution of minimum cost.

    The canonical LP formulation of $\mcrap$ is as follows,

\begin{equation}
\label{LP-mc-rap}
\boxed{
\begin{aligned}
\mcraplp\qquad
&\begin{aligned}[t]
    \min\quad & Z\\[3pt]
    \text{s.t.}\quad
    & x(\delta_L(S)) \geq 1
    && \forall S\in\mathcal W,\\
    & \sum_{a\in L} c_i(a)x_a \leq Z
    && \forall i\in[\ell],\\
    & x,Z \geq 0.
\end{aligned}
\end{aligned}
}
\end{equation}
\end{definition}

A key observation is that, under the natural cyclic ordering of the vertices of $G$, every minimum cut corresponds to a contiguous interval. This allows us to distinguish between links covering a cut from the “left” and those covering it from the “right”. Consequently, for every cut, at least half of its fractional coverage comes from links covering it from one of these two directions. We exploit this observation to separately reduce the coverage of cuts from the left and the right to two distinct $\mcutap$ instances. The idea of identifying the dominant direction of coverage was previously used in \cite{CTZ21} to round fractional solutions for the unweighted ring augmentation problem. However, extending this approach to the weighted setting, particularly with $\ell$ distinct cost functions, is nontrivial. In this setting, a simple rounding procedure for fractional $\mcrap$ solutions was previously unknown, and we address this challenge by reducing it to $\mcutap$.
We start by establishing some notation.

\paragraph{Setting.}
Consider a cycle $G=(V,E)$ with the vertices in $V$ ordered as $v_0, v_1, \cdots, v_{n-1}$ where each $v_i$ is connected to $v_{i+1}$ for $0 \leq i < n-1$ and $v_{n-1}$ is connected to $v_0$. We call $r \coloneq v_0$ the root of this cycle. Every
min-cut of $G$ has a unique representation as an interval,
\[
    S=[s,t]:=\{v_s,v_{s+1},\ldots,v_t\},\qquad 1\leq s\leq t\leq n-1,
\]
namely the side of the cut that does not contain the root. We are given a family $\mathcal W$ of minimum cuts of $G$, a multiset $L$ of links, and $\ell$ non-negative cost functions $\{c_i:L\to\mathbb R_{\geq 0}\}_{i\in[\ell]}$. We represent each link $a\in L$ as $a=(v_{t_a},v_{h_a})$, where $t_a<h_a$, and refer to $t_a$ and $h_a$ as its \emph{tail} and \emph{head}, respectively.

A link $a\in L$ \emph{covers} a cut $S$ if it crosses $S$, i.e. exactly one of its endpoints belongs to $S$. For any $A\subseteq L$, let $x(A):=\sum_{a\in A}x_a$ and $\mathrm{cost}_i(A):=\sum_{a\in A}c_i(a)$ for each $i\in[\ell]$. We further define $\mathrm{cost}(A):=\max_{i\in[\ell]}\mathrm{cost}_i(A)$.

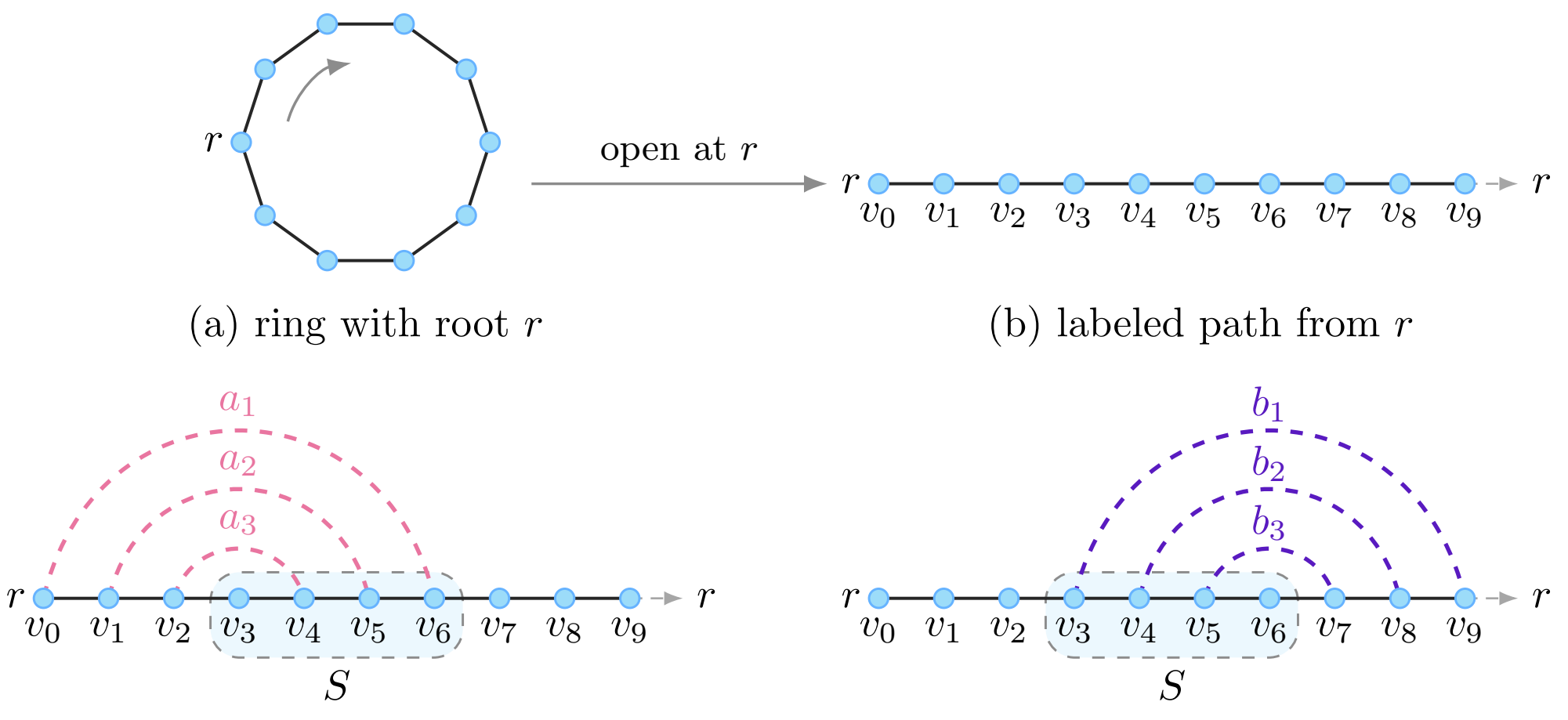
\begin{figure}[!tbp]
\centering
\begingroup
\definecolor{mcVertexFill}{rgb}{0.61338,0.86345,0.97537}
\definecolor{mcVertexDraw}{rgb}{0.4,0.7,1.0}
\definecolor{mcInk}{rgb}{0.15,0.15,0.15}
\definecolor{mcMagenta}{rgb}{0.91472,0.46078,0.62707}
\definecolor{mcPurple}{rgb}{0.34749,0.1,0.75250}
\definecolor{mcOrange}{rgb}{0.85,0.425,0}
\definecolor{mcGreen}{rgb}{0,0.6,0}
\definecolor{mcGray}{rgb}{0.55,0.55,0.55}
\definecolor{mcPale}{rgb}{0.92587,0.97171,0.99509}
\tikzset{
  mcfigure/.style={x=1cm,y=1cm,font=\small,>=Latex},
  baseedge/.style={draw=mcInk,line width=.75pt},
  vertex/.style={circle,draw=mcVertexDraw,fill=mcVertexFill,line width=.55pt,minimum size=4.6pt,inner sep=0pt,outer sep=0pt},
  linkmag/.style={draw=mcMagenta,line width=1pt,dashed},
  linkpurple/.style={draw=mcPurple,line width=1pt,dashed},
  linkorange/.style={draw=mcOrange,line width=1pt,densely dotted},
  linkgray/.style={draw=mcGray,line width=.85pt,dashed},
  pathgreen/.style={draw=mcGreen,line width=1.6pt},
  markedge/.style={draw=mcMagenta,line width=1.6pt},
  contractedge/.style={draw=black!28,line width=1.6pt},
  region/.style={draw=mcGray,dashed,fill=mcPale,line width=.55pt,rounded corners=7pt},
  block/.style={draw=mcGray,fill=mcPale,line width=.55pt,rounded corners=4pt,minimum height=.65cm,inner sep=5pt},
  selectedblock/.style={block,draw=mcMagenta,line width=1pt},
  guide/.style={draw=black!35,line width=.5pt,densely dashed},
  maparrow/.style={-{Latex[length=2mm]},draw=mcGray,line width=.65pt},
  paneltitle/.style={font=\small,align=center},
  smallnote/.style={font=\footnotesize,align=center},
  endpointlabel/.style={font=\small,inner sep=2pt}
}
\long\def\mcReductionFit#1{%
  \sbox0{#1}%
  \ifdim\wd0>\linewidth
    \resizebox{\linewidth}{!}{\usebox0}%
  \else
    \usebox0%
  \fi
}
\mcReductionFit{%
\begin{tikzpicture}[mcfigure]
  \def\mcStep{0.55}
  \def\mcDx{7.05}
  \def\mcBy{3.5}

  \foreach \j in {0,...,9} {
    \coordinate (ring\j) at ({2.72+1.05*cos(180-36*\j)},
                             {3.85+1.05*sin(180-36*\j)});
  }
  \draw[baseedge] (ring0)--(ring1)--(ring2)--(ring3)--(ring4)--(ring5)
    --(ring6)--(ring7)--(ring8)--(ring9)--cycle;
  \draw[maparrow] ({2.72+0.68*cos(165)},{3.85+0.68*sin(165)})
    arc[start angle=165,end angle=100,radius=0.68];
  \foreach \j in {0,...,9} {\node[vertex] at (ring\j) {};}
  \node[endpointlabel,left=2pt] at (ring0) {$r$};
  \node[paneltitle] at (2.72,2.3) {(a) ring with root $r$};

  \draw[maparrow] (4.12,\mcBy) -- (6.62,\mcBy)
    node[midway,above,smallnote] {open at $r$};

  \foreach \i in {0,...,9} {\coordinate (pb\i) at ({\mcDx+\i*\mcStep},\mcBy);}
  \draw[baseedge] (pb0)--(pb9);
  \draw[guide,-{Latex[length=1.6mm]}] (pb9) -- ++(0.45,0)
    node[endpointlabel,right=1pt] {$r$};
  \foreach \i in {0,...,9} {
    \node[vertex] at (pb\i) {};
    \node[endpointlabel,below=2.5pt] at (pb\i) {$v_{\i}$};
  }
  \node[endpointlabel,left=2pt] at (pb0) {$r$};
  \node[paneltitle] at ({\mcDx+2.72},2.3) {(b) labeled path from $r$};

  \foreach \i in {0,...,9} {\coordinate (pc\i) at ({\i*\mcStep},0);}
  \draw[region] ({3*\mcStep-0.24},-0.5) rectangle ({6*\mcStep+0.24},0.22);
  \draw[baseedge] (pc0)--(pc9);
  \draw[guide,-{Latex[length=1.6mm]}] (pc9) -- ++(0.45,0)
    node[endpointlabel,right=1pt] {$r$};
  \draw[linkmag] (pc0) .. controls +(0.45,1.89) and +(-0.45,1.89) .. (pc6);
  \draw[linkmag] (pc1) .. controls +(0.30,1.23) and +(-0.30,1.23) .. (pc5);
  \draw[linkmag] (pc2) .. controls +(0.16,0.56) and +(-0.16,0.56) .. (pc4);
  \foreach \i in {0,...,9} {
    \node[vertex] at (pc\i) {};
    \node[endpointlabel,below=2.5pt] at (pc\i) {$v_{\i}$};
  }
  \node[endpointlabel,left=2pt] at (pc0) {$r$};
  \node[endpointlabel,text=mcMagenta] at ({3*\mcStep},1.64) {$a_1$};
  \node[endpointlabel,text=mcMagenta] at ({3*\mcStep},1.14) {$a_2$};
  \node[endpointlabel,text=mcMagenta] at ({3*\mcStep},0.64) {$a_3$};
  \node[endpointlabel,below=1pt] at ({4.5*\mcStep},-0.5) {$S$};
  \node[paneltitle] at (2.72,-1.25) {(c) $S$ covered from the left};

  \foreach \i in {0,...,9} {\coordinate (pd\i) at ({\mcDx+\i*\mcStep},0);}
  \draw[region] ({\mcDx+3*\mcStep-0.24},-0.5)
    rectangle ({\mcDx+6*\mcStep+0.24},0.22);
  \draw[baseedge] (pd0)--(pd9);
  \draw[guide,-{Latex[length=1.6mm]}] (pd9) -- ++(0.45,0)
    node[endpointlabel,right=1pt] {$r$};
  \draw[linkpurple] (pd3) .. controls +(0.45,1.89) and +(-0.45,1.89) .. (pd9);
  \draw[linkpurple] (pd4) .. controls +(0.30,1.23) and +(-0.30,1.23) .. (pd8);
  \draw[linkpurple] (pd5) .. controls +(0.16,0.56) and +(-0.16,0.56) .. (pd7);
  \foreach \i in {0,...,9} {
    \node[vertex] at (pd\i) {};
    \node[endpointlabel,below=2.5pt] at (pd\i) {$v_{\i}$};
  }
  \node[endpointlabel,left=2pt] at (pd0) {$r$};
  \node[endpointlabel,text=mcPurple] at ({\mcDx+6*\mcStep},1.64) {$b_1$};
  \node[endpointlabel,text=mcPurple] at ({\mcDx+6*\mcStep},1.14) {$b_2$};
  \node[endpointlabel,text=mcPurple] at ({\mcDx+6*\mcStep},0.64) {$b_3$};
  \node[endpointlabel,below=1pt] at ({\mcDx+4.5*\mcStep},-0.5) {$S$};
  \node[paneltitle] at ({\mcDx+2.72},-1.25) {(d) $S$ covered from the right};
\end{tikzpicture}%
}
\endgroup
\caption{Left and right coverage of the min-cut $S=[3,6]=\{v_3,\dots,v_6\}$.}
\label{fig:reduction-left-right}
\end{figure}

\begin{theorem}
\label{thm:reduction}
Let $x$ be a feasible solution to $\mcraplp$ with a multiset of links $L$ and cost functions $\{c_i:L\to\mathbb R_{\geq 0}\}_{i\in[\ell]}$. Suppose further that $x_a\leq 1/20$ for every $a\in L$. Then, in polynomial time, we can construct, for each $j\in\{1,2\}$, a $\mcutap$ instance $(T_j,U_j)$, a bijection $\varphi_j:U_j\to L$, cost functions $\{c_i^{(j)}:U_j\to\mathbb R_{\geq 0}\}_{i\in[\ell]}$, and a vector $z^{(j)}\in\mathbb R_{\geq 0}^{U_j}$ satisfying the following properties:
\begin{enumerate}
    \item $c^{(j)}_i(u)=c_i(\varphi_j(u))$ for every $i\in[\ell]$ and
    $u\in U_j$.
    \item $z^{(j)}_u\leq10\,x_a$ for every
    $a\in L$ where $a = \varphi_j(u)$.
    \item $z^{(j)}$ satisfies the covering constraints of
    $\mcutaplp$ on $(T_j,U_j)$, and for every $i\in[\ell]$, $\sum_{u\in U_j}c^{(j)}_i(u)z^{(j)}_u \leq 10\sum_{a\in L}c_i(a)x_a$.
    \item if $R_j\subseteq U_j$ is a feasible integral solution of
    $(T_j,U_j)$ for $j\in\{1,2\}$, then $\varphi_1(R_1)\cup\varphi_2(R_2)$
    covers every cut in $\mathcal W$, and
    $c_i(\varphi_1(R_1)\cup\varphi_2(R_2))\leq
    c^{(1)}_i(R_1)+c^{(2)}_i(R_2)$ for every $i\in[\ell]$.
\end{enumerate}
\end{theorem}

Note that we can assume without loss of generality the assumption on $x_a \leq \frac{1}{20}$ for all $a \in L$ holds. Given any arbitrary fractional solution $x$, we can simply achieve this by splitting every link into
parallel copies. Let $ L$ be the multiset that contains $20$ copies
of every link $a\in L$, each with the same costs as $a$ under all objectives. For each copy $b$ of a link $a$, we set, $ x_b:=\tfrac1{20}x_{a}$.

\begin{proof}[Proof of \cref{thm:reduction}]
Consider a min-cut $S = [s,t]$ of $G$. The links $a \in L$ that cover the cut corresponding to $[s,t]$ have one of the following two forms,
\begin{enumerate}
    \item $a=v_iv_j$ where $i \in [0,s-1]$ and $j \in [s,t]$. Let $\mathcal{L}(s,t)$ denote the set of such links. In this case, we say $a$ covers $S$ from the {\em left}.
    \item $a=v_iv_j$ where $i \in [s,t]$ and $j \in [t+1,n-1]$. Let $\mathcal{R}(s,t)$ denote the set of such links. In this case, we say $a$ covers $S$ from the {\em right}.
\end{enumerate}
See \cref{fig:reduction-left-right} for an illustration.

The following simple observation follows from the feasibility of $x$.

\begin{observation}
\label{obs:two-sides}
For every $S=[s,t]\in\mathcal W$,
$\max\{x(\mathcal L(s,t)),x(\mathcal R(s,t))\}\geq \frac12$.
\end{observation}

\begin{proof}
Since $\mathcal L(s,t)$ and $\mathcal R(s,t)$ are disjoint and $x$ is a feasible solution to $\mcraplp$ for $G$, we have, $x(\delta_L(S)) = x(\mathcal{L}(s,t)) + x(\mathcal{R}(s,t)) \geq 1$.
Hence, for every $S\in\mathcal W$, $\max\{x(\mathcal L(s,t)),x(\mathcal R(s,t))\}\geq \frac12$.
\end{proof}

We partition the family of min-cuts $\mathcal W$ into two disjoint subfamilies $\mathcal L$ and $\mathcal R$ such that $x(\mathcal L(s,t))\geq \frac12$ for every $S=[s,t]\in\mathcal L$, and $x(\mathcal R(s,t))\geq \frac12$ for every $S=[s,t]\in\mathcal R$. We construct two disjoint $\mcutap$ instances for covering the min-cuts in each $\mathcal{L}$ and $\mathcal{R}$. Without loss of generality we describe this construction for cuts in $\mathcal{L}$ only. Simply reversing the ordering of the vertices in $G$ makes the construction for $\mathcal{R}$ identical to $\mathcal{L}$. 

Before proceeding with the construction of the $\mcutap$ instances, we introduce three orderings on the links in $L$.
Recall that a link $a=(v_{t_a},v_{h_a})$ has tail $v_{t_a}$ and head
$v_{h_a}$, where $t_a\leq h_a$. Fix an arbitrary total order $\preceq_0$
on the labelled links in $L$, where we treat parallel copies as distinct links, this ordering is used simply to break the ties between parallel links.
 Now, define $\preceq_1$ and $\preceq_2$ as follows: For all pair of links
$a,b\in L$,
\begin{align*}
    a\preceq_1 b
    &\quad\Longleftrightarrow\quad
    t_a<t_b\ \text{ or }\ (t_a=t_b\text{ and }a\preceq_0 b),\\
    a\preceq_2 b
    &\quad\Longleftrightarrow\quad
    h_a<h_b\ \text{ or }\ (h_a=h_b\text{ and }a\preceq_0 b).
\end{align*}
We denote $\preceq_1$ as the tail ordering and $\preceq_2$ as the head ordering respectively. When ordering a subset of $L$, we use the restriction of the corresponding order on $L$.

Let $m:=|L|$ and write the links in
the order of $\preceq_1$ as $a_1  \preceq_1 a_2 \preceq_1 \ldots \preceq_1 a_m$. For each $i\in\{0,\ldots,m\}$, define $A_i:=\{a_1,\ldots,a_i\}$, with $A_0:=\emptyset$. We then index the links in $A_i$ according to $\preceq_2$ as $b^{(i)}_1\preceq_2 b^{(i)}_2\preceq_2\cdots\preceq_2 b^{(i)}_i$.

Our construction proceeds iteratively. In each iteration $i\geq 1$, we maintain a partition $\mathcal P_i$ of $A_i$. For each $a\in A_i$, let $\mathcal P_i(a)$ denote the unique part containing $a$. We first describe the construction of these partitions and establish their structural properties, before using them to construct the tree.

\paragraph{Constructing the partitions.} We initialize $\mathcal P_0:=\{\emptyset\}$ and $\mathcal P_1:=\{\{a_1\}\}$, where the empty part in $\mathcal P_0$ is used only for initialization. All parts of $\mathcal P_i$ are nonempty for $i\geq 1$.

Suppose that for some $1\leq i<m$, $\mathcal P_i$ has been constructed.
Let $j\in[i+1]$ be the index such that $b^{(i+1)}_j=a_{i+1}$, i.e. the position of $a_{i+1}$ when sorting the links in $A_{i+1}$ according to $\preceq_2$. Because the ordering is fixed, the list for $A_{i+1}$ is obtained from the list for $A_i$ by inserting $a_{i+1}$ at position $j$. Define the set $H$ as $H=\mathcal P_i(b^{(i+1)}_{j-1})$ if $j>1$, and $H=\mathcal P_i(b^{(i+1)}_{2})$ if $j=1$.
Thus, $H$ contains either the link immediately preceding $a_{i+1}$ or the link immediately following it when sorting $A_{i+1}$ according to $\preceq_2$. Next, we extend the construction with $a_{i+1}$ as follows.

\begin{itemize}[leftmargin=*]
    \item \textbf{Case 1: $x(H\cup\{a_{i+1}\})<\frac14$.}
    Replace $H$ by $H\cup\{a_{i+1}\}$ and set $\mathcal P_{i+1} = (\mathcal P_i\setminus\{H\}) \cup\{H\cup\{a_{i+1}\}\}$.

    \item \textbf{Case 2: $x(H\cup\{a_{i+1}\})\geq\frac14$.}
    Order the links in $H\cup\{a_{i+1}\}$ as
    $d_1\preceq_2\cdots\preceq_2 d_p$, and let $q$ be the smallest index such that $x(\{d_1,\ldots,d_q\})\geq\frac1{10}$.
    Partition $H\cup\{a_{i+1}\}$ into $H_1=\{d_1,\ldots,d_q\}$ and $H_2 = (H\cup\{a_{i+1}\})\setminus H_1$.
    
    The minimality of $q$ and the assumption $x_{d_q}\leq\frac1{20}$
    gives
    \begin{equation}
        \label{eq:H_1-Upper-Lower-Bound}
        \frac1{10}\leq x(H_1)
        <\frac1{10}+\frac1{20}=\frac3{20}.
    \end{equation}
    This implies, $x(H_2) = x(H\cup\{a_{i+1}\})-x(H_1) > \frac14-\frac3{20}=\frac1{10}$.
    Both $H_1, H_2$ are nonempty and satisfy $x(H_1), x(H_2) \geq \frac{1}{10}$. Moreover, every link in
    $H_1$ precedes every link in $H_2$ under the ordering $\preceq_2$. In this case, we obtain $\mathcal{P}_{i+1}$ by replacing $H$ with $H_1, H_2$ in $\mathcal{P}_i$: $\mathcal P_{i+1} = (\mathcal P_i\setminus\{H\})\cup\{H_1,H_2\}$.
\end{itemize}
In both cases, $\mathcal{P}_{i+1}$ is a partition on the elements of $A_{i+1}$ (see \cref{fig:reduction-partition-update}).


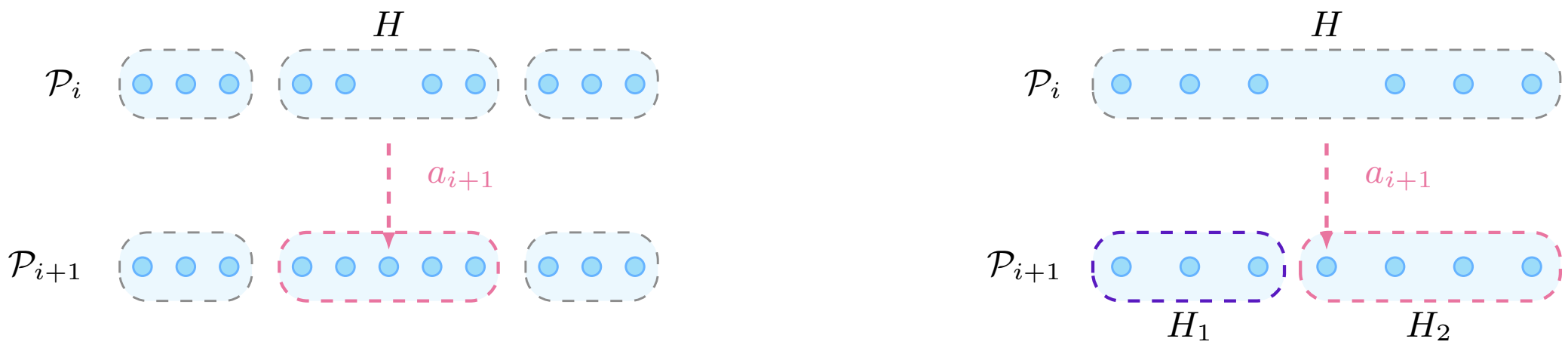
\begin{figure}[!tbp]
\centering
\begingroup
\definecolor{mcVertexFill}{rgb}{0.61338,0.86345,0.97537}
\definecolor{mcVertexDraw}{rgb}{0.4,0.7,1.0}
\definecolor{mcInk}{rgb}{0.15,0.15,0.15}
\definecolor{mcMagenta}{rgb}{0.91472,0.46078,0.62707}
\definecolor{mcPurple}{rgb}{0.34749,0.1,0.75250}
\definecolor{mcOrange}{rgb}{0.85,0.425,0}
\definecolor{mcGreen}{rgb}{0,0.6,0}
\definecolor{mcGray}{rgb}{0.55,0.55,0.55}
\definecolor{mcPale}{rgb}{0.92587,0.97171,0.99509}
\tikzset{
  mcfigure/.style={x=1cm,y=1cm,font=\small,>=Latex},
  baseedge/.style={draw=mcInk,line width=.75pt},
  vertex/.style={circle,draw=mcVertexDraw,fill=mcVertexFill,line width=.55pt,minimum size=4.6pt,inner sep=0pt,outer sep=0pt},
  linkmag/.style={draw=mcMagenta,line width=1pt,dashed},
  linkpurple/.style={draw=mcPurple,line width=1pt,dashed},
  linkorange/.style={draw=mcOrange,line width=1pt,densely dotted},
  linkgray/.style={draw=mcGray,line width=.85pt,dashed},
  pathgreen/.style={draw=mcGreen,line width=1.6pt},
  markedge/.style={draw=mcMagenta,line width=1.6pt},
  contractedge/.style={draw=black!28,line width=1.6pt},
  region/.style={draw=mcGray,dashed,fill=mcPale,line width=.55pt,rounded corners=7pt},
  block/.style={draw=mcGray,fill=mcPale,line width=.55pt,rounded corners=4pt,minimum height=.65cm,inner sep=5pt},
  selectedblock/.style={block,draw=mcMagenta,line width=1pt},
  guide/.style={draw=black!35,line width=.5pt,densely dashed},
  maparrow/.style={-{Latex[length=2mm]},draw=mcGray,line width=.65pt},
  paneltitle/.style={font=\small,align=center},
  smallnote/.style={font=\footnotesize,align=center},
  endpointlabel/.style={font=\small,inner sep=2pt}
}
\long\def\mcReductionFit#1{%
  \sbox0{#1}%
  \ifdim\wd0>\linewidth
    \resizebox{\linewidth}{!}{\usebox0}%
  \else
    \usebox0%
  \fi
}
\begin{subfigure}[t]{.48\linewidth}
\centering
\vspace{0pt}
\mcReductionFit{%
\begin{tikzpicture}[mcfigure]
  \path[use as bounding box] (-1,-.70) rectangle (4.60,2.35);
  \node[smallnote,anchor=east] at (-.38,1.60) {$\mathcal P_i$};
  \node[smallnote,anchor=east] at (-.38,0) {$\mathcal P_{i+1}$};

  \foreach \y in {0,1.60} {
    \draw[region] (-.20,\y-.30) rectangle (.96,\y+.30);
    \draw[region] (3.36,\y-.30) rectangle (4.52,\y+.30);
  }
  \draw[region] (1.20,1.30) rectangle (3.12,1.90);
  \draw[region,draw=mcMagenta,line width=.8pt]
    (1.20,-.30) rectangle (3.12,.30);
  \node[smallnote] at (2.16,2.13) {$H$};

  \foreach \x in {0,.38,.76,1.40,1.78,2.54,2.92,3.56,3.94,4.32} {
    \node[vertex] at (\x,1.60) {};
    \node[vertex] at (\x,0) {};
  }
  \draw[linkmag,-{Latex[length=1.6mm]}] (2.16,1.08) -- (2.16,.14);
  \node[smallnote,text=mcMagenta,anchor=west] at (2.36,.77) {$a_{i+1}$};
  \node[vertex] at (2.16,0) {};
\end{tikzpicture}%
}
\caption{Insertion without a split.}
\label{fig:reduction-update-no-split}
\end{subfigure}\hfill%
\begin{subfigure}[t]{.48\linewidth}
\centering
\vspace{0pt}
\mcReductionFit{%
\begin{tikzpicture}[mcfigure]
  \path[use as bounding box] (-1,-.70) rectangle (4.60,2.35);
  \node[smallnote,anchor=east] at (-.38,1.60) {$\mathcal P_i$};
  \node[smallnote,anchor=east] at (-.38,0) {$\mathcal P_{i+1}$};

  \draw[region] (-.25,1.30) rectangle (3.85,1.90);
  \node[smallnote] at (1.80,2.13) {$H$};
  \draw[region,draw=mcPurple,line width=.8pt]
    (-.25,-.30) rectangle (1.43,.30);
  \draw[region,draw=mcMagenta,line width=.8pt]
    (1.57,-.30) rectangle (3.85,.30);

  \foreach \x in {0,.60,1.20,2.40,3,3.60} {
    \node[vertex] at (\x,1.60) {};
    \node[vertex] at (\x,0) {};
  }
  \draw[linkmag,-{Latex[length=1.6mm]}] (1.80,1.08) -- (1.80,.14);
  \node[smallnote,text=mcMagenta,anchor=west] at (2,.77) {$a_{i+1}$};
  \node[vertex] at (1.80,0) {};
  \node[smallnote] at (.60,-.52) {$H_1$};
  \node[smallnote] at (2.70,-.52) {$H_2$};
\end{tikzpicture}%
}
\caption{Insertion followed by a split.}
\label{fig:reduction-update-split}
\end{subfigure}
\endgroup
\caption{An illustration of insertion of the element $a_{i+1}$ to the partitions. For simplicity assume all links have $x$-mass $\frac{1}{25}$. In (a), the mass of $H$ increases from $4/25$ to $5/25<1/4$, so we simply add $a_{i+1}$ to $H$. In (b), the mass of $H$ increases from $6/25$ to $7/25\ge1/4$ and thus, we have to split $H$. The shortest prefix of mass at least $1/10$ contains the first three links. Thus $q=3$, giving $x(H_1)=3/25$ and $x(H_2)=4/25$.}
\label{fig:reduction-partition-update}
\end{figure}

\paragraph{Properties of the partitions.} Now, we establish several useful properties of these partitions. The underlying structure of our construction is crucial in encoding these partitions as a $\mcutap$ instance.
We first show that, for every $i\geq 1$, the parts of $\mathcal P_i$ form consecutive intervals of $A_i$ under the head ordering $\preceq_2$.

\begin{claim}
\label{clm:partitions-are-intervals}
For every $i\in[m]$ and $P\in\mathcal P_i$, there exist
$1\leq r\leq q\leq i$ such that $P=\{b^{(i)}_r,b^{(i)}_{r+1},\ldots,b^{(i)}_q\}$.
Namely, each $P \in \mathcal{P}_i$ is a consecutive interval of links in $A_i$ under the head ordering ($\preceq_2$).
\end{claim}

\begin{proof}
The proof is by induction on $i$. The statement holds trivially for $i=1$ as $\mathcal{P}_1 =\{\{a_1\}\}$.
Suppose the claim holds for $i$, and consider iteration $i+1$. Let $H$ be the part to which $a_{i+1}$ is added. By construction, $H$ contains a link immediately preceding or succeeding $a_{i+1}$ in the head ordering of $A_{i+1}$. Since $H$ is an interval in $A_i$, adding $a_{i+1}$ preserves its interval structure, while all other parts remain unchanged and consecutive in $A_{i+1}$.
In the first case of our construction, $H\cup\{a_{i+1}\}$ remains an interval. In the second case, $H$ is split into two consecutive intervals $H_1$ and $H_2$. Thus, the interval property is preserved in both cases.
\end{proof}

By \cref{clm:partitions-are-intervals}, we can index the parts of $\mathcal P_i$ as $P^{(i)}_1,\ldots,P^{(i)}_{|\mathcal P_i|}$ such that every link in $P^{(i)}_r$ precedes every link in $P^{(i)}_q$ under $\preceq_2$ whenever $r<q$. The following claim bounds the $x$-mass of each part.

\begin{claim}
\label{clm:upper-lower-bound-parition}
For every $i\in[m]$ and $r\in[|\mathcal P_i|]$, $x(P^{(i)}_r)<\frac14$. Moreover, if $|\mathcal P_i| \geq 2$, then $x(P^{(i)}_r)\geq\frac1{10}$.
\end{claim}
\begin{proof}
We prove both bounds by induction on $i$. The base case follows as $x_{a_1}\leq 1/20$, $\mathcal{P}_1 = \{ \{ a_1 \} \}$. Assume the claim holds for $\mathcal{P}_i$ and let $H$ be the set in $\mathcal{P}_i$ to which $a_{i+1}$ will be added. Set $B:=H\cup\{a_{i+1}\}$. 

All other parts of $\mathcal{P}_i$ remain unchanged and thus, by our induction the bounds are preserved on them.
If $x(B)<1/4$, the bounds are preserved for $B$ too as $x(B)\geq x(H)$. Otherwise, by induction, $\frac1 4 \leq x(B)<\frac14+\frac1{20}=\frac3{10}$.
In this case, $B$ is split into $H_1,H_2$, let the elements of $B$ be ordered as $d_1 \preceq_2 \cdots \preceq_2 d_p$. Recall that we let $H_1 = \{ d_1, \cdots, d_q \}$ where $q$ is the smallest index where $x(\{ d_1, \cdots, d_q\}) \geq \frac{1}{10}$. Therefore, by definition of $q$, $x(H_1) \geq \frac{1}{10}$. Moreover, $x(H_1 \setminus \{ d_q \}) < \frac{1}{10}$. Since for any link $a$, we have $x_a \leq \frac{1}{20}$, this implies $x(H_1) < \frac{1}{10} + \frac{1}{20} = \frac{3}{20} < \frac{1}{4}$. 

Now, consider $H_2$, since $H_2 = B \setminus H_1$, $x(H_2) = x(B) - x(H_1) \geq \frac{1}{4} - \frac{3}{20} = \frac{1}{10}$. Moreover, by our induction, $x(H) < \frac{1}{4}$, since $B = H \cup \{ a_{i+1} \}$, $x(B) < \frac{1}{4} + \frac{1}{20} = \frac{3}{10}$. Putting this together with $x(H_1) \geq \frac{1}{10}$ gives, $x(H_2) < \frac{3}{10} - \frac{1}{10} = \frac{1}{5} < \frac{1}{4} $. Thus, all newly added sets to $\mathcal{P}_{i+1}$ satisfy the required lower bound and upper bound. 

If the only set in $\mathcal{P}_i$ was $H$, then $\mathcal{P}_{i+1} = \{H_1, H_2 \}$ and thus, all sets in $\mathcal{P}_{i+1}$ satisfy the required bounds. Meanwhile, if $\mathcal{P}_i \setminus \{ H\} \neq \emptyset$, then by our induction, all sets in $\mathcal{P}_i$ should satisfy the required lower bound, since both $H_1, H_2$ also did so, all sets in  $\mathcal{P}_{i+1}$ satisfy the guaranteed bounds.
\end{proof}

The following claim shows that for every cut $S=[s,t]\in\mathcal L$, some part $P\in\mathcal P_i$ satisfies $P\subseteq\mathcal L(s,t)$. Together with the preceding claim, this ensures that our construction captures a subset of the links covering $S$ from the left with $x$-mass bounded below by a positive constant, even if it does not capture all of $\mathcal L(s,t)$.

\begin{claim}
\label{clm:partition-covers-cut}
For each min-cut $S=[s,t]\in\mathcal L$, let $i(s):=|\{a\in L:t_a<s\}|$ be the number of links whose tails precede $s$. Then there exists a part $P\in\mathcal P_{i(s)}$ such that $P\subseteq\mathcal L(s,t)$ and $x(P)\geq\frac1{10}$. Moreover, $i(s) \geq 1$.
\end{claim}


\begin{figure}[!tbp]
\centering
\begingroup
\definecolor{mcVertexFill}{rgb}{0.61338,0.86345,0.97537}
\definecolor{mcVertexDraw}{rgb}{0.4,0.7,1.0}
\definecolor{mcInk}{rgb}{0.15,0.15,0.15}
\definecolor{mcMagenta}{rgb}{0.91472,0.46078,0.62707}
\definecolor{mcPurple}{rgb}{0.34749,0.1,0.75250}
\definecolor{mcOrange}{rgb}{0.85,0.425,0}
\definecolor{mcGreen}{rgb}{0,0.6,0}
\definecolor{mcGray}{rgb}{0.55,0.55,0.55}
\definecolor{mcPale}{rgb}{0.92587,0.97171,0.99509}
\tikzset{
  mcfigure/.style={x=1cm,y=1cm,font=\small,>=Latex},
  baseedge/.style={draw=mcInk,line width=.75pt},
  vertex/.style={circle,draw=mcVertexDraw,fill=mcVertexFill,line width=.55pt,minimum size=4.6pt,inner sep=0pt,outer sep=0pt},
  linkmag/.style={draw=mcMagenta,line width=1pt,dashed},
  linkpurple/.style={draw=mcPurple,line width=1pt,dashed},
  linkorange/.style={draw=mcOrange,line width=1pt,densely dotted},
  linkgray/.style={draw=mcGray,line width=.85pt,dashed},
  pathgreen/.style={draw=mcGreen,line width=1.6pt},
  markedge/.style={draw=mcMagenta,line width=1.6pt},
  contractedge/.style={draw=black!28,line width=1.6pt},
  region/.style={draw=mcGray,dashed,fill=mcPale,line width=.55pt,rounded corners=7pt},
  block/.style={draw=mcGray,fill=mcPale,line width=.55pt,rounded corners=4pt,minimum height=.65cm,inner sep=5pt},
  selectedblock/.style={block,draw=mcMagenta,line width=1pt},
  guide/.style={draw=black!35,line width=.5pt,densely dashed},
  maparrow/.style={-{Latex[length=2mm]},draw=mcGray,line width=.65pt},
  paneltitle/.style={font=\small,align=center},
  smallnote/.style={font=\footnotesize,align=center},
  endpointlabel/.style={font=\small,inner sep=2pt}
}
\long\def\mcReductionFit#1{%
  \sbox0{#1}%
  \ifdim\wd0>\linewidth
    \resizebox{\linewidth}{!}{\usebox0}%
  \else
    \usebox0%
  \fi
}
\mcReductionFit{%
\begin{tikzpicture}[mcfigure]
  \draw[draw=mcMagenta,dashed,fill=mcMagenta!5,line width=.65pt,rounded corners=5pt]
    (.325,-.72) rectangle (9.475,.91);
  \draw[region] (-.35,-.39) rectangle (2.95,.39);
  \draw[region,draw=mcMagenta,line width=1pt] (3.25,-.39) rectangle (6.55,.39);
  \draw[region] (6.85,-.39) rectangle (10.15,.39);

  \draw[baseedge,-{Latex[length=1.8mm]}] (-.55,0) -- (10.45,0);
  \foreach \x in {0,.65,1.3,1.95,2.6,3.6,4.25,4.9,5.55,6.2,7.2,7.85,8.5,9.15,9.8} {
    \node[vertex] at (\x,0) {};
  }
  \node[smallnote] at (1.30,.66) {$P_p$};
  \node[smallnote] at (4.90,.66) {$P^\star$};
  \node[smallnote] at (8.50,.66) {$P_{p+2}$};

  \draw[guide] (.325,-.91) -- (.325,1.00);
  \draw[guide] (9.475,-.91) -- (9.475,1.00);
  \node[smallnote,above=2pt] at (.325,1.00) {$s$};
  \node[smallnote,above=2pt] at (9.475,1.00) {$t$};
\end{tikzpicture}%
}
\endgroup
\caption{The sets in the partition $\mathcal{P}_{i(s)}$ that intersect $[s,t]$ ordered in their head order. Each link has $x$-mass $1/25$, and each part has mass $1/5<1/4$. The whole middle part $P^\star=P_{p+1}$ lies in $L(s,t)$ and has mass $1/5\ge1/10$.}
\label{fig:reduction-interior-block}
\end{figure}
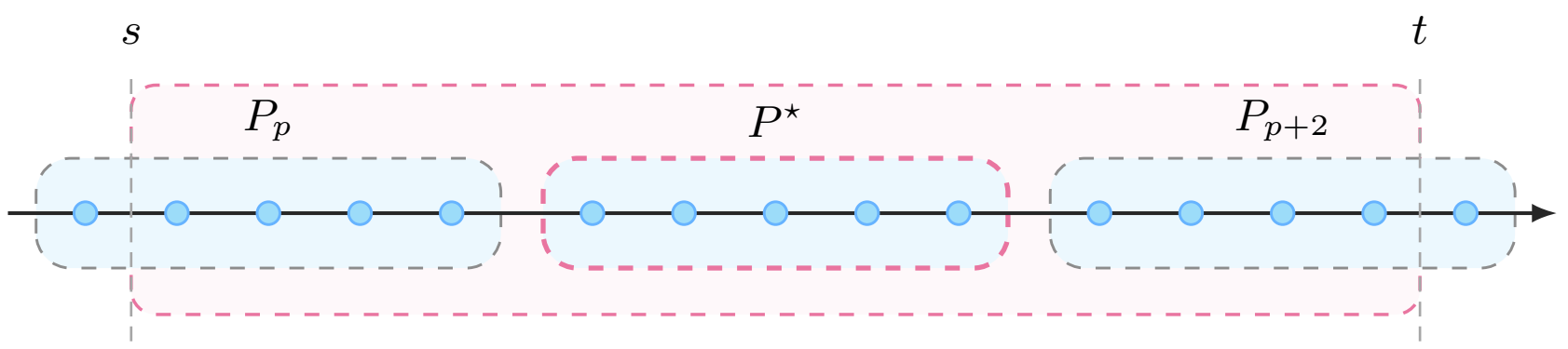

\begin{proof}


Define $i:=i(s)$. Since $\preceq_1$ orders links by their tail, we have $A_i=\{a\in L:t_a<s\}$ and $\mathcal L(s,t)=\{a\in A_i:s\leq h_a\leq t\}$. Moreover, since $x(\mathcal L(s,t))\geq\frac12$ and $x_a\leq\frac1{10}$ for every $a\in L$, we have $i\geq5$.

Observe that $\mathcal L(s,t)$ is an interval in the head ordering $\preceq_2$ on $A_i$ as $\mathcal L(s,t)$ is exactly the set of links in $A_i$ with their head in $[s,t]$. Let $P_p^{(i)},\ldots,P_q^{(i)}$ be the parts intersecting $\mathcal L(s,t)$. As $x(\mathcal L(s,t))\geq\frac12$ and each part has $x$-mass less than $\frac14$ by \cref{clm:upper-lower-bound-parition}, we have $q\geq p+2$.
For any $p<r<q$, the interval property (\cref{clm:partitions-are-intervals}) implies $P_r^{(i)}\subseteq\mathcal L(s,t)$. Moreover, $|\mathcal P_i|\geq3$, so $x(P_r^{(i)})\geq\frac1{10}$ by \cref{clm:upper-lower-bound-parition}.
\end{proof}

\begin{definition}
    \label{def:Pst}
    For each $S=[s,t]\in\mathcal L$, let $i(s):=|\{a\in L:t_a<s\}|$. Let $P(s,t)$ be an arbitrary part in $\mathcal{P}_{i(s)}$ that satisfies the guarantees in \cref{clm:partition-covers-cut}.
\end{definition}

\paragraph{Constructing the tree.}
We construct a rooted tree $T$ with layers indexed by $0,\ldots,m$, vertex set $V(T)=\{(i,P):0\leq i\leq m,\ P\in\mathcal P_i\}$, and root $(0,\emptyset)$. The index $i$ distinguishes occurrences of the same part across different layers.

For each $0\leq i<m$, connect $(i,P)$ to $(i+1,P')$ if and only if $P'\setminus\{a_{i+1}\}\subseteq P$, and call $(i,P)$ the parent of $(i+1,P')$. By construction, each unchanged part $P$ is connected to its copy at layer $i+1$, while $H$ has one child in the first case and two in the second case. Thus, every vertex other than the root has a unique parent. We denote the parent of a vertex $w$ by $\operatorname{par}(w)$.

\begin{observation}
\label{clm:partition-hierarchy}
For any link $a=a_q\in L$ and $q\leq i\leq j\leq m$, the vertex $(i,\mathcal P_i(a))$ is an ancestor of $(j,\mathcal P_j(a))$ in $T$.
\end{observation}

\begin{proof}
For each $i\leq k<j$, the part $\mathcal P_{k+1}(a)$ inherits $a$ from $\mathcal P_k(a)$. Thus,
$(k,\mathcal P_k(a))$ is the parent of
$(k+1,\mathcal P_{k+1}(a))$, proving the claim.
\end{proof}


\begin{figure}[!tbp]
\centering
\begingroup
\definecolor{mcVertexFill}{rgb}{0.61338,0.86345,0.97537}
\definecolor{mcVertexDraw}{rgb}{0.4,0.7,1.0}
\definecolor{mcInk}{rgb}{0.15,0.15,0.15}
\definecolor{mcMagenta}{rgb}{0.91472,0.46078,0.62707}
\definecolor{mcPurple}{rgb}{0.34749,0.1,0.75250}
\definecolor{mcOrange}{rgb}{0.85,0.425,0}
\definecolor{mcGreen}{rgb}{0,0.6,0}
\definecolor{mcGray}{rgb}{0.55,0.55,0.55}
\definecolor{mcPale}{rgb}{0.92587,0.97171,0.99509}
\tikzset{
  mcfigure/.style={x=1cm,y=1cm,font=\small,>=Latex},
  baseedge/.style={draw=mcInk,line width=.75pt},
  vertex/.style={circle,draw=mcVertexDraw,fill=mcVertexFill,line width=.55pt,minimum size=4.6pt,inner sep=0pt,outer sep=0pt},
  linkmag/.style={draw=mcMagenta,line width=1pt,dashed},
  linkpurple/.style={draw=mcPurple,line width=1pt,dashed},
  linkorange/.style={draw=mcOrange,line width=1pt,densely dotted},
  linkgray/.style={draw=mcGray,line width=.85pt,dashed},
  pathgreen/.style={draw=mcGreen,line width=1.6pt},
  markedge/.style={draw=mcMagenta,line width=1.6pt},
  contractedge/.style={draw=black!28,line width=1.6pt},
  region/.style={draw=mcGray,dashed,fill=mcPale,line width=.55pt,rounded corners=7pt},
  block/.style={draw=mcGray,fill=mcPale,line width=.55pt,rounded corners=4pt,minimum height=.65cm,inner sep=5pt},
  selectedblock/.style={block,draw=mcMagenta,line width=1pt},
  guide/.style={draw=black!35,line width=.5pt,densely dashed},
  maparrow/.style={-{Latex[length=2mm]},draw=mcGray,line width=.65pt},
  paneltitle/.style={font=\small,align=center},
  smallnote/.style={font=\footnotesize,align=center},
  endpointlabel/.style={font=\small,inner sep=2pt}
}
\long\def\mcReductionFit#1{%
  \sbox0{#1}%
  \ifdim\wd0>\linewidth
    \resizebox{\linewidth}{!}{\usebox0}%
  \else
    \usebox0%
  \fi
}
\mcReductionFit{%
\begin{tikzpicture}[mcfigure]
\coordinate (v01) at (0.0000,-0.0000);
\coordinate (v11) at (0.0000,-0.5800);
\coordinate (v21) at (0.0000,-1.1600);
\coordinate (v31) at (0.0000,-1.7400);
\coordinate (v41) at (0.0000,-2.3200);
\coordinate (v51) at (-2.8000,-2.9000);
\coordinate (v52) at (2.8000,-2.9000);
\coordinate (v61) at (-2.8000,-3.4800);
\coordinate (v62) at (2.8000,-3.4800);
\coordinate (v71) at (-2.8000,-4.0600);
\coordinate (v72) at (2.8000,-4.0600);
\coordinate (v81) at (-2.8000,-4.6400);
\coordinate (v82) at (1.4000,-4.6400);
\coordinate (v83) at (4.2000,-4.6400);
\coordinate (v91) at (-2.8000,-5.2200);
\coordinate (v92) at (1.4000,-5.2200);
\coordinate (v93) at (4.2000,-5.2200);
\coordinate (v101) at (-4.2000,-5.8000);
\coordinate (v102) at (-1.4000,-5.8000);
\coordinate (v103) at (1.4000,-5.8000);
\coordinate (v104) at (4.2000,-5.8000);
\coordinate (v111) at (-4.2000,-6.3800);
\coordinate (v112) at (-1.4000,-6.3800);
\coordinate (v113) at (1.4000,-6.3800);
\coordinate (v114) at (4.2000,-6.3800);
\coordinate (v121) at (-4.2000,-6.9600);
\coordinate (v122) at (-1.4000,-6.9600);
\coordinate (v123) at (1.4000,-6.9600);
\coordinate (v124) at (4.2000,-6.9600);
\node[smallnote] at (-6.80,.46) {layer $i$};
\node[smallnote] at (-5.50,.46) {inserted};
\node[smallnote,text=mcGray] at (-6.80,-0.0000) {$0$};
\node[smallnote,text=mcGray] at (-6.80,-0.5800) {$1$};
\node[smallnote,text=mcGray] at (-5.50,-0.5800) {$a_{1}$};
\node[smallnote,text=mcGray] at (-6.80,-1.1600) {$2$};
\node[smallnote,text=mcGray] at (-5.50,-1.1600) {$a_{2}$};
\node[smallnote,text=mcGray] at (-6.80,-1.7400) {$3$};
\node[smallnote,text=mcGray] at (-5.50,-1.7400) {$a_{3}$};
\node[smallnote,text=mcGray] at (-6.80,-2.3200) {$4$};
\node[smallnote,text=mcGray] at (-5.50,-2.3200) {$a_{4}$};
\node[smallnote,text=mcGray] at (-6.80,-2.9000) {$5$};
\node[smallnote,text=mcGray] at (-5.50,-2.9000) {$a_{5}$};
\node[smallnote,text=mcGray] at (-6.80,-3.4800) {$6$};
\node[smallnote,text=mcGray] at (-5.50,-3.4800) {$a_{6}$};
\node[smallnote,text=mcGray] at (-6.80,-4.0600) {$7$};
\node[smallnote,text=mcGray] at (-5.50,-4.0600) {$a_{7}$};
\node[smallnote,text=mcGray] at (-6.80,-4.6400) {$8$};
\node[smallnote,text=mcGray] at (-5.50,-4.6400) {$a_{8}$};
\node[smallnote,text=mcGray] at (-6.80,-5.2200) {$9$};
\node[smallnote,text=mcGray] at (-5.50,-5.2200) {$a_{9}$};
\node[smallnote,text=mcGray] at (-6.80,-5.8000) {$10$};
\node[smallnote,text=mcGray] at (-5.50,-5.8000) {$a_{10}$};
\node[smallnote,text=mcGray] at (-6.80,-6.3800) {$11$};
\node[smallnote,text=mcGray] at (-5.50,-6.3800) {$a_{11}$};
\node[smallnote,text=mcGray] at (-6.80,-6.9600) {$12$};
\node[smallnote,text=mcGray] at (-5.50,-6.9600) {$a_{12}$};
\draw[baseedge] (v01)--(v11);
\draw[baseedge] (v11)--(v21);
\draw[baseedge] (v21)--(v31);
\draw[baseedge] (v31)--(v41);
\draw[baseedge] (v41)--(v51);
\draw[baseedge] (v41)--(v52);
\draw[baseedge] (v51)--(v61);
\draw[baseedge] (v52)--(v62);
\draw[baseedge] (v61)--(v71);
\draw[baseedge] (v62)--(v72);
\draw[baseedge] (v71)--(v81);
\draw[baseedge] (v72)--(v82);
\draw[baseedge] (v72)--(v83);
\draw[baseedge] (v81)--(v91);
\draw[baseedge] (v82)--(v92);
\draw[baseedge] (v83)--(v93);
\draw[baseedge] (v91)--(v101);
\draw[baseedge] (v91)--(v102);
\draw[baseedge] (v92)--(v103);
\draw[baseedge] (v93)--(v104);
\draw[baseedge] (v101)--(v111);
\draw[baseedge] (v102)--(v112);
\draw[baseedge] (v103)--(v113);
\draw[baseedge] (v104)--(v114);
\draw[baseedge] (v111)--(v121);
\draw[baseedge] (v112)--(v122);
\draw[baseedge] (v113)--(v123);
\draw[baseedge] (v114)--(v124);
\draw[linkmag] (v11) .. controls (-5.55,-.90) and (-5.40,-6.20) .. (v121)
  node[pos=.49,right=3pt,smallnote,text=mcMagenta,fill=white,inner sep=1pt] {$\widetilde u_{a_2}$};
\draw[linkpurple] (v52) .. controls (6.10,-3.12) and (5.90,-6.35) .. (v124)
  node[pos=.49,right=2pt,smallnote,text=mcPurple,fill=white,inner sep=1pt] {$\widetilde u_{a_6}$};
\draw[linkorange] (v81) .. controls (-.08,-4.96) and (.05,-6.33) .. (v122)
  node[pos=.61,right=2pt,smallnote,text=mcOrange,fill=white,inner sep=1pt] {$\widetilde u_{a_9}$};
\node[vertex] at (v01) {};
\node[vertex] at (v11) {};
\node[vertex] at (v21) {};
\node[vertex] at (v31) {};
\node[vertex] at (v41) {};
\node[vertex] at (v51) {};
\node[vertex] at (v52) {};
\node[vertex] at (v61) {};
\node[vertex] at (v62) {};
\node[vertex] at (v71) {};
\node[vertex] at (v72) {};
\node[vertex] at (v81) {};
\node[vertex] at (v82) {};
\node[vertex] at (v83) {};
\node[vertex] at (v91) {};
\node[vertex] at (v92) {};
\node[vertex] at (v93) {};
\node[vertex] at (v101) {};
\node[vertex] at (v102) {};
\node[vertex] at (v103) {};
\node[vertex] at (v104) {};
\node[vertex] at (v111) {};
\node[vertex] at (v112) {};
\node[vertex] at (v113) {};
\node[vertex] at (v114) {};
\node[vertex] at (v121) {};
\node[vertex] at (v122) {};
\node[vertex] at (v123) {};
\node[vertex] at (v124) {};
\node[smallnote,right=4pt] at (v01) {$(0,\varnothing)$};
\node[smallnote,above right=3pt,inner sep=1pt] at (v41) {$\{a_{2},a_{1},a_{4},a_{3}\}$};
\node[smallnote,below right=2pt,inner sep=1pt] at (v51) {$\{a_2,a_5\}$};
\node[smallnote,below left=2pt,inner sep=1pt] at (v52) {$\{a_1,a_4,a_3\}$};
\node[smallnote,below=5pt] at (v121) {$\{a_{7},a_{11},a_{2}\}$};
\node[smallnote,below=5pt] at (v122) {$\{a_{9},a_{5},a_{10}\}$};
\node[smallnote,below=5pt] at (v123) {$\{a_{1},a_{12},a_{8}\}$};
\node[smallnote,below=5pt] at (v124) {$\{a_{4},a_{3},a_{6}\}$};
\end{tikzpicture}%
}
\endgroup
\caption{The history of partitions and their constructed respective tree for tail order $a_1,\ldots,a_{12}$ and head order $a_7,\allowbreak a_{11},\allowbreak a_2,\allowbreak a_9,\allowbreak a_5,\allowbreak a_{10},\allowbreak a_1,\allowbreak a_{12},\allowbreak a_8,\allowbreak a_4,\allowbreak a_3,\allowbreak a_6$. All $x_a=1/20$. When a part reaches five links, its mass is $1/4$ and has to split. The colored links show $\widetilde u_a=\operatorname{par}(B(a))F(a)$ for $a\in\{a_2,a_6,a_9\}$.}
\label{fig:reduction-partition-tree}
\end{figure}
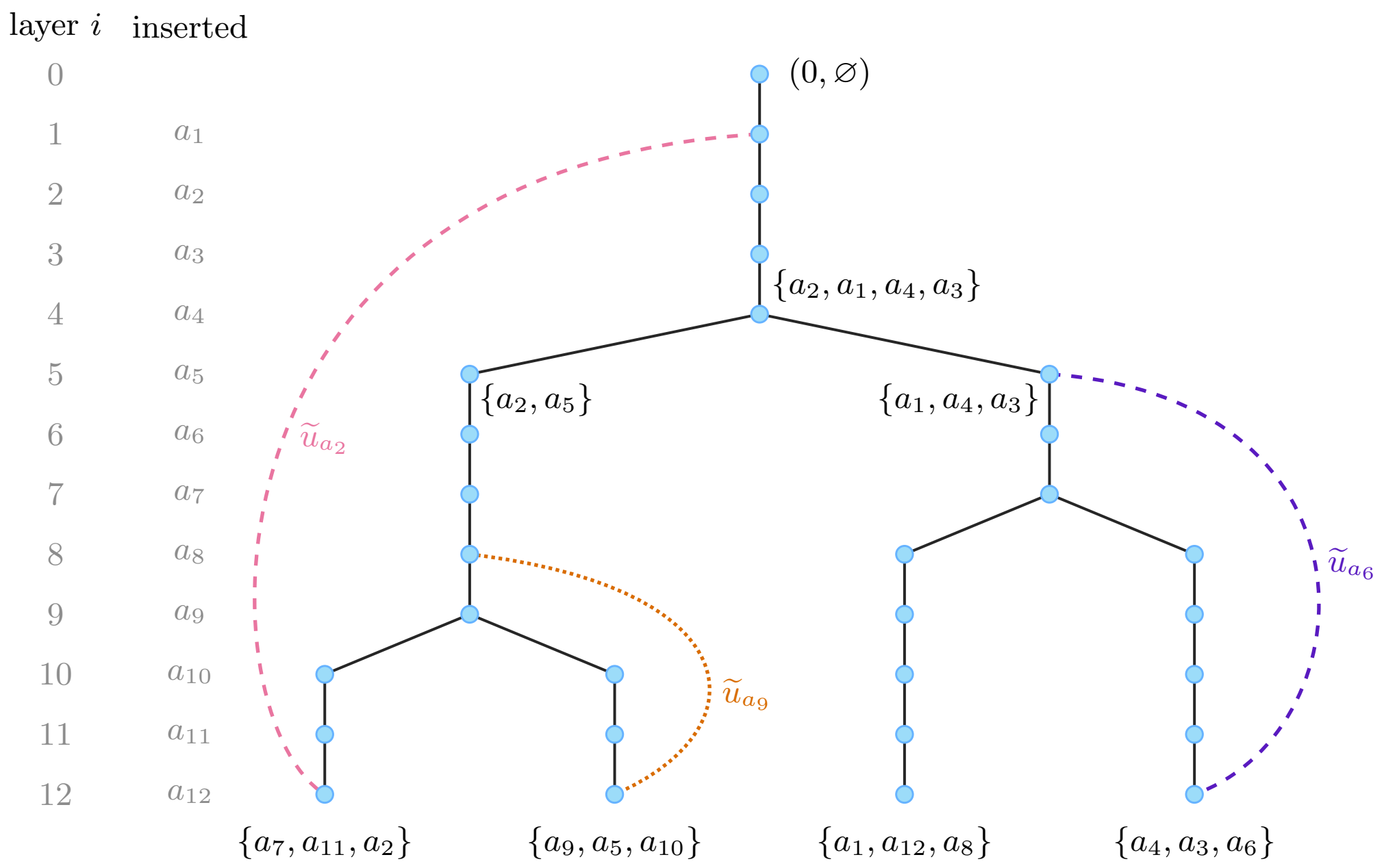

\paragraph{Encoding the cuts and the links.}

For each $i\geq1$ and $P\in\mathcal P_i$, let $e_{i,P}$ denote the edge joining $(i,P)$ to its parent. Define $F:=\{e_{i(s),P(s,t)}:[s,t]\in\mathcal L\}\subseteq E(T)$, where $P(s,t)$ and $i(s)$ are defined as in \cref{def:Pst}. These edges are well-defined since $i(s)\geq1$ by \cref{clm:partition-covers-cut}. Thus, $F$ encodes the cuts in $\mathcal L$ that we aim to cover. Note that multiple cuts may correspond to the same edge, which appears only once in $F$. 

For each link $a=a_q\in L$, define $B(a):=(q,\mathcal P_q(a))$ and $F(a):=(m,\mathcal P_m(a))$, corresponding to the first and final parts containing $a$, respectively. Associate with $a$ the uplink $\widetilde u_a:=\operatorname{par}(B(a))F(a)$, which is well-defined by \cref{clm:partition-hierarchy}. The partition tree and examples of these uplinks are illustrated in \cref{fig:reduction-partition-tree}.

We obtain $T_1$ by contracting all edges in $E(T)\setminus F$, retaining only those encoding cuts in $\mathcal L$. The resulting graph remains a rooted tree, and each uplink remains an uplink.
For each $a\in L$, let $u_a$ denote the image of $\widetilde u_a$ under the contractions, and define $U_1:=\{u_a:a\in L\}$, viewed as a multiset. We treat $U_1$ as a multiset to distinguish distinct copies of links whose endpoints coincide after contraction in the tree. This induces a bijection $\varphi_1:U_1\to L$, where, for every $a\in L$ and $i\in[\ell]$, we define
\[
\varphi_1(u_a)=a,\qquad
c_i^{(1)}(u_a)=c_i(a),\qquad
z_{u_a}^{(1)}=\min(1, 10x_a).
\] 
The construction of $(T_2,U_2)$ for $\mathcal R$ is symmetric, and $\varphi_2$, $c_i^{(2)}$, and $z^{(2)}$ are defined analogously.

\paragraph{Verifying the reduction.}
The first two properties of \cref{thm:reduction} follow from the
definitions. The following claim gives the remaining ingredients needed to complete the proof. We only prove the properties for the tree $T_1$ constructed for the min-cuts in $\mathcal{L}$, by the symmetry of our construction, the proof for $T_2$ also follows. Let $T$ be the tree $T_1$ before contracting edges outside of $F$. Recall that for a tree $T'$ and link $u$, $P_{T'}(u)$ denotes the unique path between the endpoints of $u$ in $T'$.

\begin{claim}
\label{clm:uplinks-cover-parts}
For every $a = a_q \in L$, $q \leq i \leq m$, and $P\in\mathcal P_i$, $e_{i,P}\in P_T(\widetilde u_a)$ if and only if $a \in P$. In other words, covering the min-cut corresponding to the edge $e_{i,P}$ in $T$ is equivalent to picking an uplink corresponding to one of the links in $P$.
\end{claim}

\begin{proof}
By iteratively applying \cref{clm:partition-hierarchy}, the path $P_T(\widetilde u_a)$ consists precisely of the edges $e_{k,\mathcal P_k(a)}$ for $q\leq k\leq m$. Thus, for $i \geq q$, $e_{i,P}\in P_T(\widetilde u_a)$ if and only if $P=\mathcal P_i(a)$, equivalently, $a\in P$.

\end{proof}

Note that contracting edges other than $e_{i,P}$ does not change whether the endpoints of a link lie in different
components of $T\setminus e_{i,P}$. Thus, \cref{clm:uplinks-cover-parts} implies that if $e_{i,P}\in F$, then, $\{a\in L:e_{i,P}\in P_{T_1}(u_a)\}=P$.


By \cref{clm:uplinks-cover-parts,clm:partition-covers-cut}, any feasible integral solution to the $\mcutap$ instance $(T_1,U_1)$ maps under $\varphi_1$ to a set of links covering all cuts in $\mathcal L$.
We next show that $z^{(1)}$ is a feasible fractional solution to this $\mcutap$ instance. Consider any edge $e\in E(T_1)$. By construction, $e=e_{i(s),P(s,t)}$ for some $[s,t]\in\mathcal L$. Thus, by \cref{clm:uplinks-cover-parts}, for  $e = e_{i(s),P(s,t)}$, an uplink $u_a \in U_1$ covers the cut corresponding to $e$ if and only if $a \in P(s,t)$. Therefore,

\begin{equation}
    \label{eq:reduction-cut-cover-1}
    \sum_{u_a\in U_1:\,e\in P_{T_1}(u_a)}z_{u_a}^{(1)}  =  \sum_{a\in P(s,t)}z_{u_a}^{(1)}. 
\end{equation}

Now, recall that $z_{u_a}^{(1)} = \min(1, 10 x_a)$, therefore, if for at least one $a\in P(s,t)$, $z_{u_a}^{(1)} = 1$, then it trivially holds that $\sum_{a\in P(s,t)}z_{u_a}^{(1)} \geq 1$. Otherwise,
\begin{equation}
\label{eq:reduction-cut-cover-2}
    \sum_{a\in P(s,t)}z_{u_a}^{(1)} = 10\sum_{a\in P(s,t)}x_ a\geq 1
\end{equation}

where the last inequality follows as $x(P(s,t))\geq\frac1{10}$ by \cref{clm:partition-covers-cut}. Combining \eqref{eq:reduction-cut-cover-1}, \eqref{eq:reduction-cut-cover-2} implies that $z^{(1)}$ is a feasible fractional $\mcutap$ solution for this instance.

Moreover, for each $i\in[\ell]$, the bijection
$\varphi_1$ gives
\[
    \sum_{u_a\in U_1}c_i^{(1)}(u_a)z_{u_a}^{(1)}
    \leq\sum_{a\in L}c_i(a)10x_a
    \leq10\sum_{a\in L}c_i(a)x_a.
\]
This proves the third property of our construction. 

For the fourth property, let $S_1\subseteq U_1$
be feasible integral solution for the $\mcutap$ instance on $T_1$ with uplinks $U_1$. Fix a cut $S=[s,t]\in\mathcal L$. Since $S_1$ covers the edge $e_{i(s),P(s,t)}$, by \cref{clm:uplinks-cover-parts}, it contains an uplink $u_a$ such that $a \in P(s,t)$ covering that edge. This implies $a \in \varphi_1(S_1)$, meanwhile, by \cref{clm:partition-covers-cut}, $P(s,t) \subseteq \mathcal{L}(s,t)$, and thus $a \in \varphi_1(S_1)$ covers the cut $[s,t] \in \mathcal{L}$.

Symmetrically, consider a feasible solution $S_2$ to the $\mcutap$ instance on $T_2$ with uplinks $U_2$, every cut in $\mathcal R$ is covered by a link of $\varphi_2(S_2)$. 

Since $\mathcal{W} = \mathcal{L} \cup \mathcal{R}$, the set
$\varphi_1(S_1)\cup\varphi_2(S_2)$ covers every cut in $\mathcal W$.
Finally, for any $i \in [\ell]$, the cost of this solution under cost function $c_i$ is simply bounded as,
\begin{align*}
    c_i\bigl(\varphi_1(S_1)\cup\varphi_2(S_2)\bigr)
    \leq
    \sum_{u\in S_1}c_i(\varphi_1(u))
    +\sum_{u\in S_2}c_i(\varphi_2(u))
    =c_i^{(1)}(S_1)+c_i^{(2)}(S_2).
\end{align*}

\paragraph{Runtime Analysis for the Construction} It remains to note that the construction takes polynomial time. Since
$\mathcal P_i$ is a partition of $i$ links into nonempty parts for $i\geq1$, each $i$ satisfies $|\mathcal{P}_i| \leq i \leq m$. Since the tree has $m+1$ layers at most,
\[
    |V(T)|
    =1+\sum_{i=1}^m|\mathcal P_i|
    \leq1+\frac{m(m+1)}2 = O(m^2).
\]
All partition updates, choices of $P(s,t)$, and contractions can be performed in polynomial time as well. The same holds for the construction for $\mathcal R$ which concludes the proof.

\end{proof}
\section{Rounding a Fractional $\mcutap$ Solution} 
\label{sec:rounding}
In this section, we describe the Uplink Tree Augmentation problem ($\utap$) and explain how a fractional solution to $\utap$ can be rounded to an integral one efficiently. Consider a tree $T$ rooted at a vertex $r$ with a set of uplinks $L^\uparrow$.

Since the min-cuts of $T$ are the edges of $T$, the LP relaxation of $\mcutap$ is as follows:
\begin{equation*}
\boxed{
\begin{aligned}
\mcutaplp\qquad
&\begin{aligned}[t]
    \min \quad & Z \\[3pt]
    \text{s.t.}\quad
     & \sum_{a: e \in P(a)} x_a \geq 1 && \forall e \in E(T) \\
     & \sum_{a \in L^{\uparrow}} x_a c_i(a) \leq Z && \forall i \in [\ell]\\
     & Z, x \geq 0
\end{aligned}
\end{aligned}
}
\end{equation*}

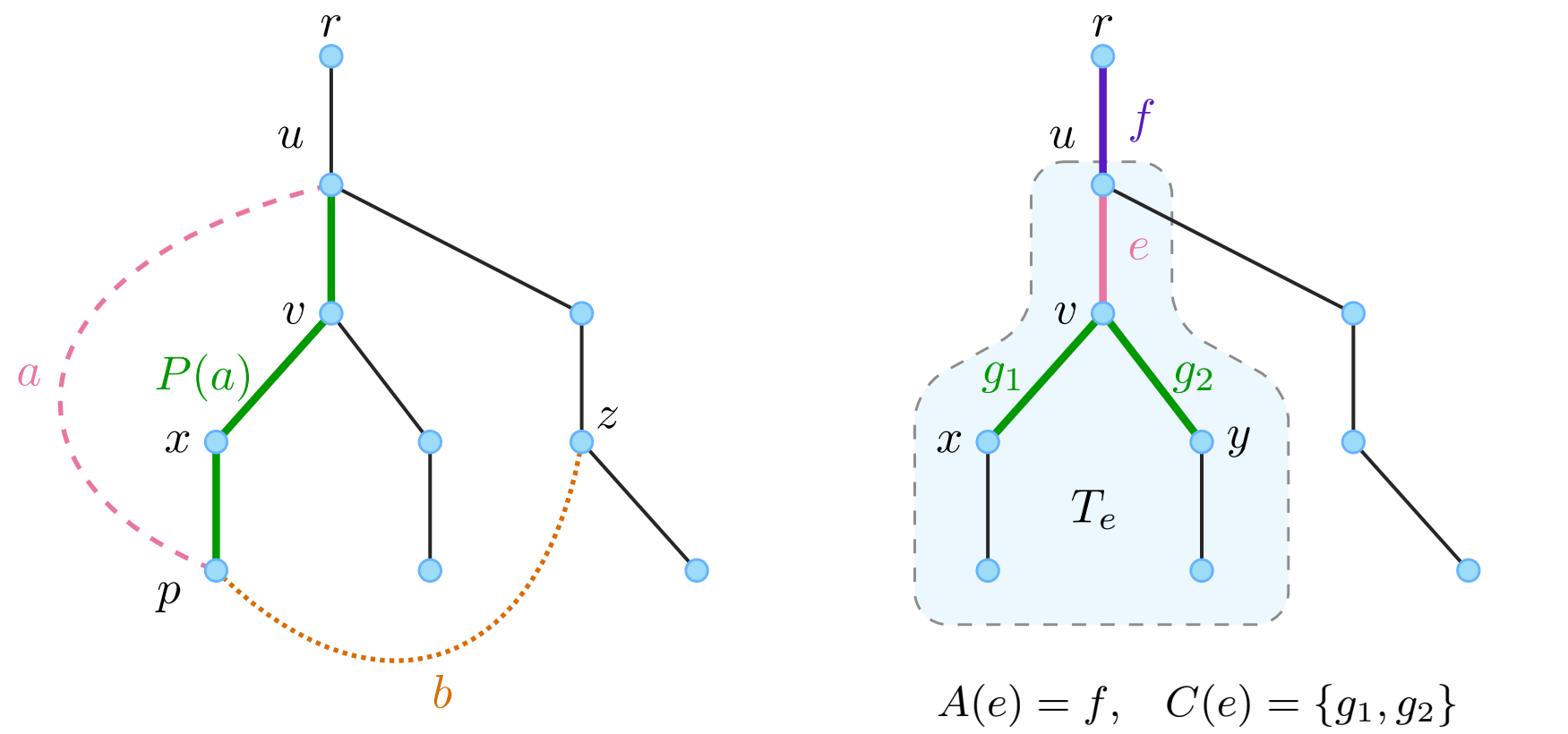
\begin{figure}[!tbp]
\centering
\begingroup
\definecolor{mcVertexFill}{rgb}{0.61338,0.86345,0.97537}
\definecolor{mcVertexDraw}{rgb}{0.4,0.7,1.0}
\definecolor{mcInk}{rgb}{0.15,0.15,0.15}
\definecolor{mcMagenta}{rgb}{0.91472,0.46078,0.62707}
\definecolor{mcPurple}{rgb}{0.34749,0.1,0.75250}
\definecolor{mcOrange}{rgb}{0.85,0.425,0}
\definecolor{mcGreen}{rgb}{0,0.6,0}
\definecolor{mcGray}{rgb}{0.55,0.55,0.55}
\definecolor{mcPale}{rgb}{0.92587,0.97171,0.99509}
\tikzset{
  mcfigure/.style={x=1cm,y=1cm,font=\small,>=Latex},
  baseedge/.style={draw=mcInk,line width=.75pt},
  vertex/.style={circle,draw=mcVertexDraw,fill=mcVertexFill,line width=.55pt,minimum size=4.6pt,inner sep=0pt,outer sep=0pt},
  linkmag/.style={draw=mcMagenta,line width=1pt,dashed},
  linkpurple/.style={draw=mcPurple,line width=1pt,dashed},
  linkorange/.style={draw=mcOrange,line width=1pt,densely dotted},
  linkgray/.style={draw=mcGray,line width=.85pt,dashed},
  linkfaint/.style={draw=black!32,line width=.9pt,dashed},
  pathgreen/.style={draw=mcGreen,line width=1.6pt},
  markedge/.style={draw=mcMagenta,line width=1.6pt},
  markpurple/.style={draw=mcPurple,line width=1.6pt},
  contractedge/.style={draw=black!28,line width=1.6pt},
  region/.style={draw=mcGray,dashed,fill=mcPale,line width=.55pt,rounded corners=7pt},
  block/.style={draw=mcGray,fill=mcPale,line width=.55pt,rounded corners=4pt,minimum height=.65cm,inner sep=5pt},
  selectedblock/.style={block,draw=mcMagenta,line width=1pt},
  guide/.style={draw=black!35,line width=.5pt,densely dashed},
  maparrow/.style={-{Latex[length=2mm]},draw=mcGray,line width=.65pt},
  paneltitle/.style={font=\small,align=center},
  smallnote/.style={font=\footnotesize,align=center},
  endpointlabel/.style={font=\small,inner sep=2pt}
}
\long\def\mcReductionFit#1{%
  \sbox0{#1}%
  \ifdim\wd0>\linewidth
    \resizebox{\linewidth}{!}{\usebox0}%
  \else
    \usebox0%
  \fi
}
\mcReductionFit{%
\begin{tikzpicture}[mcfigure]
  \def\mcDx{5.7}
  \def\mcTreeNodes{r/0/0, u/0/-0.95, v/0/-1.90, s/1.85/-1.90,
    x/-0.85/-2.85, y/0.73/-2.85, z/1.85/-2.85,
    p/-0.85/-3.80, q/0.73/-3.80, t/2.70/-3.80}
  \def\mcTreeEdges{r/u, u/v, u/s, v/x, v/y, x/p, y/q, s/z, z/t}
  \foreach \n/\px/\py in \mcTreeNodes {\coordinate (A\n) at (\px,\py);}
  \foreach \n/\px/\py in \mcTreeNodes {\coordinate (B\n) at ({\px+\mcDx},\py);}

  \foreach \i/\j in \mcTreeEdges {\draw[baseedge] (A\i)--(A\j);}
  \draw[pathgreen] (Au)--(Av)--(Ax)--(Ap);
  \draw[linkmag] (Au) .. controls (-2.45,-1.55) and (-2.55,-3.15) .. (Ap)
    node[pos=.5,left=2pt,endpointlabel,text=mcMagenta] {$a$};
  \draw[linkorange] (Ap) .. controls (0.30,-4.95) and (1.55,-4.60) .. (Az)
    node[pos=.5,below=2pt,endpointlabel,text=mcOrange] {$b$};
  \path (Av)--node[endpointlabel,left=2pt,text=mcGreen] {$P(a)$} (Ax);
  \foreach \n/\px/\py in \mcTreeNodes {\node[vertex] at (A\n) {};}
  \node[endpointlabel,above=2pt] at (Ar) {$r$};
  \node[endpointlabel] at (-0.30,-0.60) {$u$};
  \node[endpointlabel,left=3pt] at (Av) {$v$};
  \node[endpointlabel,left=3pt] at (Ax) {$x$};
  \node[endpointlabel] at (-1.20,-4.00) {$p$};
  \node[endpointlabel,above right=1pt] at (Az) {$z$};
  \node[paneltitle] at (0.17,-5.45) {(a) uplinks and cross links};

  \draw[region] ({\mcDx-0.53},-0.78)--({\mcDx+0.51},-0.78)
    --({\mcDx+0.51},-1.98)--({\mcDx+1.37},-2.45)--({\mcDx+1.37},-4.20)
    --({\mcDx-1.39},-4.20)--({\mcDx-1.39},-2.45)--({\mcDx-0.53},-1.98)--cycle;
  \foreach \i/\j in \mcTreeEdges {\draw[baseedge] (B\i)--(B\j);}
  \draw[markpurple] (Br)--(Bu);
  \draw[markedge] (Bu)--(Bv);
  \draw[pathgreen] (Bv)--(Bx) (Bv)--(By);
  \path (Br)--node[endpointlabel,right=3pt,text=mcPurple] {$f$} (Bu);
  \path (Bu)--node[endpointlabel,right=3pt,text=mcMagenta] {$e$} (Bv);
  \path (Bv)--node[endpointlabel,left=2pt,text=mcGreen] {$g_1$} (Bx);
  \path (Bv)--node[endpointlabel,right=2pt,text=mcGreen] {$g_2$} (By);
  \node[endpointlabel] at ({\mcDx-0.06},-3.35) {$T_e$};
  \foreach \n/\px/\py in \mcTreeNodes {\node[vertex] at (B\n) {};}
  \node[endpointlabel,above=2pt] at (Br) {$r$};
  \node[endpointlabel] at ({\mcDx-0.30},-0.60) {$u$};
  \node[endpointlabel,left=3pt] at (Bv) {$v$};
  \node[endpointlabel,left=3pt] at (Bx) {$x$};
  \node[endpointlabel,right=3pt] at (By) {$y$};
  \node[smallnote] at ({\mcDx+0.70},-4.80)
    {$A(e)=f$,\quad $C(e)=\{g_1,g_2\}$};
  \node[paneltitle] at ({\mcDx+0.70},-5.45) {(b) parent and children of an edge};
\end{tikzpicture}%
}
\endgroup
\caption{Our notation for rooted trees, illustrated on one tree $T$ rooted at
$r$.
(a)~The pink dashed link $a=up$ is an
uplink, because $u$ is an ancestor of $p$. The green edges are the tree
edges of the path $P(a)$ joining the endpoints of $a$. The orange dotted link $b=pz$ is a cross link, since neither endpoint is an ancestor of the other.
(b)~Consider the edge $e$ highlighted in pink. The purple edge $f=A(e)$ is the
unique edge immediately above $e$, and the green edges $g_1$ and $g_2$ are the
edges immediately below it, so $C(e)=\{g_1,g_2\}$. The shaded region is the
subtree $T_e$, which consists of $e$ together with all edges below it. Here
$f\in\delta_T(r)$, so $f$ itself has no parent edge.}
\label{fig:utap-definitions}
\end{figure}

For each edge $e = uv \in E(T)$, where $u$ is the parent of $v$, $C(e)$ denotes the set of child edges of $e$, i.e., all edges incident to $v$ other than $e$. Similarly, for each edge $e \in E(T) \setminus \delta_T(r)$, $A(e)$ denotes the unique parent edge of $e$, i.e., the unique edge satisfying $e \in C(A(e))$. We let $T_e$ denote the subtree consisting of $e$ and all its descendant edges (see \cref{fig:utap-definitions}).

\subsection{Polyhedral Properties and Dual Structure of UTAP} \label{sec:dual-prop}
We begin by studying the polyhedral structure of $\utap$. Given a nonnegative cost function $c : L^{\uparrow} \rightarrow \mathbb{R}{\geq 0}$, consider the following natural LP relaxation of $\utap$, and its dual:

\noindent
\begin{minipage}[t]{0.49\linewidth}
\centering
\small
\begin{equation*}
\boxed{
\begin{gathered}
\utaplp \\[3pt]
\begin{aligned}
\min \quad & \sum_{a \in L^\uparrow} x_a c(a) \\
\text{s.t.}\quad
& \sum_{a:e\in P(a)} x_a \geq 1
&& \forall e \in E(T) \\
& x \geq 0
\end{aligned}
\end{gathered}
}
\end{equation*}
\end{minipage}%
\hfill
\begin{minipage}[t]{0.49\linewidth}
\centering
\small
\begin{equation*}
\boxed{
\begin{gathered}
\dutaplp \\[3pt]
\begin{aligned}
\max \quad & \sum_{e\in E(T)} y_e \\
\text{s.t.}\quad
& \sum_{e\in P(a)} y_e \leq c(a)
&& \forall a\in L^\uparrow \\
& y_e \geq 0
&& \forall e\in E(T)
\end{aligned}
\end{gathered}
}
\end{equation*}
\end{minipage}
\par\addvspace{\baselineskip}

It is well known that the constraint matrix of $\utaplp$ is totally unimodular, as can be shown using the Ghouila-Houri criterion \cite{AdjiashviliTAP2017}. Consequently, the feasible polyhedron of $\utap$ is integral.
We further provide a precise characterization of the structure of optimal solutions to $\dutaplp$.

\begin{lemma}
\label{lem:UTAP-Dual}
Let $T$ be a tree rooted at $r$, let $L^{\uparrow}$
be a set of uplinks, and let
$c:L^{\uparrow}\rightarrow\mathbb{R}_{\geq 0}$
be a cost function. Assume the instance is feasible.
For each $e\in E(T)$, let
$O_e=\optutap(T_e,c)$ denote the minimum cost of covering all edges of $T_e$ using links from $L^{\uparrow}$. The solution $y^*$ defined by $y_e^* = O_e-\sum_{g\in C(e)}O_g$ for every $e \in E(T)$, is an optimal solution for $\dutaplp(T, c)$.
\end{lemma}

Before proceeding with the proof, the following remark explains what covering all edges in $T_e$ exactly means.

\begin{figure}[!tbp]
\centering
\begingroup
\definecolor{mcVertexFill}{rgb}{0.61338,0.86345,0.97537}
\definecolor{mcVertexDraw}{rgb}{0.4,0.7,1.0}
\definecolor{mcInk}{rgb}{0.15,0.15,0.15}
\definecolor{mcMagenta}{rgb}{0.91472,0.46078,0.62707}
\definecolor{mcPurple}{rgb}{0.34749,0.1,0.75250}
\definecolor{mcOrange}{rgb}{0.85,0.425,0}
\definecolor{mcGreen}{rgb}{0,0.6,0}
\definecolor{mcGray}{rgb}{0.55,0.55,0.55}
\definecolor{mcPale}{rgb}{0.92587,0.97171,0.99509}
\tikzset{
  mcfigure/.style={x=1cm,y=1cm,font=\small,>=Latex},
  baseedge/.style={draw=mcInk,line width=.75pt},
  vertex/.style={circle,draw=mcVertexDraw,fill=mcVertexFill,line width=.55pt,minimum size=4.6pt,inner sep=0pt,outer sep=0pt},
  linkmag/.style={draw=mcMagenta,line width=1pt,dashed},
  linkpurple/.style={draw=mcPurple,line width=1pt,dashed},
  linkorange/.style={draw=mcOrange,line width=1pt,densely dotted},
  linkgray/.style={draw=mcGray,line width=.85pt,dashed},
  linkfaint/.style={draw=black!32,line width=.9pt,dashed},
  pathgreen/.style={draw=mcGreen,line width=1.6pt},
  markedge/.style={draw=mcMagenta,line width=1.6pt},
  markpurple/.style={draw=mcPurple,line width=1.6pt},
  contractedge/.style={draw=black!28,line width=1.6pt},
  region/.style={draw=mcGray,dashed,fill=mcPale,line width=.55pt,rounded corners=7pt},
  block/.style={draw=mcGray,fill=mcPale,line width=.55pt,rounded corners=4pt,minimum height=.65cm,inner sep=5pt},
  selectedblock/.style={block,draw=mcMagenta,line width=1pt},
  guide/.style={draw=black!35,line width=.5pt,densely dashed},
  maparrow/.style={-{Latex[length=2mm]},draw=mcGray,line width=.65pt},
  paneltitle/.style={font=\small,align=center},
  smallnote/.style={font=\footnotesize,align=center},
  endpointlabel/.style={font=\small,inner sep=2pt}
}
\long\def\mcReductionFit#1{%
  \sbox0{#1}%
  \ifdim\wd0>\linewidth
    \resizebox{\linewidth}{!}{\usebox0}%
  \else
    \usebox0%
  \fi
}
\mcReductionFit{%
\begin{tikzpicture}[mcfigure]
  \def\mcDx{7.0}
  \def\mcTreeNodes{r/0/0, u/0/-0.95, v/0/-1.90, s/1.85/-1.90,
    x/-0.85/-2.85, y/0.73/-2.85, z/1.85/-2.85,
    p/-0.85/-3.80, q/0.73/-3.80, t/2.70/-3.80}
  \def\mcTreeEdges{r/u, u/v, u/s, v/x, v/y, x/p, y/q, s/z, z/t}
  \def\mcGoneEdges{r/u, u/s, s/z, z/t}
  \def\mcQuoNodes{U/0/0, v/0/-0.95,
    x/-0.85/-1.90, y/0.73/-1.90, p/-0.85/-2.85, q/0.73/-2.85}
  \def\mcQuoEdges{U/v, v/x, v/y, x/p, y/q}
  \foreach \n/\px/\py in \mcTreeNodes {\coordinate (A\n) at (\px,\py);}
  \foreach \n/\px/\py in \mcQuoNodes {\coordinate (C\n) at ({\px+\mcDx},\py);}

  \draw[region] (-0.53,-0.78)--(0.51,-0.78)--(0.51,-1.98)--(1.37,-2.45)
    --(1.37,-4.20)--(-1.39,-4.20)--(-1.39,-2.45)--(-0.53,-1.98)--cycle;
  \foreach \i/\j in \mcTreeEdges {\draw[baseedge] (A\i)--(A\j);}
  \foreach \i/\j in \mcGoneEdges {\draw[contractedge] (A\i)--(A\j);}
  \draw[markedge] (Au)--(Av);
  \draw[linkmag] (Au) .. controls (-2.45,-1.55) and (-2.55,-3.15) .. (Ap)
    node[pos=.5,right=2pt,endpointlabel,text=mcMagenta] {$a$};
  \draw[linkpurple] (Ar) .. controls (1.55,-1.15) and (1.45,-3.35) .. (Aq)
    node[pos=.5,right=2pt,endpointlabel,text=mcPurple] {$b$};
  \draw[linkgray] (Ar) .. controls (2.85,-1.10) and (3.05,-3.20) .. (At)
    node[pos=.5,right=2pt,endpointlabel,text=mcGray] {$c$};
  \draw[linkorange] (Av) .. controls (0.20,-2.85) and (0.35,-3.40) .. (Aq)
    node[pos=.5,left=3pt,endpointlabel,text=mcOrange] {$d$};
  \path (Au)--node[endpointlabel,right=3pt,text=mcMagenta] {$e$} (Av);
  \node[endpointlabel] at (-0.30,-3.50) {$T_e$};
  \foreach \n/\px/\py in \mcTreeNodes {\node[vertex] at (A\n) {};}
  \node[endpointlabel,above=2pt] at (Ar) {$r$};
  \node[endpointlabel] at (-0.30,-0.60) {$u$};
  \node[endpointlabel,left=3pt] at (Av) {$v$};
  \node[endpointlabel,below=3pt] at (Ap) {$p$};
  \node[endpointlabel,below=3pt] at (Aq) {$q$};
  \node[endpointlabel,below=3pt] at (At) {$t$};
  \node[paneltitle] at (0.41,-4.85) {(a) the original instance};

  \draw[maparrow] (3.25,-2.20)--(4.60,-2.20)
    node[midway,above,smallnote] {contract\\$E(T)\setminus T_e$};

  \foreach \i/\j in \mcQuoEdges {\draw[baseedge] (C\i)--(C\j);}
  \draw[markedge] (CU)--(Cv);
  \draw[linkmag] (CU)
    .. controls ({\mcDx-1.95},-0.55) and ({\mcDx-2.05},-2.55) .. (Cp)
    node[pos=.5,left=2pt,endpointlabel,text=mcMagenta] {$a$};
  \draw[linkpurple] (CU)
    .. controls ({\mcDx+2.05},-0.55) and ({\mcDx+2.15},-2.80) .. (Cq)
    node[pos=.5,right=2pt,endpointlabel,text=mcPurple] {$b$};
  \draw[linkorange] (Cv)
    .. controls ({\mcDx+0.20},-1.90) and ({\mcDx+0.35},-2.45) .. (Cq)
    node[pos=.5,left=3pt,endpointlabel,text=mcOrange] {$d$};
  \draw[linkgray] (CU)
    .. controls ({\mcDx+0.25},1.10) and ({\mcDx+1.15},0.60) .. (CU);
  \node[endpointlabel,text=mcGray] at ({\mcDx+0.80},0.68) {$c$};
  \path (CU)--node[endpointlabel,right=3pt,text=mcMagenta] {$e$} (Cv);
  \foreach \n/\px/\py in \mcQuoNodes {\node[vertex] at (C\n) {};}
  \node[endpointlabel] at ({\mcDx-0.38},0.26) {$u^{\star}$};
  \node[endpointlabel,left=3pt] at (Cv) {$v$};
  \node[endpointlabel,below=3pt] at (Cp) {$p$};
  \node[endpointlabel,below=3pt] at (Cq) {$q$};
  \node[paneltitle] at ({\mcDx+0.04},-4.85) {(b) the contracted instance $O_e$};
\end{tikzpicture}%
}
\endgroup
\caption{The instance, obtained by contracting everything outside $T_e$.
(a)~The original instance with the subtree $T_e$ shaded.
(b)~The resulting instance for $T_e$. The contraction identifies $u$ and every vertex outside $T_e$ into the single vertex $u^{\star}$, and each
link is transferred accordingly. The gray link $c=rt$ has both
endpoints identified with $u^{\star}$ and becomes a loop, drawn here only for
reference.}
\label{fig:utap-subtree-contraction}
\end{figure}
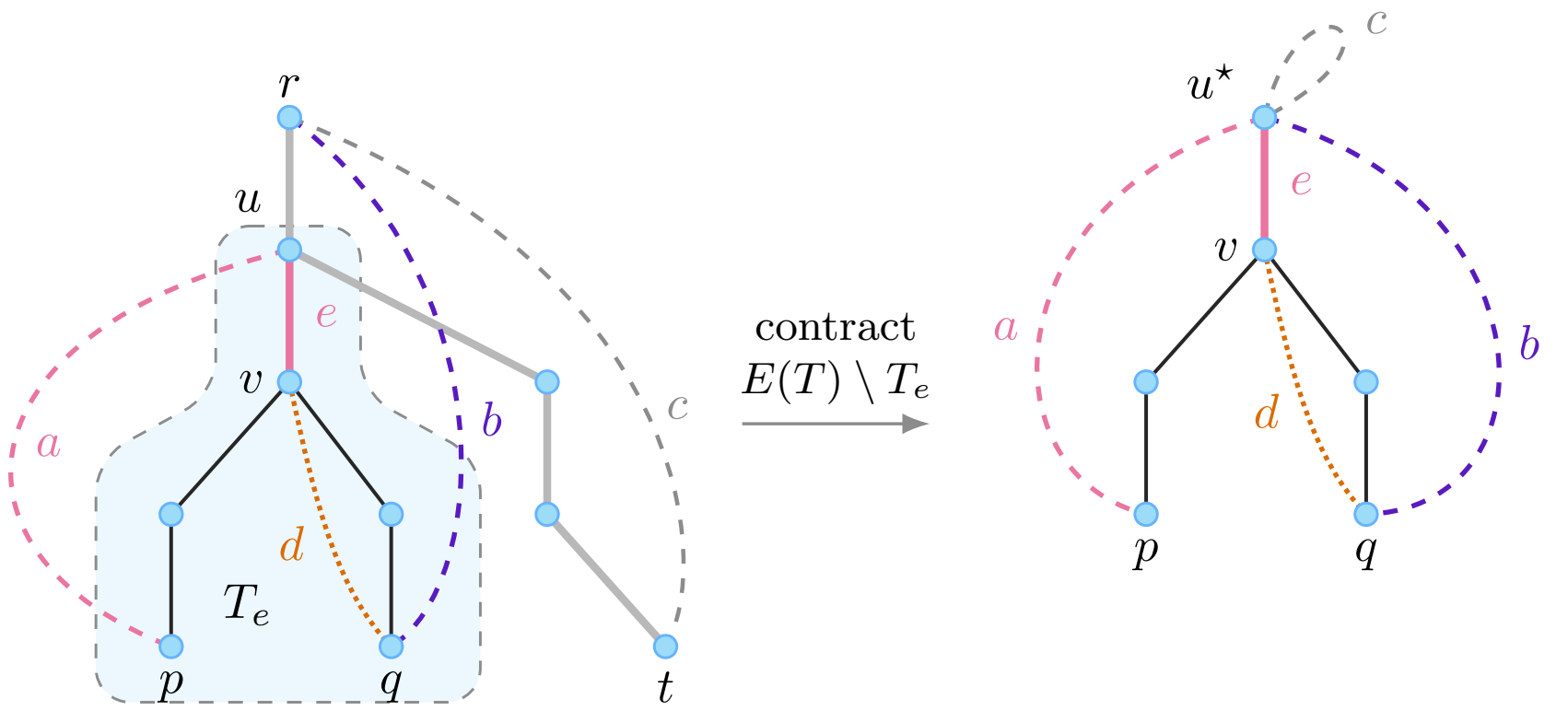

\begin{remark}
    Covering all edges of 
$T_e$ can be thought of as solving the $\utap$ instance obtained by contracting
$E(T)\setminus E(T_e)$ and inheriting the available links and their
costs. In particular, the root of this contracted tree is the image of the previous root $r$ and links entering $T_e$ from above remain available
with their upper endpoints mapped to the root of the contracted tree (see \cref{fig:utap-subtree-contraction} for more details).
\end{remark}

\begin{proof}[Proof of \cref{lem:UTAP-Dual}]
We prove that $y^*$ is dual feasible and achieves the optimal primal objective value.

\emph{Non-negativity.}
Fix an edge $e\in E(T)$, and consider an
optimal set of links covering $T_e$.
Since all links are uplinks, no link can
cover edges in two distinct subtrees
$T_g$ and $T_f$ for $g,f\in C(e)$.
Consequently, the links covering these
subtrees must have total cost at least
$\sum_{g\in C(e)}O_g$.
It follows that $O_e\geq\sum_{g\in C(e)}O_g$, and hence $y_e^*\geq 0$.

\emph{Dual constraints.}
Fix an uplink $a\in L^{\uparrow}$, and
let $e_1,\ldots,e_k$ be the edges of $P(a)$,
ordered from top to bottom.
Define $B(a)=
\left(\bigcup_{i=1}^{k}C(e_i)\right)
\setminus P(a)$. By the definition of $y^*$, the terms corresponding to consecutive edges of $P(a)$ cancel, giving $\sum_{e\in P(a)}y_e^*    =\sum_{i=1}^{k} (O_{e_i}-\sum_{g\in C(e_i)}O_g) = O_{e_1}-\sum_{g\in B(a)}O_g$.

Observe that $T_{e_1}$ can be covered
by selecting $a$ together with an optimal
cover of $T_g$ for each $g\in B(a)$.
Therefore, $O_{e_1}\leq c(a)+\sum_{g\in B(a)}O_g$. Combining the two inequalities yields $\sum_{e\in P(a)}y_e^*\leq c(a)$,
establishing dual feasibility.

\emph{Optimality.}
Finally, summing $y_e^*$ over all tree
edges, we obtain
\begin{align*}
\sum_{e\in E(T)}y_e^*
=\sum_{e\in E(T)}
(O_e-\sum_{g\in C(e)}O_g) =\sum_{e\in\delta_T(r)}O_e
=\optutap(T,c).
\end{align*}
The second equality follows by
telescoping: every edge not incident to
$r$ contributes once positively and once
negatively. The last equality holds
because uplinks cannot cover edges
belonging to distinct subtrees rooted
at the children of $r$. Thus, these
subtrees can be covered independently.

Since $y^*$ is dual feasible and its
objective value equals the optimal
integral primal value, it is an optimal
dual solution by weak duality.
\end{proof}

We next establish a useful structural property of $\utap$ under randomized rounding. In particular, we characterize how optimal dual solutions change after a single round of randomized rounding, showing that the expected dual weight of each edge $e \in E(T)$ decreases approximately in proportion to its probability of remaining uncovered. The following definition will be useful in bounding the cost of extending a partial solution to a feasible solution for $\utap$.
\begin{definition}
    Given a set $S \subseteq L^\uparrow$, the residual cost function $c^S$ with respect to $S$ is the cost function that agrees with $c$ on $L^\uparrow \setminus S$ and assigns cost zero to links in $S$. Formally, $c^S(a) = 0$ if $a \in S$ and $c^S(a) = c(a)$ if $a \notin S$.
\end{definition}

\begin{lemma}
\label{lem:utap-shrinkage}
Given a tree $T$, rooted at a vertex $r$, a set of uplinks $L^\uparrow$, and a cost function $ c : L^\uparrow \rightarrow \mathbb R_{\geq 0}$, let $x$ be a feasible solution to $\utaplp(T,c)$ and let $\gamma \geq 1$. Set $\hat x_a = \min(1, \gamma x_a)$. Sample each $a \in L^\uparrow$ independently with probability $\hat x_a$ and let $S$ be the set of sampled links. Then,
    \begin{align*}
        \E[\optutap(T, c^S)] \leq 2 \gamma  e^{-\gamma} \sum_{a \in L^\uparrow} c(a) x_a.
    \end{align*}
    Equivalently, the expected minimum cost of augmenting $T\cup S$ to a $2$-edge-connected graph is at most $2\gamma e^{-\gamma} \sum_{a\in L^\uparrow}c(a)x_a$.
\end{lemma}
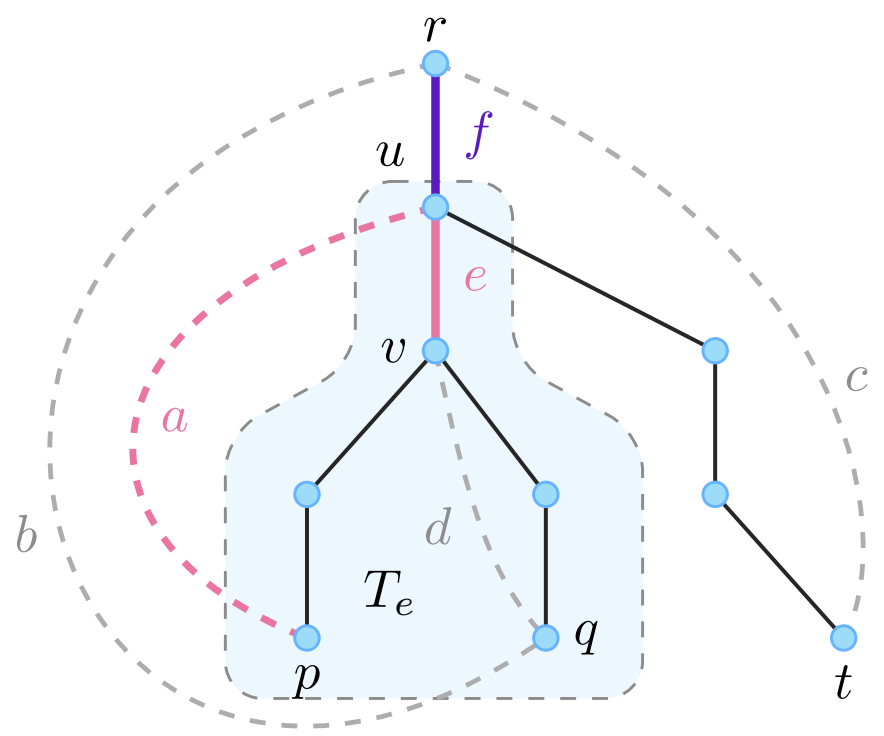
\begin{proof}
    For any set $H \subseteq L^\uparrow$, let $y^H$ be the optimal dual solution to the $\utap$ problem on $T$ with cost $c^H$ as defined in \cref{lem:UTAP-Dual}. Moreover, let $O^H_e$ be the cost of the optimal $\utap$ solution for $T_e$ under cost function $c^H$.
    
    By integrality of the $\utap$ polytope and linearity of expectation, $\E[\optutap(T, c^S)] = \sum_{e\in E} \E \left[y^S_e\right]$. For each $e \in E$, define $L(e) = \{a \in L^\uparrow : e \in P(a) \}$, i.e. $L(e)$ contains the links in $L^\uparrow$ that cover the cut corresponding to the edge $e$. Let $Z_e$ be the indicator random variable that $L(e) \cap S = \emptyset$; indicating whether $e$ is uncovered by the links in $S$.

\begin{figure}[!tbp]
\centering
\begingroup
\definecolor{mcVertexFill}{rgb}{0.61338,0.86345,0.97537}
\definecolor{mcVertexDraw}{rgb}{0.4,0.7,1.0}
\definecolor{mcInk}{rgb}{0.15,0.15,0.15}
\definecolor{mcMagenta}{rgb}{0.91472,0.46078,0.62707}
\definecolor{mcPurple}{rgb}{0.34749,0.1,0.75250}
\definecolor{mcOrange}{rgb}{0.85,0.425,0}
\definecolor{mcGreen}{rgb}{0,0.6,0}
\definecolor{mcGray}{rgb}{0.55,0.55,0.55}
\definecolor{mcPale}{rgb}{0.92587,0.97171,0.99509}
\tikzset{
  mcfigure/.style={x=1cm,y=1cm,font=\small,>=Latex},
  baseedge/.style={draw=mcInk,line width=.75pt},
  vertex/.style={circle,draw=mcVertexDraw,fill=mcVertexFill,line width=.55pt,minimum size=4.6pt,inner sep=0pt,outer sep=0pt},
  linkmag/.style={draw=mcMagenta,line width=1pt,dashed},
  linkpurple/.style={draw=mcPurple,line width=1pt,dashed},
  linkorange/.style={draw=mcOrange,line width=1pt,densely dotted},
  linkgray/.style={draw=mcGray,line width=.85pt,dashed},
  linkfaint/.style={draw=black!32,line width=.9pt,dashed},
  pathgreen/.style={draw=mcGreen,line width=1.6pt},
  markedge/.style={draw=mcMagenta,line width=1.6pt},
  markpurple/.style={draw=mcPurple,line width=1.6pt},
  contractedge/.style={draw=black!28,line width=1.6pt},
  region/.style={draw=mcGray,dashed,fill=mcPale,line width=.55pt,rounded corners=7pt},
  block/.style={draw=mcGray,fill=mcPale,line width=.55pt,rounded corners=4pt,minimum height=.65cm,inner sep=5pt},
  selectedblock/.style={block,draw=mcMagenta,line width=1pt},
  guide/.style={draw=black!35,line width=.5pt,densely dashed},
  maparrow/.style={-{Latex[length=2mm]},draw=mcGray,line width=.65pt},
  paneltitle/.style={font=\small,align=center},
  smallnote/.style={font=\footnotesize,align=center},
  endpointlabel/.style={font=\small,inner sep=2pt}
}
\long\def\mcReductionFit#1{%
  \sbox0{#1}%
  \ifdim\wd0>\linewidth
    \resizebox{\linewidth}{!}{\usebox0}%
  \else
    \usebox0%
  \fi
}
\mcReductionFit{%
\begin{tikzpicture}[mcfigure]
  \def\mcTreeNodes{r/0/0, u/0/-0.95, v/0/-1.90, s/1.85/-1.90,
    x/-0.85/-2.85, y/0.73/-2.85, z/1.85/-2.85,
    p/-0.85/-3.80, q/0.73/-3.80, t/2.70/-3.80}
  \def\mcTreeEdges{r/u, u/v, u/s, v/x, v/y, x/p, y/q, s/z, z/t}
  \foreach \n/\px/\py in \mcTreeNodes {\coordinate (\n) at (\px,\py);}

  \draw[region] (-0.53,-0.78)--(0.51,-0.78)--(0.51,-1.98)--(1.37,-2.45)
    --(1.37,-4.20)--(-1.39,-4.20)--(-1.39,-2.45)--(-0.53,-1.98)--cycle;
  \foreach \i/\j in \mcTreeEdges {\draw[baseedge] (\i)--(\j);}
  \draw[markpurple] (r)--(u);
  \draw[markedge] (u)--(v);

  \draw[linkfaint] (r) .. controls (-4.30,-0.92) and (-2.55,-6.10) .. (q)
    node[pos=.5,left=2pt,endpointlabel,text=black!45] {$b$};
  \draw[linkfaint] (r) .. controls (2.85,-1.10) and (3.05,-3.20) .. (t)
    node[pos=.5,right=2pt,endpointlabel,text=black!45] {$c$};
  \draw[linkfaint] (v)
    .. controls (0.20,-2.85) and (0.35,-3.40) .. (q)
    node[pos=.5,left=3pt,endpointlabel,text=black!45] {$d$};
  \draw[linkmag,line width=1.3pt] (u)
    .. controls (-2.45,-1.55) and (-2.55,-3.15) .. (p)
    node[pos=.5,right=2pt,endpointlabel,text=mcMagenta] {$a$};

  \path (r)--node[endpointlabel,right=3pt,text=mcPurple] {$f$} (u);
  \path (u)--node[endpointlabel,right=3pt,text=mcMagenta] {$e$} (v);
  \node[endpointlabel] at (-0.30,-3.50) {$T_e$};
  \foreach \n/\px/\py in \mcTreeNodes {\node[vertex] at (\n) {};}
  \node[endpointlabel,above=2pt] at (r) {$r$};
  \node[endpointlabel] at (-0.30,-0.60) {$u$};
  \node[endpointlabel,left=3pt] at (v) {$v$};
  \node[endpointlabel,below=3pt] at (p) {$p$};
  \node[endpointlabel,right=3pt] at (q) {$q$};
  \node[endpointlabel,below=3pt] at (t) {$t$};
  \node[smallnote] at (0.10,-4.85)
    {$L(e)=\{a,b\}$,\quad $L(f)=\{b,c\}$\\[1pt]
     $L(e)\setminus L(f)=\{a\}$,\quad $L(f)\setminus L(e)=\{c\}$};
\end{tikzpicture}%
}
\endgroup
\caption{The link $a$ covers $e$ but not its parent edge $f=A(e)$. The four
available uplinks are $a=up$, $b=rq$, $c=rt$ and $d=vq$. Link $b$ covers both $e$ and $f$; link $c$ covers $f$ but no edge of $T_e$ at all; and link $d$ covers neither. The pink link $a$ is
therefore the only link of $L(e)\setminus L(f)$, and because it covers $e$ but
not $f$, its topmost edge is $e$. No link can cover $e$ from above without also
covering $f$.}
\label{fig:utap-link-difference}
\end{figure}
   
    By \cref{lem:UTAP-Dual}, $y^S_e = O^S_e - \sum_{g \in C(e)} O^S_g$. If $S \cap L(e) \neq \emptyset$, some link in $S$ covers $e$ at zero cost. Thus, covering $T_e$ is equivalent to covering its child subtrees, implying $O^S_e = \sum_{g \in C(e)} O^S_g$ and hence $y^S_e = 0$. Otherwise, $S \cap L(e) = \emptyset$, and therefore $y^S_e = y^{S \setminus L(e)}_e$.
    Therefore, for every realization of $S$, $y_e^S = Z_e \cdot y_e^{S\setminus L(e)}$. Observe that $Z_e$ depends only on the sampling outcomes of links in $L(e)$, whereas $y_e^{S\setminus L(e)}$ depends only on those of links in $L^\uparrow\setminus L(e)$. Since links are sampled independently, these two random variables are independent. Consequently,
    \begin{equation}
        \label{eq:opt-S-bound-1}
        \E[\optutap(T, c^S)] = \sum_{e\in E} \E \left[y^S_e\right] 
        =\sum_{e\in E} \E \left[Z_e\right] \E \left[ y_e^{S \setminus L(e)}\right] 
    \end{equation}

    Now, consider some $e \in E$. If there is some $a \in L(e)$ with $\hat x_a =1$, then $\E[Z_e] = 0$. Otherwise,
    \begin{align*}
        \E[Z_e] = \Prb[S \cap L(e) = \emptyset] = \prod_{a \in L(e)} (1-\hat x_a) 
        \leq \prod_{a \in L(e)} e^{-\hat x_a}
        = e^{- \sum_{a\in L(e)} \hat x_a} 
        = e^{-\gamma  \sum_{a\in L(e)}  x_a} 
        \leq e^{-\gamma }
    \end{align*}
    where the first inequality uses $1-x \leq e^{-x}$ for all $x \in \mathbb R$ and the last inequality uses that $x$ was a feasible solution for $\utaplp(T, c)$. In both cases, we have $\E[Z_e] \leq e^{-\gamma }$. Substituting this in equation \eqref{eq:opt-S-bound-1} and by the non-negativity of dual variables,
    \begin{equation}
        \label{eq:final-opt-S-Bound}
        \E[\optutap(T, c^S)] = \sum_{e\in E} \E \left[Z_e\right] \E \left[ y_e^{S \setminus L(e)}\right] \leq e^{-\gamma } \sum_{e\in E}  \E \left[ y_e^{S \setminus L(e)}\right] 
    \end{equation}
    By the definition of $y_e^{S \setminus L(e)}$, 
    \begin{equation} \label{eq:opt-S-bound-2}
        \sum_{e\in E}  \E [ y_e^{S \setminus L(e)}]  = \sum_{e\in E}  \E [O_e^{S \setminus L(e)} - \sum_{g\in C(e)}O_{g}^{S\setminus L(e)}]
    \end{equation}
    Rearranging the terms and using the fact that
    each $e\in E(T)\setminus\delta_T(r)$
    has a unique parent edge $A(e)$, we obtain
    \begin{align*}
    \sum_{e\in E(T)}
    \E\left[
    O_e^{S\setminus L(e)}
    -\sum_{g\in C(e)}O_g^{S\setminus L(e)}
    \right]
    &=
    \sum_{e\in E(T)}\E[O_e^{S\setminus L(e)}]
    -\sum_{e\in E(T)\setminus\delta_T(r)}
    \E[O_e^{S\setminus L(A(e))}]\\
    &=
    \sum_{e\in\delta_T(r)}
    \E[O_e^{S\setminus L(e)}]
    +\sum_{e\in E(T)\setminus\delta_T(r)}
    \E\left[
    O_e^{S\setminus L(e)}
    -O_e^{S\setminus L(A(e))}
    \right].
    \end{align*}
    The first equality follows by reindexing the double sum: every edge $e \in E(T)\setminus\delta_T(r)$ belongs to the child set of exactly one edge, namely $C(A(e))$. The second equality follows by partitioning $E(T)$ into edges incident to $r$ and those not incident to $r$, and combining the corresponding terms.
    Next, we bound each term of the equation separately. Firstly, note that $O_e^{S \setminus L(e)} \leq O_e$. Thus, $\sum_{e \in \delta_T(r)} \E \left[ O_e^{S \setminus L(e)}\right] \leq \sum_{e \in \delta_T(r)} O_e = \optutap(T, c)$
    where the equality follows as in an $\utap$ instance, to cover all edges, we must cover each $T_e$ for $e \in \delta_T(r)$ disjointly. Since $x$ is feasible for $\utaplp(T,c)$ and the LP relaxation is integral, we have $\optutap(T,c) =\utaplp(T,c) \leq \sum_{a\in L^\uparrow}c(a)x_a$. Combining this with the preceding bound yields
    \begin{equation}
    \label{eq:opt-S-bound-first-term}
        \sum_{e\in\delta_T(r)}
        \E\left[O_e^{S\setminus L(e)}\right]
        \leq \optutap(T,c)
        \leq \sum_{a\in L^\uparrow}c(a)x_a \leq \sum_{a \in L^\uparrow} c(a) \hat x_a.
    \end{equation}
    Recall without loss of generality, we can assume $x \le 1$; hence, $ x_a \leq \hat{x}_a$ for all $a\in L^\uparrow$.
    Next, fix some $e \in E \setminus \delta_T(r)$, let $f := A(e)$. We give an upper bound for $\E \left[ O_e^{S\setminus L(e)} - O_e^{S\setminus L(f)}\right]$. Consider a link $a \in L^\uparrow$ such that $a \in L(f)$. Since $a$ is an uplink, either $a$ does not cover any edge in $T_e$, or $a \in L(e)$ (see~\cref{fig:utap-link-difference}). 

    Consider any realization of $S$, the two cost functions $c^{S \setminus L(e)}$ and $c^{S \setminus L(f)}$ agree on all link that cover $T_e$ except for the links in $S \cap (L(e) \setminus L(f))$, note that the links in $L(f)\setminus L(e)$ do not cover any edges in $T_e$. To cover $T_e$, one can pick all links in $S \cap (L(e) \setminus L(f))$ and then pick all the links that the optimal covering of $T_e$ under the cost function $c^{S \setminus L(f)}$ takes. The cost of this solution with respect to $c^{S\setminus L(e)}$ is $O_e^{S \setminus L(f)} + \sum_{a \in S \cap (L(e) \setminus L(f))} c(a)$. Consequently,
    \begin{equation}
        \label{eq-opt-S-bound-Second-Term-individual}
        \E \left[ O_e^{S\setminus L(e)} - O_e^{S\setminus L(f)} \right] \leq \E \left[ \sum_{a \in S \cap (L(e) \setminus L(f))} c(a)  \right] = \sum_{a \in L(e) \setminus L(f)} c(a) \hat x_a 
    \end{equation}

    Summing \eqref{eq-opt-S-bound-Second-Term-individual} over all $e \in E\setminus \delta_T(r)$,
    \begin{equation}
        \label{eq-opt-S-bound-Second-Term}
        \sum_{e \in E \setminus \delta_T(r)} \E \left[ O_e^{S\setminus L(e)} - O_e^{S\setminus L(A(e))} \right] \leq \sum_{e \in E \setminus \delta_T(r)} \, \sum_{a \in L(e) \setminus L(A(e))} c(a) \hat x_a \leq \sum_{a \in L^\uparrow} c(a) \hat x_a
    \end{equation}
    The last inequality follows because, for each $a \in L^\uparrow$, the condition $a \in L(e) \setminus L(A(e))$ can hold only when $e$ is the topmost edge of $P(a)$. Thus, each link contributes to at most one such term.

    Substituting \eqref{eq:opt-S-bound-first-term} and \eqref{eq-opt-S-bound-Second-Term} in \eqref{eq:opt-S-bound-2} gives,
    \begin{align*}
        \sum_{e\in E}  \E \left[ y_e^{S \setminus L(e)}\right]  = \sum_{e\in E}  \E \left[ O_e^{S \setminus L(e)} - \sum_{g\in C(e)}O_{g}^{S\setminus L(e)} \right] \leq 2 \left( \sum_{a \in L^\uparrow} c(a) \hat x_a \right)
    \end{align*}
    Applying this to \eqref{eq:final-opt-S-Bound} yields,
    \begin{align*}
        \E[\optutap(T, c^S)] \leq e^{-\gamma} \sum_{e\in E}  \E \left[ y_e^{S \setminus L(e)}\right] \leq  2e^{-\gamma } \left( \sum_{a \in L^\uparrow} c(a) \hat x_a \right) \leq 2 \gamma  e^{-\gamma }  \left( \sum_{a \in L^\uparrow} c(a) x_a \right)
    \end{align*}
    where the last inequality follows from $\hat x_a \leq \gamma  x_a$ for all $a \in L^\uparrow$ and the non-negativity of the cost function. This concludes the proof.
\end{proof}

\subsection{Randomized Rounding with Completion for $\mcutap$}\label{sec:rand-compl}
Recall the $\mcutaplp$ relaxation of $\mcutap$.

\mcutapLPBox

Let $B$ be an upper bound on the link costs, i.e., $c_i(a) \leq B$ for all $a \in L^\uparrow$ and $i \in [\ell]$. We later show how to preprocess the instance to obtain a sufficiently tight bound $B$.

In this section, we utilize the polyhedral structure of the $\utap$ solutions to give a $O(\log \ell)$ approximation algorithm for $\mcutap$. More formally, we prove the following result:
\begin{theorem}
    \label{thm:rounding}
    Consider a $\mcutap$ instance on tree $T$, set of uplinks $L^\uparrow$, and cost functions $\{c_i : L^\uparrow \rightarrow \mathbb R_{\geq 0} \}_{i\in [\ell]}$. Let $B$ be a value such that for all $a \in L^\uparrow$ and $i \in [\ell]$, $c_i(a) \leq B$. Given a feasible solution $x$ to $\mcutaplp(T,\{c_1,\cdots,c_\ell\})$, there exists an algorithm running in polynomial time that constructs a solution $S \subseteq L^\uparrow$ where $T \cup S$ is $2$-edge-connected and $\mathbb E[\mathrm{cost}(S)] \leq O(\log \ell) (\mathrm{cost}(x)+B)$.
\end{theorem}

\paragraph{Description of the algorithm.} For all $a \in L^\uparrow$, let $\hat x_a \coloneq \min(\lceil  \log \ell + 1 \rceil x_a, 1)$. Our algorithm proceeds in two stages, initially, we randomly sample each $a \in L^\uparrow$ to be in $S$ with probability $\hat{x}_a$. Then, we augment $S$ by a set $H$ such that $T \cup S \cup H$ becomes $2$-edge-connected.

To this end, define the aggregate cost function $c: L^\uparrow \rightarrow \mathbb{R}_{\geq 0}$ by $c(a) = \sum_{i \in [\ell]} c_i(a)$ for every $a \in L^\uparrow$. Given the sampled set $S$, let $H \subseteq L^\uparrow \setminus S$ be a minimum-cost set of additional links, with respect to $c$, such that $T \cup S \cup H$ is $2$-edge-connected. Equivalently, $c(H) = \optutap(T,c^S)$. Such a set $H$ can be computed in polynomial time by solving the integral LP formulation of $\utap$ with cost function $c^S$. We bound the costs of $S$ and $H$ separately.

\paragraph{Bounding the Cost of Randomized Rounding} We show the expected cost of $S$ is at most a factor of $O(\log \ell)$ larger than the cost of $x$ and $B$.
\begin{claim}
    \label{clm:random-rounding-bound}
    Given a fractional solution $x$ to $\mcutaplp\left(T,\{c_i\}_{i \in [\ell]}\right)$, let $B> 0$ be a number such that for all $i \in [\ell]$ and $a \in L^\uparrow$, $c_i(a) \leq B$. 
    
    Let $S$ be obtained from $x$ by sampling each $a \in L^\uparrow$ with probability $\hat x_a = \min(\lceil  \log \ell + 1 \rceil x_a, 1)$ independently, then,
    \begin{align*}
        \E \left[ \max_{i \in [\ell]} \left( \sum_{a\in S} c_i(a) \right) \right] \leq \lceil \log \ell + 1 \rceil B + 2\lceil \log \ell + 1 \rceil \mathrm{cost}(x) 
    \end{align*}
\end{claim}
\begin{proof}
    For all $a \in L^\uparrow$, let $X_a$ be the indicator random variable of $a \in S$. 
    Let $C_i=\sum_{a\in L^\uparrow}c_i(a)X_a$ denote the total sampled cost under $c_i$. By the log-sum-exp inequality, for every realization of $S$, $\max_{i\in[\ell]}C_i \leq B\log\left(\sum_{i\in[\ell]}e^{C_i/B}\right)$. Taking expectations and applying Jensen's inequality, we obtain
    \begin{equation}
        \label{eq:S-bound-expectation-main}
        \E\left[\max_{i\in[\ell]}C_i\right] \leq B\log\left(\sum_{i\in[\ell]}\E\left[\exp\left(\frac{C_i}{B}\right)\right]\right).
    \end{equation}    
    Fix any $i\in[\ell]$. By the definition of $C_i$
    and the independence of $\{X_a\}_{a\in L^\uparrow}$, we have
    \begin{equation}
    \label{eq:expected-S-first}
    \begin{aligned}
    \E\left[\exp\left(\frac{C_i}{B}\right)\right]
    =\E\left[\exp\left(\sum_{a\in L^\uparrow}\frac{c_i(a)}{B}X_a\right)\right]
    =\prod_{a\in L^\uparrow}
    \E\left[
    \exp\left(\frac{c_i(a)}{B}X_a\right)
    \right].
    \end{aligned}
    \end{equation}
    Now, for each $a \in L^\uparrow$, $\E \left[ \exp \left( \frac{c_i(a) }{B} X_a \right) \right] = \exp\left( \frac{c_i(a) }{B} \right) \hat x_a + (1-\hat{x}_a)$. Since by assumption $\frac{c_i(a) }{B} \leq 1$, using $e^z \leq 1+ 2z$ for $z \in [0,1]$,
    \begin{equation*}
        \E \left[ \exp \left( \frac{c_i(a) }{B} X_a \right) \right] \leq \left(1 + \frac{2c_i(a)}{B} \right) \hat x_a +1 - \hat{x}_a = 1+\frac{2c_i(a)}{B} \hat x_a \leq \exp\left( \frac{2c_i(a)}{B} \hat x_a \right)
    \end{equation*}
    where the last inequality uses $1+z \leq e^z$ for all $ z\in \mathbb R$. 
    Substituting this back in \eqref{eq:expected-S-first} gives,
    \begin{align*}
        \label{eq:expected-S-second}
        \E \left[  \exp \left( \frac{C_i}{B} \right) \right] 
        \leq \prod_{a\in L^\uparrow} \exp\left( \frac{2c_i(a)}{B} \hat x_a \right) 
        &= \exp\left( \frac{2\sum_{a \in L^\uparrow} c_i(a) \hat{x}_a}{B} \right) \\ 
        &\leq \exp\left( \frac{2 \lceil \log \ell + 1 \rceil\sum_{a \in L^\uparrow} c_i(a) {x}_a}{B} \right)
    \end{align*}
    where the last inequality follows as for all $a \in L^\uparrow$, $\hat{x}_a \leq \lceil \log \ell + 1 \rceil x_a$. 
    Finally, since $\sum_{a \in L^\uparrow} c_i(a) {x}_a \leq \mathrm{cost}(x)$ we obtain,
    \begin{equation}
        \label{eq:individual-term-bound-S-expected}
        \E \left[  \exp \left( \frac{C_i}{B} \right) \right] \leq \exp\left(2 \lceil \log \ell + 1 \rceil \frac{\mathrm{cost}(x)}{B} \right).
    \end{equation}
    Substituting \eqref{eq:individual-term-bound-S-expected} in \eqref{eq:S-bound-expectation-main} yields,
    \begin{align*}
        \E \left[\max_{i \in [\ell]} C_i\right] \leq B \log\left( \sum_{i \in [\ell]}  \exp\left(2 \lceil \log \ell + 1 \rceil \frac{\mathrm{cost}(x)}{B} \right) \right) &= B \log\left( \ell \exp\left(2 \lceil \log \ell + 1 \rceil \frac{\mathrm{cost}(x)}{B} \right) \right) \\ &\leq \lceil \log \ell + 1 \rceil B + 2\lceil \log \ell + 1 \rceil \mathrm{cost}(x) 
    \end{align*}
    which concludes the proof of this claim.
\end{proof}

\paragraph{Bounding the Completion Cost.}
Next, we bound the expected cost of $H$ under the aggregate cost function $c = \sum_{i\in[\ell]} c_i$ we defined earlier.
\begin{observation}
    \label{obs:mc-to-nonmc}
    For a $\mcutap$ instance with tree $T$ and cost function $\{ c_i\}_{i \in [\ell]}$, let $c$ be the cost function such that for all $a \in L^\uparrow$, $c(a) = \sum_{i \in [\ell]} c_i(a)$. Then,
    \begin{align*}
        \optmcutap\left(T, \{ c_i\}_{i \in [\ell]}\right) \leq \optutap(T, c) \leq \ell  \optmcutap\left(T, \{ c_i\}_{i \in [\ell]}\right).
    \end{align*}
More generally, for any set of links $H \subseteq L^\uparrow$, $\max_{i \in [\ell]} c_i(H) \leq c(H) \leq \ell  \max_{i \in [\ell]} c_i(H)$.
\end{observation}
\begin{proof}
    Consider a set $H \subseteq L^\uparrow$, since the costs are non-negative, $c_j(H) \leq c(H)$. Moreover, $c_j(H) \leq \max_{i \in [\ell]} c_i(H)$ for all $j \in [\ell]$. This implies, $\max_{i \in [\ell]} c_i(H) \leq c(H) = \sum_{i \in [\ell]} c_i(H) \leq \ell \max_{i \in [\ell]} c_i(H)$.

    Now, let $O$ be the optimal $\mcutap$ solution with costs $\{ c_i\}_{i \in [\ell]}$ and $A$ be the optimal $\utap$ solution with  cost $c$, both on the tree $T$. We have,
    \begin{align*}
        \optmcutap\left(T, \{ c_i\}_{i \in [\ell]}\right) = \max_{i\in [\ell]} c_i(O) \leq \max_{i \in[\ell]} c_i(A) \leq c(A) = \optutap(T, c),
    \end{align*}
    where the first inequality holds as $O$ is the optimal $\mcutap$ solution.
    Furthermore,
    \begin{align*}
        \optutap(T, c) = c(A) \leq c(O) \leq \ell \max_{i \in [\ell]} c_i(O) = \ell \optmcutap\left(T, \{ c_i\}_{i \in [\ell]}\right),
    \end{align*}
    where the first inequality holds as $A$ is the optimal $\utap$ solution.
\end{proof}
We are ready to bound the cost of $H$,
\begin{claim}
    \label{clm:patching-bound}
    Consider a $\mcutap$ instance with tree $T$ and cost function $\{ c_i\}_{i \in [\ell]}$, and let $c$ be the aggregate cost function. Let $S$ be defined as in \cref{clm:random-rounding-bound} and let $H$ be the optimal $\utap$ solution for $T$ under cost function $c^S$. Then,
    \begin{align*}
      \E\left[\max_{ i \in [\ell]} c^S_i(H) \right] \leq 2 \lceil \log \ell + 1 \rceil \mathrm{cost}(x).
    \end{align*}
\end{claim}

\begin{proof}
By \cref{obs:mc-to-nonmc}, $\E\left[\max_{ i \in [\ell]} c^S_i(H) \right] \leq \E\left[\optutap(T, c^S)\right]$. Since $S$ is a randomly sampled subset of $L^\uparrow$, where each link $a \in L^\uparrow$ is included independently with probability $\hat x_a = \min\left(1, \lceil \log \ell + 1 \rceil x_a\right)$, applying \cref{lem:utap-shrinkage} we obtain,
\begin{align*}
    \E\left[\optutap(T, c^S)\right] \leq 2 \lceil \log \ell + 1 \rceil e^{-\lceil \log \ell + 1 \rceil} \left( \sum_{a \in L^\uparrow} c(a) x_a \right) 
    &\leq \frac{2  \lceil \log \ell + 1 \rceil }{\ell} \left( \sum_{a \in L^\uparrow} c(a) x_a \right) \\ 
    &\leq   \frac{2  \lceil \log \ell + 1 \rceil }{\ell} \left( \ell \cdot \max_{i \in [\ell] }  \sum_{a \in L^\uparrow} c_i(a) x_a \right)  \\
    &= 2 \lceil \log \ell + 1 \rceil \mathrm{cost}(x)
\end{align*}
Together with \cref{obs:mc-to-nonmc}, $\E\left[\max_{ i \in [\ell]} c^S_i(H) \right] \leq \E\left[\optutap(T, c^S)\right] \leq 2  \lceil \log \ell + 1 \rceil  \mathrm{cost}(x)$.
\end{proof}

\begin{proof}[Proof of \cref{thm:rounding}]

We can finally show our algorithm indeed achieves the desired bound on its cost,
\begin{align*}
    \E\left[\max_{ i \in [\ell]} c_i(S \cup H) \right] \leq \E\left[\max_{ i \in [\ell]} \left(c_i(S) + c_i^S( H)\right) \right] \leq \E\left[\max_{ i \in [\ell]} c_i(S) \right]  + \E\left[ \max_{i \in [\ell]} c_i^S( H) \right] 
\end{align*}
Applying \cref{clm:random-rounding-bound,clm:patching-bound} shows $\E\left[\max_{ i \in [\ell]} c_i(S \cup H) \right]  \leq 4 \lceil \log \ell + 1 \rceil \mathrm{cost}(x) + \lceil \log \ell + 1 \rceil  B$.
\end{proof}

\section{A tight Approximation Guarantee for $\mccap$}
\label{sec:analysis-mccap}

In this section, we combine the reduction from \cref{sec:mccap-to-mcutap} with the algorithm developed in \cref{sec:rounding} to obtain an $O(\log \ell)$-approximation algorithm for $\mccap$. We further establish a matching hardness result, showing that our approximation guarantee is essentially optimal under a standard complexity assumption. 

\subsection{An $O(\log \ell)$-approximation for $\mccap$}
In this section, we establish one of our main results, which is an $O(\log \ell)$-approximation algorithm for $\mccap$.
\begin{theorem}\label{thm:mc-cap}
    Given a graph $G = (V,E)$, a set of links $L$, and $\ell$ cost functions $\{  c_i : L \rightarrow \mathbb R_{\geq 0}\}_{i \in [\ell]}$. The $\mccap$ problem on $G$ with links $L$ and cost function $\{c_i \}_{i \in [\ell]}$  admits an $O(\log \ell)$ approximation algorithm in polynomial time. 
\end{theorem}

\begin{proof}
    Before describing the algorithm, we establish some notation. For any number $r$, we define $L_r$ as the set of all links $a \in L$ such that the cost of $a$ in any criterion is at most $r$. Formally, $L_r = \{ a \in L \mid c_i(a) \leq r, \, \, \forall i \in [\ell] \}$. Moreover, let $R$ be the set of all costs that links in $L$ can acquire, i.e. $R = \{ c_i(a) \mid  i \in [\ell],a\in L \}$.

    \paragraph{Description of the Algorithm} We consider the elements of $R$ in decreasing order of their weight. For each $r \in R$, we solve the LP relaxation of $\mccap$ on $G$ restricted to links in $L_r$, and let $x^{(r)}$ denote an optimal solution. If the LP is infeasible, we terminate, as no smaller upper bound on the maximum link cost can yield a feasible solution. We feed $x^{(r)}$ to our reduction in \cref{thm:reduction} to obtain two fractional $\mcutap$ solutions $z^{(1)}, z^{(2)}$ for trees $T_1, T_2$. By \cref{thm:reduction}, $\mathrm{cost}\left(z^{(i)}\right) \leq 10 \cdot \mathrm{cost}(x^{(r)})$ for any $i \in \{ 1, 2 \}$.
    
    Next, we apply the rounding algorithm of \cref{thm:rounding} to $z^{(1)}$ and $z^{(2)}$, obtaining link sets $H_1^{(r)}$ and $H_2^{(r)}$, respectively. For each $i \in \{1,2\}$, let $S_i^{(r)} \subseteq L_r$ denote the corresponding set of original links obtained by mapping $H_i^{(r)}$ back through the reduction. By \cref{thm:reduction}, $S^{(r)}_1 \cup S^{(r)}_2$ is a feasible solution for the $\mccap$ on $G$. Moreover, for any $i \in \{ 1,2 \}$, the cost of each $S^{(r)}_i$ is bounded by, 
    \begin{align*}
        \mathbb E\left[\mathrm{cost}\left(S_i^{(r)}\right) \right] \leq  \lceil \log \ell + 1 \rceil \left( 4\cdot \mathrm{cost}\left(z^{(i)}\right) +  r\right)
    \end{align*}
    where we have used \cref{thm:rounding} and the fact that the cost of every link in $L_r$ is bounded by $r$.
    Thus, the $\mccap$ solution $S^{(r)}_1 \cup S^{(r)}_2$ simply has expected cost,
    \begin{align}
        \mathbb E\left[\mathrm{cost}\left(S^{(r)}_1 \cup S^{(r)}_2\right) \right] &\leq  \lceil \log \ell + 1 \rceil \left(4  \cdot \mathrm{cost}\left(z^{(1)}\right) + 4  \cdot \mathrm{cost}\left(z^{(2)}\right) + 2 r \right) \notag  \\ \label{eq:mccap-bound} &\leq \lceil \log \ell + 1 \rceil \left( 80 \cdot \mathrm{cost}\left(x^{(r)}\right) + 2 r\right)
    \end{align}
    Finally, the algorithm returns the minimum-cost solution among all those constructed during the procedure.

    \paragraph{Analysis of the Approximation Ratio}  Let $O^*$ denote an optimal solution to $\mccap$ on $G$ with links $L$. Let $r^*$ denote the maximum cost assigned to any link in $O^*$ by any of the cost functions $c_1,\ldots,c_\ell$ (Note that the case $r^* = 0$ is trivial). 
    Clearly, $r^* \leq \mathrm{cost}(O^*)$. Moreover, $O^*$ is a feasible solution to the $\mccap$ LP relaxation when the set of allowed links is restricted to $L_{r^*}$. This implies, $\mathrm{cost}\left(x^{(r^*)}\right) \leq \mathrm{cost}(O^*)$. Putting these together in \eqref{eq:mccap-bound} gives,
    \begin{align*}
        \mathbb E\left[\mathrm{cost}\left(S^{(r^*)}_1 \cup S^{(r^*)}_2\right) \right] \leq \lceil \log \ell + 1 \rceil \left( 80 \cdot \mathrm{cost}\left(x^{(r^*)}\right) + 2 r^*\right) \leq 82 \lceil \log \ell + 1 \rceil \cdot \mathrm{cost}(O^*)
    \end{align*}

    Thus,
    \begin{align*}
        \mathbb E\left[\min_{r \in R} \mathrm{cost}\left(S^{(r)}_1 \cup S^{(r)}_2\right) \right] \leq 
        \mathbb E\left[\mathrm{cost}\left(S^{(r^*)}_1 \cup S^{(r^*)}_2\right) \right] \leq 82 \lceil \log \ell + 1 \rceil \cdot \mathrm{cost}(O^*)
    \end{align*}
    Thus, our algorithm returns an $82 \lceil \log \ell + 1 \rceil$-approximation for $\mccap$ in expectation.

    Note that the algorithm clearly runs in polynomial time as $|R| \leq \ell \cdot |L|$. Moreover, both the reduction from $\mccap$ to $\mcutap$ and the rounding procedure for $\mcutap$ run in polynomial time.
 \end{proof}

Interestingly, the canonical LP formulation of $\mccap$ has an integrality gap of $\Theta(\ell)$, even when $G$ consists of a single edge. This highlights the importance of incorporating the maximum link cost into the performance guarantee of our algorithm in \cref{sec:rounding}.

\begin{remark}
    \label{rem:integrality-gap}
    The canonical LP for $\mccap$ (\mccaplp) has integrality
    gap $\Theta(\ell)$. To see the lower bound, let $G$ consist of a
    single edge $uv$, and let $L = \{a_1,\ldots,a_\ell\}$ be parallel
    links between $u$ and $v$, with $c_i(a_j) = 1$ if $i=j$ and $0$
    otherwise. The optimal integral cost is $1$, whereas the feasible
    fractional solution $x_{a_i} = 1/\ell$ for all $i \in [\ell]$
    has cost $1/\ell$, yielding a gap of $\Omega(\ell)$.

    For the upper bound, define $c(a) = \sum_{i=1}^{\ell} c_i(a)$
    for every $a \in L$. Any feasible fractional solution $x$ to
    $\mccaplp$ is also feasible for the $\mathrm{CAP}$ LP
    under $c$, with
    $c(x) \leq \ell \cdot \max_{i \in [\ell]} c_i(x)$.
    The constant integrality gap of the canonical LP formulation for $\mathrm{CAP}$
    together with \cref{obs:mc-to-nonmc}, yields an $O(\ell)$
    upper bound.
\end{remark}

A similar $\Theta(\ell)$ integrality gap for multi-criteria survivable network design was previously observed by \cite{LADDHA2022536FairNetworkDesign}.

\subsection{Hardness of Approximation for $\mccap$}\label{sec:mccap-hardness} 
In this section, we show that our $O(\log \ell)$-approximation
guarantee for $\mccap$ is nearly tight. Specifically, we prove that, for any fixed $\varepsilon > 0$, $\mccap$ admits no polynomial-time $O(\log^{1-\varepsilon}\ell)$-approximation unless every problem in $\NP$ can be solved in deterministic quasi-polynomial time. This hardness holds even when the underlying graph is a star. 

\begin{theorem}
    \label{thm:hardness-mccap}
    For any fixed $\varepsilon > 0$, $\mccap$ with $\ell$ criteria admits no polynomial-time $O(\log^{1-\varepsilon}\ell)$-approximation, even when $G$ is a star, unless $\NP \subseteq \bigcup_{c \geq 1} \mathsf{DTIME}\left(n^{(\log n)^c}\right)$.
\end{theorem}

\begin{proof}
    We give an approximation-preserving reduction from the
    min-max Robust Representatives Selection Problem
    ($\mathrm{RRSP}$)
    \cite{Deineko2013ComplexityAI,kasperski2015approximability}.

    Consider an $\mathrm{RRSP}$ instance with a set of tools
    $T$ partitioned into $T_1,\ldots,T_k$ and $\ell$ cost
    functions $c_1,\ldots,c_\ell$. The goal is to select
    exactly one tool from each $T_q$ while minimizing
    $\max_{i\in[\ell]}\sum_{t\in X}c_i(t)$.

    Construct a star $G$ with center $r$ and leaves
    $v_1,\ldots,v_k$. For each tool $t\in T_q$, introduce
    a link $a_t$ between $r$ and $v_q$, with
    $c_i(a_t)=c_i(t)$ for all $i\in[\ell]$.

    By construction, a link $a_t$ covers the minimum cut
    corresponding to the edge $rv_q$ if and only if
    $t\in T_q$. Thus, a set of links is feasible for
    $\mccap$ if and only if the corresponding tools
    include at least one from each $T_q$.
    Since costs are nonnegative, any such solution can
    be restricted to exactly one tool per group without
    increasing its cost. Hence, the two instances have
    the same optimal value, and any approximation for
    $\mccap$ yields the same approximation for
    $\mathrm{RRSP}$.

    Kasperski et al.~\cite[Theorem~3]{kasperski2015approximability}
    showed that, for any fixed $\varepsilon>0$,
    $\mathrm{RRSP}$ admits no polynomial-time
    $O(\log^{1-\varepsilon}\ell)$-approximation unless
    $\NP$ admits quasi-polynomial-time algorithms.
    Hence, the theorem follows.
\end{proof}

\section{An $O(\log k \log \ell)$-Approximation Algorithm for MC-$k$-ECSS}\label{sec:ecss}
In this section, we apply our techniques to the MC-$k$-ECSS problem. Specifically, we show that combining our reduction from $\mccap$ to $\mcutap$ and our rounding scheme for $\mcutap$ with a simple iterative procedure yields an $O(\log k \log \ell)$-approximation algorithm for MC-$k$-ECSS. 

\begin{theorem}
    \label{thm:k-ecss}
    Consider a $k$-edge-connected graph $G = (V,E)$ and $\ell$ cost functions $\{  c_i : E \rightarrow \mathbb R_{\geq 0}\}_{i \in [\ell]}$. There exists an $O(\log k \log \ell)$ approximation algorithm for the  MC-$k$-ECSS problem on $G$ with cost functions $\{ c_i \}_{i \in [\ell]}$ that runs in polynomial time.
\end{theorem}

\begin{proof}
    Our algorithm proceeds in $k$ iterations, constructing a $j$-edge-connected spanning subgraph $H^{(j)}$ in iteration $j$. We then augment $H^{(j)}$ to $H^{(j+1)}$, increasing the objective value in expectation by at most $\frac{O(\log \ell)}{k-j} \cdot \opt$. Summing over all iterations yields an $O(\log k \log \ell)$-approximation algorithm for MC-$k$-ECSS.

    Let $\opt$ denote the optimal cost of the multi-criteria $k$-edge-connected spanning subgraph of $G$ with given cost functions $c_1, \cdots, c_\ell$. Without loss of generality, we assume there is some value $B \le \opt$ such that, for all $e \in E$ and $i \in [\ell]$, $c_i(e) \leq B$. 
    This can be achieved by a similar guessing procedure to \cref{sec:analysis-mccap} which iterates over all edge weights.
    Consider the standard LP relaxation of the  MC-$k$-ECSS problem,

    \mcecssLPBox
    
    Let $x$ be an optimal solution to $\mcecsslp$. For any $W \subseteq E$, let $x^W$ denote the solution obtained from $x$ by setting $x^W(e)=0$ for all $e \in W$, while leaving all other entries unchanged. For a subgraph $H$, we write $x^H := x^{E(H)}$.

    


    We initialize $H^{(1)}$ using the $O(\log \ell/\log\log \ell)$-approximation algorithm of \cite{VondrakChekuriZenklusenMinMax} for the multi-criteria spanning tree problem, ensuring $\mathrm{cost}(H^{(1)}) \leq O(\log \ell)\opt$. Following the standard connectivity-augmentation framework (see, e.g., \cite{WilliamsonGoemansMihailVazirani1995,DoryGhaffari2023}), we construct a $(j+1)$-edge-connected spanning subgraph $H^{(j+1)}$ from $H^{(j)}$. If $H^{(j)}$ is already $(j+1)$-edge-connected, we set $H^{(j+1)}=H^{(j)}$. Otherwise, consider $z^{(j)} := x^{H^{(j)}}/(k-j)$. This is a feasible solution to the $\mccap$ LP on $H^{(j)}$ with available links $E \setminus E(H^{(j)})$: every minimum cut $S$ of $H^{(j)}$ has size $j$, and hence $x^{H^{(j)}}(\delta(S)) \geq k-j$. Thus, $z^{(j)}$ covers every minimum cut of $H^{(j)}$.

    To construct $H^{(j+1)}$, we apply our $O(\log \ell)$-approximation algorithm for $\mccap$ to round $z^{(j)}$ on $H^{(j)}$ with available links $E \setminus E(H^{(j)})$. Let $T^{(j)}$ denote the set of edges selected by the algorithm, and set $H^{(j+1)} = H^{(j)} \cup T^{(j)}$. 

    
    We now analyze the cost of the resulting solution. A direct application of \cref{thm:rounding} to each iteration separately may introduce an additional factor of $k$ in the approximation guarantee, due to its additive dependence on the maximum link cost $B$. To avoid this loss, we analyze the costs incurred across all iterations jointly.

    For each iteration $j \geq 1$, the solution $T^{(j)}$
    consists of four sets: $T^{(j)} =
    R_{\mathcal{L}}^{(j)} \cup R_{\mathcal{R}}^{(j)}
    \cup C_{\mathcal{L}}^{(j)} \cup C_{\mathcal{R}}^{(j)}$.
    Here, $R_{\mathcal{L}}^{(j)}$ and $R_{\mathcal{R}}^{(j)}$
    contain the edges selected during randomized rounding,
    while $C_{\mathcal{L}}^{(j)}$ and $C_{\mathcal{R}}^{(j)}$
    contain those added during the completion step.
    The subscripts $\mathcal{L}$ and $\mathcal{R}$ correspond
    to cuts covered from the left and right, respectively,
    by the fractional solution.
    First, we bound the cost of $C^{(j)}_{\mathcal{Q}}$ for $\mathcal{Q} \in \{\mathcal{L}, \mathcal{R} \}$. By \cref{clm:patching-bound},
    \begin{equation}
        \label{eq:ecss-patching-cost}
        \mathbb E\left[\mathrm{cost}\left(C_\mathcal{Q}^{(j)}\right)\right] \leq 20 \lceil \log \ell +1 \rceil \mathrm{cost}\left(z^{(j)}\right) \leq \left(\frac{20\lceil\log \ell +1 \rceil}{k-j}\right) \mathrm{cost}\left(x\right),
    \end{equation}
    where the last inequality holds as $x^{H^{(j)}} \leq x$ and the cost functions are non-negative. Note that the extra factor of $10$ compared to \cref{clm:patching-bound} is due to the fact that in our reduction in \cref{thm:reduction} we use a $\mcutap$ solution that scales the fractional solution up by a factor of $10$. Summing this over all choices of $j$ gives,
    \begin{align}\label{eq:ecss-cumulative-patching}
         \mathbb E\left[\mathrm{cost}\left( \bigcup_{j=1}^{k-1} C_\mathcal{L}^{(j)} \cup C_\mathcal{R}^{(j)}\right)\right] 
         &\leq \sum_{\mathcal{Q \in \{ \mathcal{L}, \mathcal{R} \}}} \sum_{j=1}^{k-1}   \mathbb E\left[\mathrm{cost}\left(C_\mathcal{Q}^{(j)}\right)\right]
         \leq 40 \lceil \log \ell + 1 \rceil H_k \cdot \mathrm{cost}(x), 
    \end{align}
    where the last line follows from \eqref{eq:ecss-patching-cost}.


    We now bound the cumulative cost of edges selected during randomized rounding. For each $1 \leq j < k$, edge $e$, and $\mathcal{Q} \in \{\mathcal{L},\mathcal{R}\}$, let $Y_{j,e,\mathcal{Q}}$ be the indicator that $e \in R_{\mathcal{Q}}^{(j)}$. Independently, define Bernoulli random variables $Z_{j,e,\mathcal{Q}}$ with success probabilities $z_{j,e,\mathcal{Q}} := \min\left\{1, \frac{10\lceil \log \ell + 1 \rceil x_e}{k-j}
    \right\}$. Let $I_{e,j}$ indicate whether $e$ is eligible for augmentation in iteration $j$, namely, $I_{e,j}=1$ if $e\notin E(H^{(j)})$ and $H^{(j)}$ is not $(j+1)$-edge-connected, and $0$ otherwise. Since $I_{e,j}$ is determined before the randomized rounding in iteration $j$, we may couple these random variables so that $Y_{j,e,\mathcal{Q}}=I_{e,j}Z_{j,e,\mathcal{Q}}$. Thus, for any $i\in[\ell]$, we have
    \begin{align*}
         \mathrm{cost}_i \left( \bigcup_{j=1}^{k-1} (R_{\mathcal{\mathcal{L}}}^{(j)} \cup R_{\mathcal{\mathcal{R}}}^{(j)}) \right)  \leq \sum_{j=1}^{k-1} \sum_{e \in E} \sum_{\mathcal{Q} \in \{\mathcal{L}, \mathcal{R} \}} c_i(e) Z_{j,e, \mathcal{Q}} I_{e,j}\leq \sum_{j=1}^{k-1} \sum_{e \in E} \sum_{\mathcal{Q} \in \{\mathcal{L}, \mathcal{R} \}} c_i(e) Z_{j,e, \mathcal{Q}},
    \end{align*}
    where the last inequality follows as the costs are non-negative and $Z_{j,e, \mathcal{Q}} I_{e,j}  \leq Z_{j,e, \mathcal{Q}}$.


For each $i \in [\ell]$, define $D_i' := \mathrm{cost}_i\left( \bigcup_{j=1}^{k-1}(R_{\mathcal{L}}^{(j)} \cup R_{\mathcal{R}}^{(j)}) \right)$ and $D_i := \sum_{j=1}^{k-1} \sum_{e \in E} \sum_{\mathcal{Q} \in \{\mathcal{L},\mathcal{R}\}} c_i(e) Z_{j,e,\mathcal{Q}}$. We couple the randomized rounding decisions with the independent Bernoulli variables $Z_{j,e,\mathcal Q}$ so that $Y_{j,e,\mathcal Q} = I_{e,j}Z_{j,e,\mathcal Q} \leq Z_{j,e,\mathcal Q}$. Consequently, since all costs are nonnegative, $D_i' \leq D_i$ holds for every realization. Thus, it suffices to bound $\E[\max_{i\in[\ell]} D_i]$. As in the proof of \cref{clm:random-rounding-bound}, applying the log-sum-exp inequality followed by Jensen's inequality yields
    \begin{equation}
        \label{eq:ecss-max-to-log-sum}
        \E \left[\max_{i \in [\ell]} D_i\right] \leq \E \left[ B \log\left( \sum_{i \in [\ell]} \exp \left( \frac{ D_i}{B} \right) \right) \right] \leq B \log\left( \sum_{i \in [\ell]} \E \left[  \exp \left(  \frac{D_i}{B} \right) \right] \right)
    \end{equation}
    Now, fix any $i \in [\ell]$. Since the random variables $\{Z_{j,e,\mathcal Q}\}_{j,e,\mathcal Q}$ are mutually independent, we have
    \begin{align}
        \label{eq:ecss-bound-indiv-exp}
        \E \left[  \exp \left( \frac{D_i}{B} \right) \right] = \E \left[  \exp \left( \sum_{j=1}^{k-1} \sum_{e \in E} \sum_{\mathcal{Q} \in \{\mathcal{L}, \mathcal{R} \}} \frac{c_i(e)}{B} Z_{j,e, \mathcal{Q}} \right) \right] =  \prod_{j=1}^{k-1}\prod_{e \in E} \prod_{\mathcal{Q} \in \{\mathcal{L}, \mathcal{R} \}} \E \left[ \exp \left( \frac{c_i(e) }{B} Z_{j,e, \mathcal{Q}} \right) \right]
    \end{align}
    Now, for any $1\le j\le k-1$, $e \in E$, and $\mathcal{Q \in \{ \mathcal{L}, \mathcal{R} \}}$,
    \begin{equation}
    \label{eq:ecss-indiv-exp-bound}
        \E \left[ \exp \left( \frac{c_i(e) }{B} Z_{j,e, \mathcal{Q}} \right) \right] =\exp \left( \frac{c_i(e) }{B} \right) z_{j,e, \mathcal{Q}} + (1- z_{j,e, \mathcal{Q}} ).
    \end{equation}
    Since by assumption $\frac{c_i(e) }{B} \leq 1$, using $e^z \leq 1+ 2z$ for $z \in [0,1]$ in \eqref{eq:ecss-indiv-exp-bound},
    \begin{equation*}
         \E \left[ \exp \left( \frac{c_i(e) }{B} Z_{j,e, \mathcal{Q}} \right) \right] \leq \left(1 + \frac{2c_i(e)}{B} \right) z_{j,e, \mathcal{Q}} +1 - z_{j,e, \mathcal{Q}} = 1+\frac{2c_i(e)}{B} z_{j,e, \mathcal{Q}} \leq \exp\left( \frac{2c_i(e)}{B} z_{j,e, \mathcal{Q}} \right)
    \end{equation*}
    where the last inequality uses $1+z \leq e^z$ for all $ z\in \mathbb R$.  Substituting this back in \eqref{eq:ecss-bound-indiv-exp} gives,
    \begin{align}
        \E \left[  \exp \left( \frac{D_i}{B} \right) \right] 
        &\leq \prod_{j=1}^{k-1}\prod_{e \in E} \prod_{\mathcal{Q} \in \{\mathcal{L}, \mathcal{R} \}} \exp\left( \frac{2c_i(e)}{B} z_{j,e, \mathcal{Q}} \right) \notag \\ 
        &= \exp\left( \frac{2\sum\limits_{j=1}^{k-1} \sum\limits_{e \in E} \sum\limits_{\mathcal{Q} \in \{\mathcal{L}, \mathcal{R} \}} c_i(e) z_{j,e, \mathcal{Q}}}{B} \right) \notag \\
        &\leq \exp\left( \frac{20 \lceil  \log \ell + 1 \rceil \sum\limits_{j=1}^{k-1} \sum\limits_{e \in E} \sum\limits_{\mathcal{Q} \in \{\mathcal{L}, \mathcal{R} \}} c_i(e) \frac{ x_e}{k-j}}{B} \right). \label{eq:ecss-expected-S-second}
    \end{align}
    where the last inequality follows from $z_{j,e, \mathcal{Q}} = \min(1, \frac{10\lceil  \log \ell + 1 \rceil x_e}{k-j})$. Note that,
    \begin{align}
        \label{eq:ecss-sum-bound}
        \sum\limits_{j=1}^{k-1} \sum\limits_{e \in E} \sum\limits_{\mathcal{Q} \in \{\mathcal{L}, \mathcal{R} \}} c_i(e) \frac{ x_e}{k-j} = \sum_{e \in E} \left(\left(\sum\limits_{\mathcal{Q} \in \{\mathcal{L}, \mathcal{R} \}}\sum_{j=1}^{k-1} \frac{1}{k-j} \right) x_e c_i(e) \right) \leq 2H_{k} \sum_{e\in E} x_e c_i(e)
    \end{align}
    Finally, substituting \eqref{eq:ecss-sum-bound} in \eqref{eq:ecss-expected-S-second} and using $\sum_{e \in E} c_i(e) {x}_e \leq \mathrm{cost}(x)$,
    \begin{equation}
        \label{eq:ecss-individual-term-bound-S-expected}
        \E \left[  \exp \left( \frac{D_i}{B} \right) \right] \leq \exp\left(40 \lceil \log \ell + 1 \rceil H_k \frac{\mathrm{cost}(x)}{B} \right).
    \end{equation}
    Applying \eqref{eq:ecss-individual-term-bound-S-expected} in \eqref{eq:ecss-max-to-log-sum} yields,
    \begin{align*}
        \E \left[\max_{i \in [\ell]} D_i'\right] 
        \leq \E \left[\max_{i \in [\ell]} D_i\right] 
        \leq B \log\left( \sum_{i \in [\ell]} \E \left[  \exp \left(  \frac{D_i}{B} \right) \right] \right)
        &\leq B \log\left( \sum_{i \in [\ell]}  \exp\left(40 \lceil \log \ell + 1 \rceil H_k \frac{\mathrm{cost}(x)}{B} \right)\right) \notag \\
        &= B \log \ell + 40 \lceil \log \ell + 1 \rceil H_k \cdot {\mathrm{cost}(x)} 
    \end{align*}
    Therefore,  
    \begin{align}
        \label{eq:ecss-cumulative-rounding}
        \mathbb E \left[  \mathrm{cost} \left( \bigcup_{j=1}^{k-1} (R_{\mathcal{\mathcal{L}}}^{(j)} \cup R_{\mathcal{\mathcal{R}}}^{(j)}) \right) \right] \leq B \log \ell + 40 \lceil \log \ell + 1 \rceil H_k \cdot {\mathrm{cost}(x)} \leq O(\log \ell \log k) \opt.
    \end{align}
     The final solution of our algorithm is $H^{(k)} = H^{(1)} \cup \left( \bigcup_{j=1}^{k-1} (R_{\mathcal{\mathcal{L}}}^{(j)} \cup R_{\mathcal{\mathcal{R}}}^{(j)}) \right) \cup \left( \bigcup_{j=1}^{k-1} (C_{\mathcal{\mathcal{L}}}^{(j)} \cup C_{\mathcal{\mathcal{R}}}^{(j)}) \right)$. Thus,
     \begin{align*}
         \mathbb E \left[  \mathrm{cost} \left( H^{(k)} \right) \right] &\leq \mathbb E \left[ \mathrm{cost}(H^{(1)}) + \mathrm{cost}\left( \bigcup_{j=1}^{k-1} (R_{\mathcal{\mathcal{L}}}^{(j)} \cup R_{\mathcal{\mathcal{R}}}^{(j)}) \right) + \mathrm{cost}\left( \bigcup_{j=1}^{k-1} (C_{\mathcal{\mathcal{L}}}^{(j)} \cup C_{\mathcal{\mathcal{R}}}^{(j)}) \right) \right] \\
         &\leq   O(\log \ell \log k) \cdot (\opt + \mathrm{cost}(x)+ B ),
     \end{align*}
     where the last line follows from \eqref{eq:ecss-cumulative-patching}, \eqref{eq:ecss-cumulative-rounding}, and that $\mathrm{cost}(H^{(1)}) \leq O(\log \ell) \cdot \opt$. Finally, since $x$ was the optimal solution to the LP relaxation of MC-$k$-ECSS and $B \leq \opt$, 
     \begin{align*}
         \mathbb E\left[ \mathrm{cost}\left(H^{(k)}\right)\right] \leq O(\log \ell \log k) \cdot \opt
     \end{align*}
     This concludes the proof of the $O(\log \ell \log k)$ approximation guarantee of our algorithm. Note that the correctness and polynomial runtime of this algorithm follow directly from the correctness and runtime analysis of our algorithm for $\mccap$.
    
\end{proof}

\section*{AI Disclosure}

AI tools were used to generate figures and assist with writing of this draft. They also contributed to the technical development of this work. In particular, during a conversation with GPT-6 Astra, the authors provided initial proof ideas, which the model developed into the partition framework, leading to a proof of the reduction from $\mccap$ to $\mcutap$.

The proof presented in this manuscript differs from the model-generated proof and the authors redeveloped all AI-generated content and wrote the entire manuscript to improve readability and exposition. Nevertheless, the final proof builds on ideas that emerged from the interaction with the model. The authors have verified the final arguments and take full responsibility for their correctness and for the content of the manuscript.

\bibliographystyle{alpha}
\bibliography{references}

\appendix
\crefalias{section}{appendix}
\crefname{section}{Appendix}{Appendices}
\Crefname{section}{Appendix}{Appendices}

\section{Reducing $\mccap$ to $\mcrap$}
\label{app:reduce-to-rap}

we show that rounding a $\mccap$ instance can be reduced to rounding a $\mcrap$ instance. Similar techniques for the single-criteria version of the Connectivity Augmentation Problem have been used previously, and this reduction is fairly well-known \cite{WaldoGrandoniAmeliCycleAugmentation,TraubZenklusenWeightedCAP}.  However, these prior applications dealt with a single objective function; although handling multiple objectives does not add any difficulty to the reduction, we state it here for the sake of completeness. Before proceeding, we need to define a cactus.

\begin{definition}
    An undirected graph $G = (V,E)$ is called a cactus if every edge $e \in E$ is exactly a part of one cycle in $G$.
\end{definition}

Dinitz, Karzanov, and Lomonosov \cite{DKL76} showed that all min-cuts of any graph $G$ can be represented by a single cactus. 

\begin{theorem}[\cite{DKL76}]
\label{thm:cactus-represent}
For any undirected graph $G$, one can construct, in polynomial time,
a cactus ${G'}$ with a mapping $\varphi: V(G) \rightarrow V(G')$ satisfying the following properties.
\begin{itemize}
    \item For any min-cut $S$ of $G$, $S = \varphi^{-1}(S')$ for some min-cut $S'$ of $G'$.
    \item For any min-cut $S'$ of $G'$, there exists a min-cut $S$ of $G$ such that $S = \varphi^{-1}(S')$.
\end{itemize}
where for a subset $S' \subseteq V(G')$, $\varphi^{-1}(S') = \{ v \in V(G) \mid \varphi(v) \in S'\}$.
\end{theorem}

Using \cref{thm:cactus-represent}, one can assume without loss of generality that the graph $G$ in a $\mccap$ instance is a cactus. To continue, we use the following Lemma proved in both \cite{DKL76}
 and \cite{CTZ21}.
 
\begin{lemma}[\cite{DKL76,CTZ21}]
    \label{lem:cap-to-rap}
       Let $G = (V,E)$ be a cactus and $\mathcal{W}$ be a family of the min-cuts of $G$. Pick some $r \in V$ as the root of $G$. Then, there exists a cycle $H = (V, E')$ where for each $S \in \mathcal{W}$, $|\delta_{E'}(S)| = 2$. I.e. the projection of all min-cuts in $\mathcal{W}$ on $H$ is a min-cut.
\end{lemma}

Finally, to see the reduction from $\mccap$ to $\mcrap$, simply apply \cref{thm:cactus-represent} and \cref{lem:cap-to-rap} to obtain a graph $H$ from our original graph $G$ where for each $W \in \mathcal{W}$, $W$ is also a min-cut of $H$. Since $H$ uses the same set of vertices as $G$, we can keep the links $L$ exactly the same; moreover, $\mathcal{W}$ is a family of min-cuts of $G$, and by definition, a family of min-cuts of $H$ too. It is straightforward to see that given a fractional solution for $\mccap$ on $G$, one can round the same solution for $\mcrap$ on $H$ to obtain a feasible integral solution for the $\mccap$ on $G$. This concludes the reduction.

\end{document}